\documentclass[a4paper,11pt]{article}
\usepackage{jcappub} 
\usepackage{lineno}

\usepackage[normalem]{ulem}
\usepackage{graphicx}
\usepackage{dcolumn}
\usepackage{bm}
\usepackage{graphics}
\usepackage{slashed}
\usepackage{amssymb}
\usepackage{natbib}
\usepackage{amsmath}
\usepackage{url}
\usepackage{here}
\usepackage[usenames,dvipsnames,svgnames,table]{xcolor}
\usepackage{color}
\usepackage{xcolor}
\usepackage{tabularx}
\usepackage{hyperref}
\usepackage{enumitem}
\usepackage{soul}

\def\ud{{\rm d}}

\usepackage[]{inputenc}
\usepackage[T1]{fontenc}
\usepackage{hyperref}
\usepackage{color}
\usepackage{lmodern}
\usepackage{fullpage}
\usepackage[thinlines]{easytable}

\newcommand{\be}{\begin{equation}}
\newcommand{\ee}{\end{equation}}
\newcommand{\bea}{\begin{eqnarray}}
\newcommand{\eea}{\end{eqnarray}}

\newcommand{\m}{\mu}
\newcommand{\n}{\nu}

\arxivnumber{2307.13041} 
\title{Well-posed evolution of field theories with anisotropic scaling: \\
the Lifshitz scalar field in a black hole space-time}

\author{M. E. Rubio${}^{1,2}$,}
\author{\'A. D. Kov\'acs${}^{1,2}$,}
\author{M. Herrero-Valea${}^{3}$,}
\author{M. Bezares${}^{1,2,4,5}$ and}
\author{E. Barausse${}^{1,2}$}

\affiliation{${}^{1}$SISSA, Via Bonomea 265, 34136 Trieste, Italy and INFN Sezione di Trieste}
\affiliation{${}^{2}$IFPU - Institute for Fundamental Physics of the Universe, Via Beirut 2, 34014 Trieste, Italy}
\affiliation{${}^{3}$Institut de Fisica d’Altes Energies (IFAE), The Barcelona Institute of Science and Technology, Campus UAB, 08193 Bellaterra (Barcelona) Spain}
\affiliation{${}^{4}$Nottingham Centre of Gravity, University of Nottingham, University Park, Nottingham, NG7 2RD, UK}
\affiliation{${}^{5}$School of Mathematical Sciences, University of Nottingham, University Park, Nottingham, NG7 2RD, UK}

\emailAdd{mrubio@sissa.it}

\abstract{Partial differential equations exhibiting an anisotropic scaling between space and time -- such as those of Ho\v rava-Lifshitz gravity -- have a dispersive nature. They contain higher-order spatial derivatives, but remain second order in time. This is inconvenient for performing long-time numerical evolutions, as standard explicit schemes fail to maintain convergence unless the time step is chosen to be very small. In this work, we develop an implicit evolution scheme that does not suffer from this drawback, and which is stable and second-order accurate. As a proof of concept, we study the numerical evolution of a Lifshitz scalar field on top of a spherically symmetric black hole space-time. We explore the evolution of a static pulse and an (approximately) ingoing wave-packet for different strengths of the Lorentz-breaking terms, accounting also for the effect of the angular momentum eigenvalue and the resulting effective centrifugal barrier. Our results indicate that the dispersive terms produce a cascade of modes that accumulate in the region in between the Killing and universal horizons, indicating a possible instability of the latter.}

\begin{document}
\maketitle
\flushbottom

\section{Introduction}
\label{sec-Intro}
In 2009, Petr Ho\v rava proposed a power-counting renormalizable ultraviolet (UV) completion of General Relativity (GR) \cite{Horava:2009uw} by endowing space-time with a preferred foliation in space-like hypersurfaces. This comes at the cost of reducing the local symmetry of the theory down to \textit{foliation preserving diffeomorphisms} (FDiff), namely
\begin{equation}\label{eq:FDiff}
    t\rightarrow t'(t), \quad x^{i}\rightarrow x'^i(t,x),
\end{equation}
where $t$ is the preferred time direction along the foliation, and $x^i$ is a chart parametrizing the orthogonal leafs.

Action functionals invariant under \eqref{eq:FDiff} allow for time and spatial derivative operators to be included \textit{independently}, thus breaking local Lorentz invariance (LLI), but also attaining a faster convergence of UV integrals in Quantum Field Theory \cite{Fujimori:2015mea,Anselmi:2007ri}. Ostrogradsky ghosts are avoided simply by fixing the number of time derivatives to two, with the action nonetheless still admitting higher spatial derivative operators. Power-counting renormalizability is achieved in four-dimensional gravity in this way by adding up to six spatial derivatives in a FDiff invariant way. 

A remarkable consequence of this construction is that, due to the different order of derivatives along distinct directions, the UV dynamics of the theory becomes invariant under an \textit{anisotropic Lifshitz scaling} with critical exponent $z$,
\begin{align}\label{eq:lifshitz_scaling}
    t\rightarrow b^z t,\quad x^i\rightarrow b x^i, \quad b\in\mathbb{R},
\end{align}
where the number of spatial derivatives in the UV is $2z$. This is the scaling symmetry  that allows for the modified power counting of UV divergences leading to renormalizability -- with $z=3$ in the case of gravity in four space-time dimensions.

Since its formulation, countless works have  explored the consequences of Ho\v rava's proposal \citep{MHV-review-HG}. A far from exhaustive list includes: understanding the infra-red (IR) dynamics of the theory \cite{Blas:2010hb,Blas:2009qj}; the structure of constraints \cite{Donnelly:2011df,Bellorin:2022qeu}; its perturbative UV dynamics \cite{Barvinsky:2015kil,Barvinsky:2017kob,Barvinsky:2019rwn,Barvinsky:2021ubv,Radkovski:2023cew, Griffin:2017wvh,Benedetti:2013pya}; the search for black-hole \cite{Barausse:2011pu,Barausse:2012qh,Barausse:2013nwa,Berglund:2012bu}, regular \cite{Lara:2021jul,Mazza:2023iwv}, and cosmological solutions \cite{Mukohyama:2010xz}; the interaction with matter fields \cite{Pospelov:2010mp,Blas:2014aca}, and low-energy signatures in observations \cite{Audren:2014hza,Barausse:2019yuk,Bettoni:2017lxf,Bonetti:2015oda,Cornish:2017jml,EmirGumrukcuoglu:2017cfa,Yagi:2013ava,Yagi:2013qpa,Franchini:2021bpt,Gupta:2021vdj}, among others. However,  despite incredible advances, the problem of classically solving and dynamically evolving equations of motion exhibiting the scaling \eqref{eq:lifshitz_scaling} is still mostly unexplored, 
except in the simplest
 situations -- such as for static and spherically symmetric space-times, or in the perturbative limit \cite{Oshita:2021onq}, where the coefficients accompanying the higher derivative terms are small enough.

Due to the presence of higher-order spatial derivatives, equations exhibiting an anisotropic scaling have a dispersive nature. While in GR the dispersion relation of massless fields takes a universal form $p^\m p_\m=\omega^2-k^2$=0, with $p^\m=(\omega,k^i)$ the four-momentum of the field\footnote{Hereinafter we use a mostly minus signature for the metric. Greek indices denote the full set of space-time components, while Latin ones are restricted to spatial directions only.}; the presence of higher spatial derivatives generically modifies it to
\begin{align}\label{eq:dispersion_generic}
    \omega^2=k^2+\Theta k^{2z},
\end{align}
where we have assumed the addition of an operator with $2z$ spatial derivatives, controlled by a dimensionful coupling $\Theta$. Standard explicit numerical methods tend to develop instabilities when evolving equations of motion exhibiting this behaviour \citep{NumRecipes}. For hyperbolic systems, a Von Neumann analysis of explicit methods -- such as the standard Runge-Kutta algorithms -- shows that the corresponding evolution is \textit{stable} as long as the dimensionless \textit{Courant number}
\begin{equation}
    C=\frac{v_{\text{m}} \Delta t}{\Delta x}
\end{equation}
satisfies $C\leq C_{\text{max}}$, where $\Delta t$ and $\Delta x$ are the time and space steps used for discretization, and $v_{\text{m}}$ is the maximum speed of propagation allowed by the equation. The value of $C_{\text{max}}$ generically depends on the method, but for explicit schemes it usually takes the value $C_{\text{max}} = 1$. 

In the dispersive case instead, a similar von Neumann analysis yields a more restrictive condition for stability, of the form
\begin{equation}\label{cfl-dispersive}
    \frac{\Delta t}{(\Delta x)^p}\leq \alpha, 
\end{equation}
where $p>1$ and $\alpha > 1$ depend both on the equation and on the scheme. This is very inconvenient if long-time simulations with high spatial resolution are sought, as the time step needed to keep the numerical error bounded has to be chosen very small.

In this work, we circumvent this issue by providing a robust numerical scheme that allows for evolving initial value problems exhibiting a UV Lifshitz scaling \eqref{eq:lifshitz_scaling} in non-trivial curved backgrounds. We do this by building an implicit numerical method, where the usual finite difference discretization of derivative operators is substituted by averages between two consecutive time steps. This results in an integration scheme whose stability analysis leads to no bounds for the time and spatial steps, allowing for long-time evolution regardless of the desired spatial resolution. We show the robustness of our methods by studying a situation of physical interest -- the scattering of a Lifshitz scalar field on top of a static, spherically symmetric, and asymptotically flat black hole space-time, which we choose to be a solution to the field equations of Ho\v rava gravity at low energies. 

This paper is organized as follows. In section \ref{sec:scalar_field} we introduce the formulation of the Lifshitz scalar field theory both in flat and curved space-times, obtaining the equations of motion and discussing their dispersive character. An implicit method for the numerical evolution of equations of this kind is introduced in section \ref{sec:numerical}, while an application to the Lifshitz scalar field around a Lorentz-violating black hole is discussed in detail in section \ref{lifshitz-field-BH-spacetime}. Finally, we show our results in section \ref{sec:results}, drawing conclusions and future research directions in section \ref{sec:conclusions}.


\section{The Lifshitz scalar field}\label{sec:scalar_field}

Field theories exhibiting the anisotropic scaling \eqref{eq:lifshitz_scaling} have been known for quite a long time. The simplest among them is the Lifshitz scalar field, used in studies of quantum phase transitions in various strongly correlated systems \cite{Ardonne:2003wa}. Its action in a $3+1$ flat space-time with $z=3$ reads
\begin{align}\label{eq:Lifshitz_flat}
    S=\frac{1}{2}\int dt d^3x \left((\partial_t \phi)^2-\phi (-\partial^2) \phi-\frac{c_2}{\Lambda^2}\phi (-\partial^2)^2 \phi -\frac{c_3}{\Lambda^4}\phi (-\partial^2)^3 \phi\right),
\end{align}
where $\partial^2=\partial_i \partial^i$, $\Lambda$ is an energy scale, all the $c_i$ are dimensionless, and we have fixed the speed of light to one, which is always possible by rescaling the spatial coordinates. Note that the action \eqref{eq:Lifshitz_flat} contains all possible parity-preserving spatial derivative operators with $z\leq 3$, while keeping two time derivatives to avoid Ostrogradsky ghosts. This structure allows for a dynamical flow of $z$ with energy, which changes from $z=1$ in the deep IR to $z=3$ in the UV, thus mimicking the behavior expected in Ho\v rava gravity \cite{Horava:2009uw}. We have omitted a possible potential for the scalar field, since our goal is to focus on the effect of the higher derivative operators.

The equation of motion obtained from \eqref{eq:Lifshitz_flat} reads
\begin{align}\label{eq:flat_eom}
   \left[ \partial_t^2 -\partial^2 +\frac{c_2}{\Lambda^2} (\partial^2)^2 -\frac{c_3}{\Lambda^4}(\partial^2)^3\right] \phi=0,
\end{align}
which leads to a modified dispersion relation of the form \eqref{eq:dispersion_generic}, that is
\begin{align}
    \omega^2 = k^2 +\frac{c_2 k^4}{\Lambda^2}+\frac{c_3 k^6}{\Lambda^4}.
\end{align}
This becomes explicitly Lorentz-violating for momentum scales $k \gtrsim\Lambda$. Stability of the degrees of freedom demands $\omega^2>0$, which in particular requires the parabola $y(\hat{x})=1+c_2 \hat{x}+c_3\hat{x}^2$ to be concave, leading to $c_3>0$. Regarding the value of $c_2$, $\omega^2$ is positive for all $c_2>0$, while for $c_2<0$ we must require the minimum of the parabola to be above zero, corresponding to $c_2^2\leq 4 c_3$. Putting both conditions together, we get $c_2>-2\sqrt{c_3}$.

In order to extend this structure to curved space-times, we couple the scalar field to Ho\v rava gravity \cite{Horava:2009uw}, which implements the anisotropic scaling \eqref{eq:lifshitz_scaling} through a foliation in co-dimension one space-like hypersurfaces. This is described by supplementing the space-time curved metric $g_{\m\n}$ with a hypersurface-orthogonal unit-norm and timelike vector $U^\m$, called the \emph{\ae ther} \cite{Jacobson:2000xp}. Time derivatives are identified with derivatives along $U^\m$, while the spatial Laplacian $\partial^2$ is replaced by $\Delta_\gamma=\gamma^{\m\n}\nabla_\m\nabla_\n$, where $\gamma^{\m\n}$ is the orthogonal projector $\gamma^{\m\n}=-g^{\m\n}+U^{\m}U^{\n}$. The action thus takes the form \cite{Barvinsky:2017mal}
\begin{align}
    S_{\phi}=\frac{1}{2}\int dt d^3 x\sqrt{|g|}\left(({\cal L}_U \phi)^2-\phi (-\Delta_\gamma) \phi-\frac{c_2}{\Lambda^2}\phi (-\Delta_\gamma)^2 \phi -\frac{c_3}{\Lambda^4}\phi (-\Delta_\gamma)^3 \phi\right),
\end{align}
where ${\cal L}_U$ is the Lie derivative along $U^\m$. The corresponding equation of motion is
\begin{align}\label{eq:equation_final}
     \left[ \square+\frac{c_2}{\Lambda^2} \Delta_\gamma^2 -\frac{c_3}{\Lambda^4}\Delta_\gamma^3\right] \phi=0,
\end{align}
where $\Box = \mathcal{L}^2_U+ a^c\nabla_c + K \mathcal{L}_U -\Delta_\gamma$ is the standard four-dimensional diffeomorphism invariant D'Alembertian, with $a_\m=U^\n\nabla_\n U_\m$ the acceleration of the \ae ther, and $K=\gamma^{\m\n}K_{\m\n}$ the trace of the extrinsic curvature of the foliation leafs $K_{\m\n}=(\mathcal{L}_U \gamma_{\m\n})/2$. This equation can be then thought of as a generalization of \eqref{eq:flat_eom} to curved space-times, but also as a useful toy model that captures many of the subtleties brought by the presence of the anisotropic scaling \eqref{eq:lifshitz_scaling}, in particular regarding the integration of the equations of motion with numerical methods.

Equation \eqref{eq:equation_final} admits a \textit{well-posed} initial value problem; that is, for an appropriate initial data set there exists (locally) a unique solution, which depends continuously on the data. This condition holds when $c_3\geq 0$ and $c_2\geq 0$ (or if $c_2<0$ and $|c_2|$ is sufficiently small). The proof of this statement, as well as a more detailed study of the Cauchy problem in Ho\v{r}ava-Lifshitz theories will be explored in more detail in a forthcoming work \cite{HL_cauchy}. Notice that these conditions include the ones required before for stability of the degrees of freedom.

A first attempt at evolving \eqref{eq:equation_final} on top of a black hole space-time was performed in \cite{Oshita:2021onq}, using a finite difference method. Although their scheme is convergent, stability is only guaranteed as long as the time step used for simulations is small enough, which can be problematic when aiming for long run-times. This can be seen from a Von Neumann analysis of the local stability of the numerical scheme \cite{NumRecipes}. Assuming that the coefficients of the equations are locally constant (both in space and time), the eigenmodes of the discretized equations take the form $\phi^n_j = \xi^{n} e^{ikj\Delta x}$, where $\phi^n_j$ is the numerical approximation of the solution at time step $t_n=n \Delta t$ and grid point $x_j:=j\Delta x$, with $\Delta t$ and $\Delta x$ the time and spatial steps used for numerical integration, and $k$ is the frequency of the mode (assuming one spacial dimension for simplicity). The \textit{amplification factor} $\xi^n$ generically depends on $k$ and controls the amplitude of the $k-$eigenmode. Since time evolution of a single eigenmode scales as some power of $\xi^n$, the discretized equations will be stable if and only if $\left|\xi^n\right|\leq 1$ for all $k$. For hyperbolic equations, this analysis implies the well known \textit{Courant-Friedrichs-Lewy} (CFL) condition, which relates $\Delta t$ and $\Delta x$ with the maximum characteristic speed in the equation. For dispersive equations such as \eqref{eq:equation_final} instead, the anisotropic scaling between space and time leads to a stronger upper-bound for the time step $\Delta t$, which scales in the form given in \eqref{cfl-dispersive}, for some $p>1$. This issue complicates the running of long simulations with high spatial resolutions, as computations quickly become very costly. Achieving stable simulations of dispersive equations of the Lifshitz kind thus requires to consider a different approach.

\section{An implicit method for equations with anisotropic scaling}
\label{sec:numerical}

In order to achieve stable, high-resolution numerical evolution, we present here a fully implicit scheme, following the spirit of the well-known Crank-Nicolson method for simple diffusion equations (see \citep{NumRecipes} for references). For the sake of generality, we will diverge for the moment from the specifics of Eq. \eqref{eq:equation_final}, and consider instead a generic partial differential equation (PDE) in $1+1$ dimensions for a scalar field $\psi(t,x)$, of the form
\begin{eqnarray} \label{master-equation}
\partial^2_t\psi &=& F\left(t,x,\psi,\partial_t\psi,\{\partial^j_x\psi\}_{j=1}^{J},\{\partial^k_x\partial_t\psi\}_{k=0}^{K}\right), \nonumber\\
\psi\mathcal{j}_{t=0} &=& \psi_0(x),\\
\partial_t\psi\mathcal{j}_{t=0} &=& \psi_1(x), \nonumber
\end{eqnarray}
where $F$ is a sufficiently smooth function, linear in $\psi$ and in all its derivatives, and $\psi_0,\,\psi_1$ are given initial conditions. The indices $j$ and $k$ span the ranges $1\leq j \leq J$, and $0\leq k \leq K$, with $J\geq 3$, and $K\geq 0$. With these conditions, Eq. \eqref{master-equation} always contains more spatial derivatives than time derivatives. Hence, it corresponds to a PDE with a dispersive nature, leading in fact to a dispersion relation of the form $\omega^2=p(k)$, where $p(k)$ is a polynomial of order greater than or equal to three. Notice that the general problem \eqref{master-equation} does include the Lifshitz field equation \eqref{eq:equation_final}.

We aim to numerically evolve Eq. \eqref{master-equation}, for which we consider a first-order reduction in time, with dynamical variables $\psi$ and $\Pi:=\partial_t\psi$; namely,
\begin{eqnarray}
   \dot{\psi} &=& \Pi \\
   \dot{\Pi} &=& F\left(t,x,\psi,\Pi,\{\partial^j_x\psi\}_{j=1}^{J},\{\partial^k_x\Pi\}_{k=0}^{K}\right).
\end{eqnarray}
We consider a uniform spatial grid of $N$ points $x_i = x_0 + (i-1)\Delta x$, $i = 1,\ldots, N$, with step $\Delta x = L/(N-1)$, being $L$ the length of the  spatial domain. We also discretize time as $t_n = n \Delta t$, $n=0,1,2,\ldots$, with time step $\Delta t$, and define the grid functions as $\psi^n_i := \psi(t_n,x_i)$. 

While for standard explicit methods, derivatives are discretized using finite difference operators, here we replace them with the average of the derivatives in two consecutive time steps. This procedure leads to a fully implicit scheme, meaning that in order to evaluate the solution at a given time step, we need to solve an algebraic equation containing the information of the solution both at the current and previous steps. More specifically, a second-order accurate approximation of the first derivative of $\psi$ at the grid point $i$, centered at a time step $n+\frac{1}{2}$, is computed as
\begin{equation}
    (D\psi)^{n+\frac{1}{2}}_i \equiv \frac{1}{2}\left[\frac{\psi^n_{i+1}-\psi^n_{i-1}}{2\Delta x} + \frac{\psi^{n+1}_{i+1}-\psi^{n+1}_{i-1}}{2\Delta x}\right].  
\end{equation}
For the second derivative, one gets
\begin{equation}
    (D^2\psi)^{n+\frac{1}{2}}_i \equiv \frac{1}{2}\left[\frac{\psi^n_{i+1}-2\psi^n_{i}+\psi^n_{i-1}}{(\Delta x)^2}+\frac{\psi^{n+1}_{i+1}-2\psi^{n+1}_{i}+\psi^{n+1}_{i-1}}{(\Delta x)^2}\right];  
\end{equation}
and similarly for higher-order derivatives. The same average is also performed for evaluating the field $\psi^{n+\frac{1}{2}}_i$, namely
\begin{equation}
    \psi^{n+\frac{1}{2}}_i\equiv \frac{1}{2}\left(\psi^n_i + \psi^{n+1}_i\right).
\end{equation}
For the time derivative, instead, we just keep the forward finite difference
\begin{equation}
    (\partial_t\psi)^{n+1}_i \equiv \frac{\psi^{n+1}_i-\psi^n_i}{\Delta t},
\end{equation}
ensuring thus second-order accuracy also in time. 

Therefore, by replacing all the averaged fields and derivatives into the previous system, we get the following scheme:
\begin{eqnarray}
   \frac{\psi^{n+1}_i-\psi^n_i}{\Delta t} &=& \Pi^{n+\frac{1}{2}}_i \label{imp-scheme1} \\
   \frac{\Pi^{n+1}_i-\Pi^n_i}{\Delta t} &=& F\left(t_{n+\frac{1}{2}},x_i,\psi^{n+\frac{1}{2}}_i,\Pi^{n+\frac{1}{2}}_i,\{(D^j\psi)^{n+\frac{1}{2}}_i\}_{j=1}^{J},\{(D^k\Pi)^{n+\frac{1}{2}}_i\}_{k=0}^{K}\right) \label{imp-scheme2},
\end{eqnarray}
where $D^j$ denotes the finite difference average operator of the $j-$th spatial derivative. Due to the linearity of $F$ with respect to derivatives, the above system can be recast in matrix form, obtaining a band-diagonal linear system for the variables evaluated at time step $n+1$, namely
\begin{equation} \label{full-matrix-implicit}
\begin{bmatrix} 
d_1 & u_{11} & 0 & \cdots & \cdots & 0 \\ 
\ell_{1} & d_2 & u_{12} & u_{22} & \cdots & \vdots \\ 
0 & 0 &  d_1 & u_{11} & \ddots & \vdots \\ 
\ell_{3} & \ell_{2} & \ell_{1} & d_2 & \ddots & 0 \\
0 & 0 &  0 & 0 & d_1 & \vdots \\ 
\vdots & \ddots & \ddots & \ddots & \ddots & \vdots \\ 
0 & \cdots & \cdots & 0 & \cdots & d_{2N} 
\end{bmatrix}
\begin{bmatrix} 
\psi^{n+1}_2 \\
\Pi^{n+1}_2 \\
\psi^{n+1}_3 \\
\Pi^{n+1}_3 \\
\vdots \\
\vdots \\
\psi^{n+1}_{N-1} \\
\Pi^{n+1}_{N-1} \\
\end{bmatrix}
=
\begin{bmatrix} 
\bar{d}_1 & \bar{u}_{11} & 0 & \cdots & \cdots & 0 \\ 
\bar{\ell}_{1} & \bar{d}_2 & \bar{u}_{12} & \bar{u}_{22} & \cdots & \vdots \\ 
0 & 0 &  \bar{d}_1 & \bar{u}_{11} & \ddots & \vdots \\ 
\bar{\ell}_{3} & \bar{\ell}_{2} & \bar{\ell}_{1} & \bar{d}_2 & \ddots & 0 \\
0 & 0 &  0 & 0 & \bar{d}_1 & \vdots \\ 
\vdots & \ddots & \ddots & \ddots & \ddots & \vdots \\ 
0 & \cdots & \cdots & 0 & \cdots & \bar{d}_{1} 
\end{bmatrix}
\begin{bmatrix} 
\psi^{n}_2 \\
\Pi^{n}_2 \\
\psi^{n}_3 \\
\Pi^{n}_3 \\
\vdots \\
\vdots \\
\psi^{n+1}_{N-1} \\
\Pi^{n+1}_{N-1} \\
\end{bmatrix}
+
\begin{bmatrix} 
b_{11} \\
b_{21} \\
0 \\
0 \\
\vdots \\
\vdots \\
b_{1N} \\
b_{2N} \\
\end{bmatrix}
,
\end{equation}
where the second term on the right-hand side contains the information on the boundary conditions, which need to be specified. The number of upper/lower diagonal bands in the matrices depends on the stencil used -- that is, on the number of neighboring points needed to approximate the derivatives at a given grid point. The coefficients $\{d_j,u_{ij},\ell_j\}$ and $\{\bar{d}_j,\bar{u}_{ij},\bar{\ell}_j\}$ are grid functions that also depend on $\Delta x$ and $\Delta t$, but not on the dynamical fields. The system \eqref{full-matrix-implicit} can be solved with standard numerical methods for band-diagonal linear systems. These usually require $\mathcal{O}(N)$ iterations at each time step, unlike Gaussian elimination, which requires $\mathcal{O}(N^3)$ iterations (see \citep{NumRecipes} for details on the corresponding algorithm). Finally, notice that the generic scheme \eqref{imp-scheme1}-\eqref{imp-scheme2} is also valid for the case in which $F$ is a non-linear function of $\psi$, which  would clearly lead to a \textit{non-linear} system of coupled equations for the grid functions $\psi^n_i$ and $\Pi^n_i$. Nevertheless, such a system can be solved e.g. by means of standard iterative numerical methods.

\section{The Lifshitz field in a black hole space-time}
\label{lifshitz-field-BH-spacetime}

As a proof of concept of the numerical method introduced in the previous section, we evolve the Lifshitz equation \eqref{eq:equation_final} on top of a spherically symmetric, static, and asymptotically flat black hole space-time \cite{Berglund:2012bu}, solution to the equations of motion of khronometric gravity in vacuum \cite{Blas:2010hb,Jacobson:2013xta}. This corresponds to the low-energy limit of Ho\v rava gravity, as discussed in appendix \ref{app:EA_gravity}. We are thus implicitly assuming that the backreaction of the scalar field onto the geometry is negligible, and that gravitational perturbations are suppressed. It remains unknown whether this is a solid assumption, precisely because a proper understanding of the gravitational dynamics in Ho\v rava gravity at all energies would require a more sophisticated version of the methods that we are pioneering here.

\subsection{The background solution}

As already commented, we describe the derivation of the space-time solution in appendix \ref{app:EA_gravity}, but report it here for practical purposes. The metric and \ae ther in Schwarzschild coordinates $\{\tau,r,\theta,\varphi\}$ read
\begin{align}\label{eq:bg_sol}
    \ud s^2=f(r)\ud \tau^2-\frac{\ud r^2}{f(r)}-r^2\ud \Omega^2, \quad   U_\m \ud x^\m=\frac{H(r)}{2A(r)}\ud \tau+\frac{1-f(r)A(r)^2}{2A(r)f(r)}\ud r
\end{align}
with 
\begin{align}
    f(r)&=1-\frac{2\mu}{r}-c_{13} \frac{r_{\text{\ae}}^4}{r^4}, \\
    A(r)&=\frac{1}{f(r)}\left(-\frac{r_{\text{\ae}}^2}{r^2}+\sqrt{f(r)+\frac{r_{\text{\ae}}^4}{r^4}}\right),\\
    H(r)&=1+f(r)A(r)^2,
\end{align}
where $\mu$ is the mass of the black hole, $\ud \Omega^2$ is the $\mathbb{S}^2$ line element, and
\begin{equation}
    r_{\text{\ae}}=\frac{\mu}{2}\left(\frac{27}{1-c_{13}}\right)^{1/4}
\end{equation}
contains the only free parameter in the action, $c_{13}$. Note that since the solution is static, there exists a Killing vector $\chi^\m=(1,0,0,0)$, with norm $\chi^2=f(r)$. The latter  flips sign at the surface $r=r_K$, where $r_K$ is the positive solution of $f(r_K)=0$, signaling the position of a Killing horizon.

Although this form of the metric is useful for solving the equations of motion, it can be problematic for numerically evolving Eq. \eqref{eq:equation_final}. In this chart of coordinates, the Laplacian in the orthogonal leafs, $\Delta_\gamma$, also contains higher time derivatives, so the corresponding equation of motion obtained from Eq. \eqref{eq:equation_final} does not fit the cases discussed in the previous section. Nevertheless, this issue can be solved by aligning the time direction with the integral curves of $U^\m$, describing Eulerian observers in the preferred frame of Ho\v rava gravity. This can be achieved simply by introducing another time coordinate $t$ (the \textit{preferred time}) satisfying
\begin{align}\label{eq:change_preferred_frame}
    \ud \tau=\ud t-\frac{1-f(r)A(r)^2}{f(r)H(r)}\ud r,
\end{align}
so that the metric and \ae ther now read
\begin{align}
   &\ud  s^2=f(r)\ud t^2-2\frac{1-f(r)A(r)^2}{H(r)}\ud t \ud r -\frac{4A(r)^2}{H(r)^2} \ud r^2-r^2\ud \Omega^2,  \\
   &U_\m \ud x^\m=\frac{H(r)}{2A(r)}\ud t \label{sol-pref}.
\end{align}
Notice that after this transformation, the metric  takes the Arnowitt-Deser-Misner (ADM) form \cite{Arnowitt:1959ah}
\begin{align}
    \ud s^2=(N^2-N_iN^i)\ud t^2-2 N_i \ud x^i dt-h_{ij}\ud x^i \ud x^j,
\end{align}
where $N,N^i$ and $h^{ij}$ are the lapse, shift and induced metric in the foliation leafs, respectively, and are given by
\begin{align}
    N=\frac{H(r)}{2A(r)},\quad N_i \ud x^i=\frac{1-f(r)A(r)^2}{H(r)} \ud r,\quad  h_{ij}\ud x^i \ud x^j=\left(\frac{2A(r)}{H(r)}\right)^2 \ud r^2+r^2\ud \Omega^2.
\end{align}

This chart of coordinates, however, has a pathology whenever $N \equiv 0$, which corresponds to $H(r_U)=1+f(r_U)A(r_U)^2=0$ for some value $r_U$ of the radial coordinate \cite{Blas:2011ni,Barausse:2011pu,Berglund:2012bu}. From the relation \eqref{eq:change_preferred_frame}, we see that this point lies at finite $r$, but it is mapped to $t\rightarrow +\infty$, signalling that the foliation cannot be globally extended smoothly beyond this point, although observers moving inwards along $U^\m$ can still cross it in finite proper time (because $\ud \tau\equiv U_\m \ud x^\m$ remains finite). Remarkably, the surface ${r=r_U}$ represents a trapping surface for all modes, \textit{regardless} of their propagation speed. This can be seen by noting that the Killing vector $\chi^\m$ is space-like in the vicinity of the point $r=r_U$, while the product $\chi\cdot U=H(r)/(2A(r))$ precisely vanishes at this point, and becomes negative in the inner region\footnote{Note importantly that this cannot be avoided by a change of coordinates, since once in the preferred frame the symmetry group of the theory is restricted to FDiff \eqref{eq:FDiff}, under which $N \rightarrow N (dt'/dt)$.}. This is enough to characterize this surface as a universal trapping surface, hence named \emph{universal horizon}~\cite{Blas:2011ni,Barausse:2011pu}. The region $r<r_U$ always lies behind the Killing horizon  -- otherwise $\chi\cdot U=0$ would not be possible, as $U^\m$ is time-like everywhere by definition. In our particular case, one has
\begin{equation}
    r_U=\frac{3\mu}{2}.   
\end{equation}
For a more detailed discussion on causality within space-times endowed with universal horizons, see \cite{Bhattacharyya:2015gwa}.

For our purposes here, the universal horizon implies a limitation. Since our dynamical equations will be evolved in preferred time, we can only cover the region $r>r_U$, as the foliation and the time coordinate $t$ do not extend into the inner region. However, from the practical point of view of an observer sitting at a large radius (the ``asymptotic infinity'') this is enough, since they cannot observe anything coming from inside the universal horizon. Note, however, that the same is not true for the Killing horizon. While the surface sitting at $f(r_K)=0$ is a trapping surface in GR, it is not the case here anymore, since causal modes can move at speeds larger than unity \cite{Cropp:2013sea}. The universal horizon is the only true trapping surface within this geometry. However, the region  between the two horizons still encodes important features of the dynamics of the system, due to the character change of the Killing vector, which is associated with the only notion of conserved energy in the system.

\subsection{Numerical Implementation}

We now provide details about the specifics of our simulations. In particular, we introduce the ansatz for the solution, as well as the initial data and boundary conditions. We also discuss the numerical scheme, following the general approach described in Section \ref{sec:numerical}.

We solve Eq. \eqref{eq:equation_final} in spherical coordinates $\{t,r^*,\theta,\phi\}$, where the radial tortoise-like coordinate $r^*$ is chosen to push the universal horizon to  infinity, corresponding to $r^*\rightarrow -\infty$. The relation between the areal radius $r$ and the new coordinate $r^*$ is given by\footnote{In principle, any transformation of the form
\[
    \frac{\text{d}r}{\text{d}r^*}=\left(\frac{H(r)}{2A(r)}\right)^p
\]
with integer exponent $p\geq 1$ equally pushes the universal horizon to $r^*\to-\infty$. We choose here $p=2$ so that the transformation decays in a smoother way when approaching the universal horizon, avoiding localized high-frequency instabilities that would otherwise require the addition of artificial dissipation in the equations.}
\be\label{tortoise_coord}
\frac{\text{d}r}{\text{d}r^*}=\left(\frac{H(r)}{2A(r)}\right)^2.
\ee
Although complicated in general, one can see that it behaves as $r_*\sim r$ for large $r$ due to asymptotic flatness of the background solution, which implies $f(r\rightarrow \infty)\rightarrow 1$ and $A(r\rightarrow \infty)\rightarrow 1$. Close to the universal horizon, we have $H(r)/(2A(r))\propto (r-r_U)$ and hence $r_*\propto -(r-r_U)^{-1}$, indeed placing $r_U$ at $r_*\rightarrow -\infty$. The condition \eqref{tortoise_coord} is an ordinary differential equation, which can be solved numerically using the standard fourth-order Runge-Kutta method.

We also define the parameters $\kappa_2=c_2/\Lambda^2$ and $\kappa_3=c_3/\Lambda^4$ for computational convenience. Thus, to ensure the stability and well-posedness of the problem, we require $\kappa_3\geq 0$ and $\kappa_2 > -2 \sqrt{\kappa_3}$. After this change, Eq.~\eqref{eq:equation_final} reads
\begin{align}\label{final-lif-eq}
     \left[ \square+\kappa_2 \Delta_\gamma^2 -\kappa_3\Delta_\gamma^3\right] \phi=0.
\end{align}

Due to the spherical symmetry of the background, we consider the following ansatz for the scalar field:
\begin{equation}\label{evo-equation-modes}
    \phi(t,r^*,\theta,\varphi) = \sum_{\ell=0}^{\infty}\sum_{m=-\ell}^{\ell}{\psi_{\ell m}(t,r^*) Y_{\ell m}(\theta,\varphi)},
\end{equation}
where $Y_{\ell m}(\theta,\varphi)$ are the spherical harmonics. Plugging this into Eq. \eqref{final-lif-eq}, we get an effective $1+1$ dimensional differential equation for every mode $\psi_{\ell m}(t,r^*)$, which is independent of $m$, and reads
\begin{equation}\label{evoeq-expl}
\left[\partial^2_t+ \zeta_{10}\partial_t + \zeta_{11}\partial_t\partial_{r^*} +
\sum_{j=1}^{6}{\zeta_{0j}\partial^j_{r^*}} + V_{\text{eff}}
\right]\psi_{\ell m}= 0.
\end{equation}
The coefficients $\zeta_{ij}$ and the effective potential $ V_{\rm eff}$ are functions of $r^*$ through the coordinate transformation \eqref{tortoise_coord}, the metric functions $A(r),\,H(r)$, and their derivatives up to fifth order. They also depend on $\kappa_2$, $\kappa_3$ and $\ell$. The explicit formulae for $\zeta_{ij}$ are given in Appendix \eqref{app-coeffs-explicit}, while we show $V_{\rm eff}$ later in Eq. \eqref{eq:Veff}. Let us note that $V_{\rm eff}$ vanishes for $\ell=0$, and corresponds to an effective centrifugal barrier, which plays an important role in the behavior of the solution close to the Killing horizon, as we will discuss later.

Following the scheme introduced in Section \ref{sec:numerical}, Eq. \eqref{evoeq-expl} admits the implicit discretization
\begin{eqnarray}
   \frac{\psi^{n+1}_i-\psi^n_i}{\Delta t} &=& \Pi^{n+\frac{1}{2}}_i \label{num-scheme-lif1}\\
   \frac{\Pi^{n+1}_i-\Pi^n_i}{\Delta t} &=& -\zeta^i_{10}\Pi^{n+\frac{1}{2}}_i - \zeta^i_{11}(D\Pi)^{n+\frac{1}{2}}_i -
\sum_{j=1}^{6}{\zeta^i_{0j}(D^j\psi)^{n+\frac{1}{2}}_i}-V^i_{\rm eff}\psi^{n+\frac{1}{2}}_i, \label{num-scheme-lif2}
\end{eqnarray}
where $\zeta^i_{jk} := \zeta_{jk}(r^*_i)$ and $V^i_{\rm eff} := V_{\rm eff}(r^*_i)$.

We implemented centered finite difference operators with second-order accuracy, with a stencil of seven grid points. In this case, the system \eqref{full-matrix-implicit} has five upper diagonals and seven lower ones. Table \ref{table-coeffs-matrices} shows the explicit form of the coefficients in the matrices.

\begin{table}\fontsize{10}{10}
\centering
 \begin{tabular}{c||c}
 \hline
 \textit{\rm{\scriptsize{Element}}} & \textit{\rm{\scriptsize{Expression}}} \\ [5pt]
 \hline\hline\\[1pt]
 $u_{11}$ & $\displaystyle\frac{1}{\Delta t}$ \\[10pt]
 $u_{12}$ & $\displaystyle \frac{\zeta_{01} (\Delta x)^5+2 \zeta_{02} (\Delta x)^4-2 \zeta_{03} (\Delta x)^3-8 \zeta_{04}
   (\Delta x)^2+5 \zeta_{05} \Delta x+30 \zeta_{06}}{4 (\Delta x)^6}$ \\[10pt]
 $u_{13}$ & $\displaystyle \frac{\zeta_{11}}{4 \Delta x}$ \\[10pt]
 $u_{14}$ & $\displaystyle \frac{\zeta_{03} (\Delta x)^3+2 \zeta_{04} (\Delta x)^2-4 \zeta_{05} \Delta x-12 \zeta_{06}}{4
   (\Delta x)^6}$ \\[10pt]
 $u_{15}$ & $\displaystyle \frac{\zeta_{05} \Delta x+2 \zeta_{06}}{4 (\Delta x)^6}$ \\[10pt]
 $d_1,\, \bar{d}_1$ & $\displaystyle \frac{1}{2}$ \\[10pt]
 $d_2$ & $\displaystyle \frac{\zeta_{10}}{2} + \frac{1}{\Delta t}$ \\[10pt]
 $\bar{d}_{2}$ & $\displaystyle -\frac{\zeta_{10}}{2} + \frac{1}{\Delta t}$ \\[10pt]
 $\ell_1$ & $\displaystyle \frac{V_{\rm eff}}{2}-\frac{\zeta_{02} (\Delta x)^4-3 \zeta_{04} (\Delta x)^2+10
   \zeta_{06}}{(\Delta x)^6}$ \\[10pt]
 $\ell_2$ & $\displaystyle - u_{13}$ \\[10pt]
 $\ell_3$ & $\displaystyle - u_{12}$ \\[10pt]
 $\ell_4$ & $\displaystyle - u_{14}$  \\[10pt]
 $\ell_5$ & $\displaystyle - u_{15}$ \\[10pt]
 \hline
 \end{tabular}
 \caption{Explicit form of the matrix elements for the scheme \eqref{num-scheme-lif1}-\eqref{num-scheme-lif2}. Each of them is a grid function, which should be evaluated at the grid point $r^*_i$. Here we have omitted the corresponding indices for the sake of notation clarity. For instance, $u_{11}$ corresponds to $u_{11}(r^*_i)$. Similarly for the coefficients $\zeta_{ij}$, which represent the grid functions $\zeta_{ij}(r^*_i)$.}
 \label{table-coeffs-matrices}
\end{table}

\subsubsection{Initial data}

As initial data for the Lifshitz field, we considered two different profiles. The first one  is a static Gaussian pulse, which we refer to as \textit{ID Type I}, given by
\begin{eqnarray}\label{inidat-type1}
    \psi(0,r^*) &=& a_0\exp{\left[-\frac{(r^*-r^*_c)^2}{\sigma^2}\right]}\,,\\
    \Pi(0,r^*) &=& 0\,, \nonumber
\end{eqnarray}
where $a_0$, $r^*_c$ and $\sigma$ are the Gaussian amplitude, mean and variance, respectively. 

The second profile that we consider, which we refer to as \textit{ID Type II}, is given by an (approximately) ingoing wave-packet
\begin{eqnarray}\label{inidat-type2}
    \psi(0,r^*) &=& \exp{\left[-\frac{(r^*-r^*_c)^2}{\sigma^2}\right]}\cos{(\omega r^*)}\,,\\
    \Pi(0,r^*) &=& -\left(\frac{2(r^*-r^*_c)}{\sigma^2}\cos{(\omega r^*)}+\omega\sin{(\omega r^*)}\right)\exp{\left[-\frac{(r^*-r^*_c)^2}{\sigma^2}\right]}\,.\nonumber
\end{eqnarray}
This corresponds to an exact ingoing wave-packet when $\kappa_2=\kappa_3=0$, satisfying $\partial \psi/\partial t|_{t=0}=\partial\psi/\partial r^*|_{t=0}$. In the Lifshitz case, it will also contain outgoing modes, but we expect those to be negligible far enough from the gravitational well, as long as the energy of the Killing energy of the wave-packet $\omega$ is small. Hereinafter, we set $\omega=1$. 

\subsubsection{Boundary conditions}

Setting boundary conditions for equations endowed with an anisotropic scaling \eqref{eq:lifshitz_scaling} is highly non-trivial, due to the different order in derivatives along distinct directions. While in the two-dimensional wave equation (regardless of the boundary conditions imposed), the general solution can always be written as the superposition of left-moving and right-moving waves,
\begin{align}\label{eq:sol_wave_eq}
    \Psi(t,r)=f_L(r-ct)+f_R(r+ ct),
\end{align}
with $f_L, f_R$ arbitrary functions and $c$ the propagation speed of the waves,  this is no longer true for the case at hand, and in particular for Eq. \eqref{final-lif-eq}. This can be seen by simply plugging the ansatz \eqref{eq:sol_wave_eq} into \eqref{eq:flat_eom}. Only when $f'_L(x)\propto f_L(x)$, and $f'_R(x)\propto f_R(x)$ -- where a prime denotes differentiation -- solutions can be decoupled, as long as $c$ takes the right value. This poses a problem for setting up a successful evolution scheme. While in the case of the wave equation one can always impose pure ingoing or outgoing boundary conditions, simply by selecting left or right movers at the appropriate boundary surface, such procedure is not possible here. This is of particular relevance at the universal horizon, which is a semi-permeable surface that only allows for ingoing modes.

One possibility to face this issue that has been recently explored in other contexts \citep{Dima2020,Dolan13} is to add a \textit{perfectly matching layer} (PML) covering a small region close to the boundaries. This introduces artificial dissipation suppressing spurious reflected waves, by modifying the kinetic term in the equations of motion. Although this can be implemented systematically when only second order derivatives are involved, we have not found a way to extend it to the case with higher derivatives.

Nevertheless, following the spirit of the PML method, we implement instead an \textit{artificial dissipative layer} (ADL), controlled by a function $L(x)$, which suppresses waves exiting the domain of interest during the numerical evolution. The layer takes the shape of a function whose value is unity within the physical domain, but decays smoothly to zero in the regions close to the boundary. We implement it in our work by replacing the numerical solution $\psi^n_i$ at each time step by $\psi^n_i \to L_i \psi^n_i$.

\begin{figure} 
\centering
\includegraphics[width=12cm]{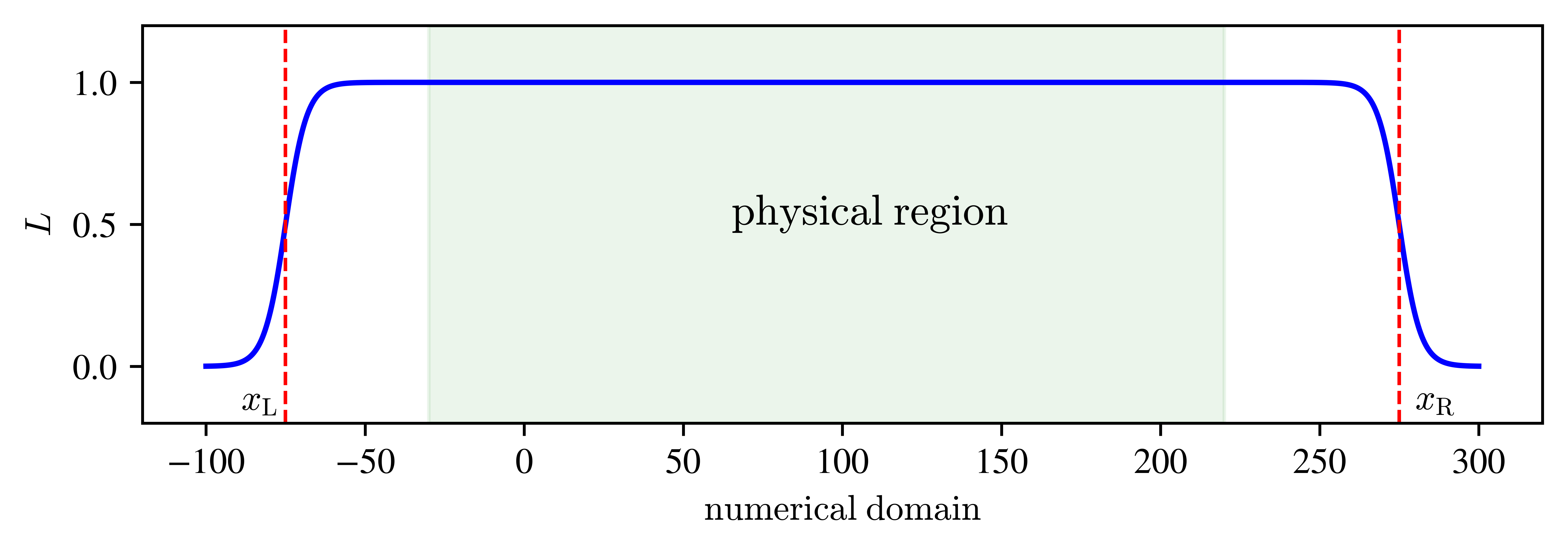}
\caption{\textit{Artificial dissipative layer}. Example of $L(x)$ with parameters $s=0.2$, $x_L=-75$, and $x_R=275$. The green-shaded rectangle represents the physical region where the evolution is unaltered. The layer aims to mimic
boundary conditions describing
waves escaping the numerical domain by artificial damping at the extremes. The red vertical dotted lines are located at $x=x_{\rm L}$ and $x=x_{\rm R}$, corresponding to the midpoints of the damping zones.}
\label{plot-layer}
\end{figure}

We choose in particular the following function for the ADL:
\begin{equation}\label{Lfunction-layer}
    L(x) = 
    \left\{
        \begin{array}{rcl}
            \displaystyle \frac{1 - \tanh\left[s\left(x-x_{\rm R}\right)\right]}{2},  & \mbox{if} & x \geq \displaystyle \frac{x_{\rm L}+x_{\rm R}}{2} \\
            \\
            \displaystyle \frac{1 - \tanh\left[s\left(x_{\rm L}-x\right)\right]}{2}, & \mbox{if} & x < \displaystyle \frac{x_{\rm L}+x_{\rm R}}{2}
        \end{array}
    \right.\,\,,
\end{equation}
where $s$ controls the slope of the function in the extremes of the numerical domain, while $x_{\rm L}$ and $x_{\rm R}$ correspond to the left and right midpoints of the decaying regions, respectively. An illustration of this function is given in Figure \ref{plot-layer}.

\section{Results}\label{sec:results}

We report here the results of our simulations, performed with different values of the coupling parameters $\kappa_2$ and $\kappa_3$ controlling the strength of the Lorentz-breaking terms in Eq. \eqref{final-lif-eq}. We also investigate the behaviour of wave modes with angular number $\ell > 0$ for fixed values of the couplings, as well as the effect of the centrifugal barrier induced by the effective potential $V_{\rm eff}$. Finally, we probe the validity of the numerical code by performing convergence tests, and show the robustness of our results against a change in the position of the ADL zone.

\subsection{Evolution of the Lifshitz field}

We start off by studying the dynamical evolution of the $\ell=m=0$ mode for the two different initial data introduced in the previous section. 

\begin{figure}
\centering
\includegraphics[width=7.5cm]{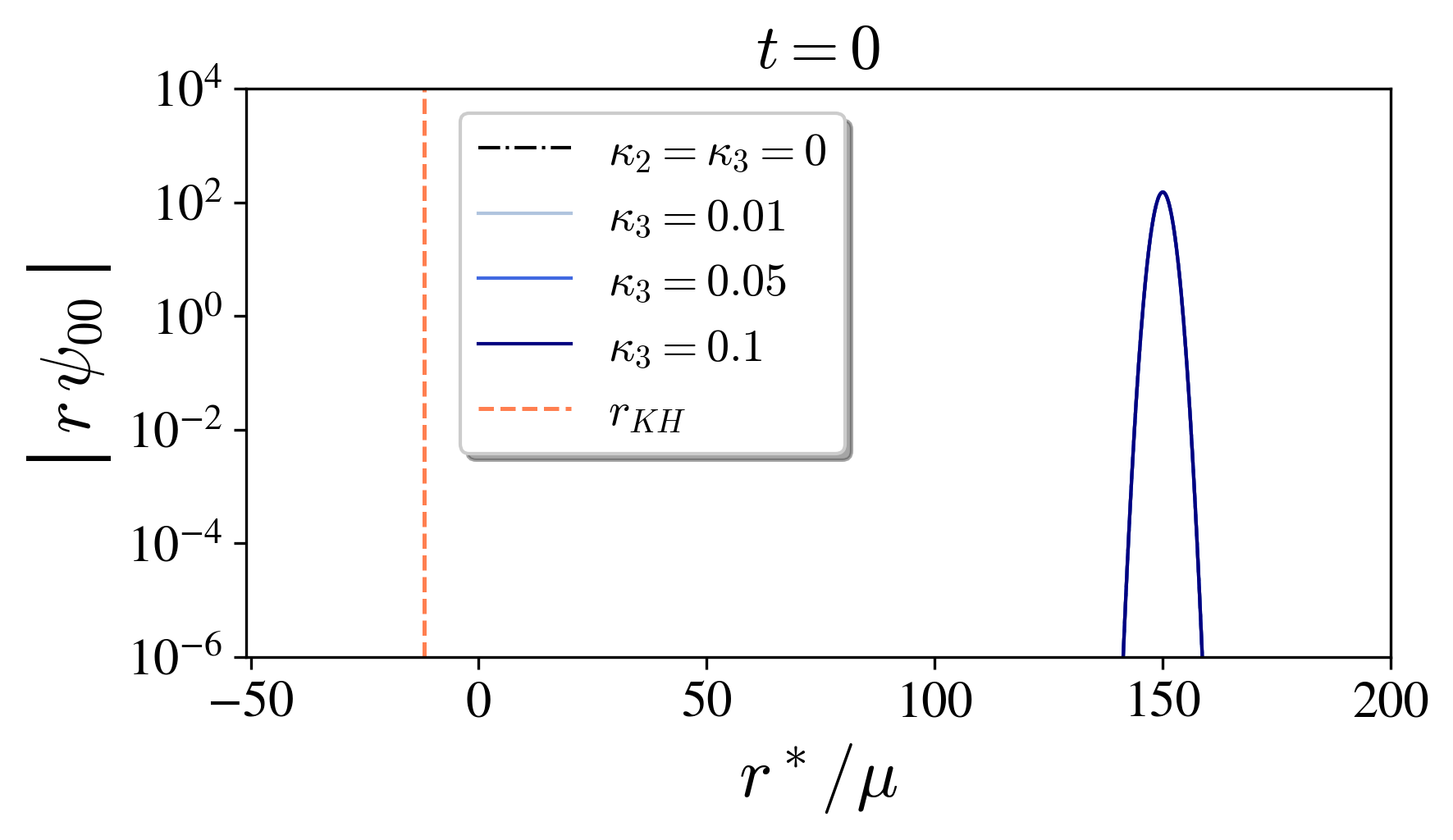}
\includegraphics[width=7.5cm]{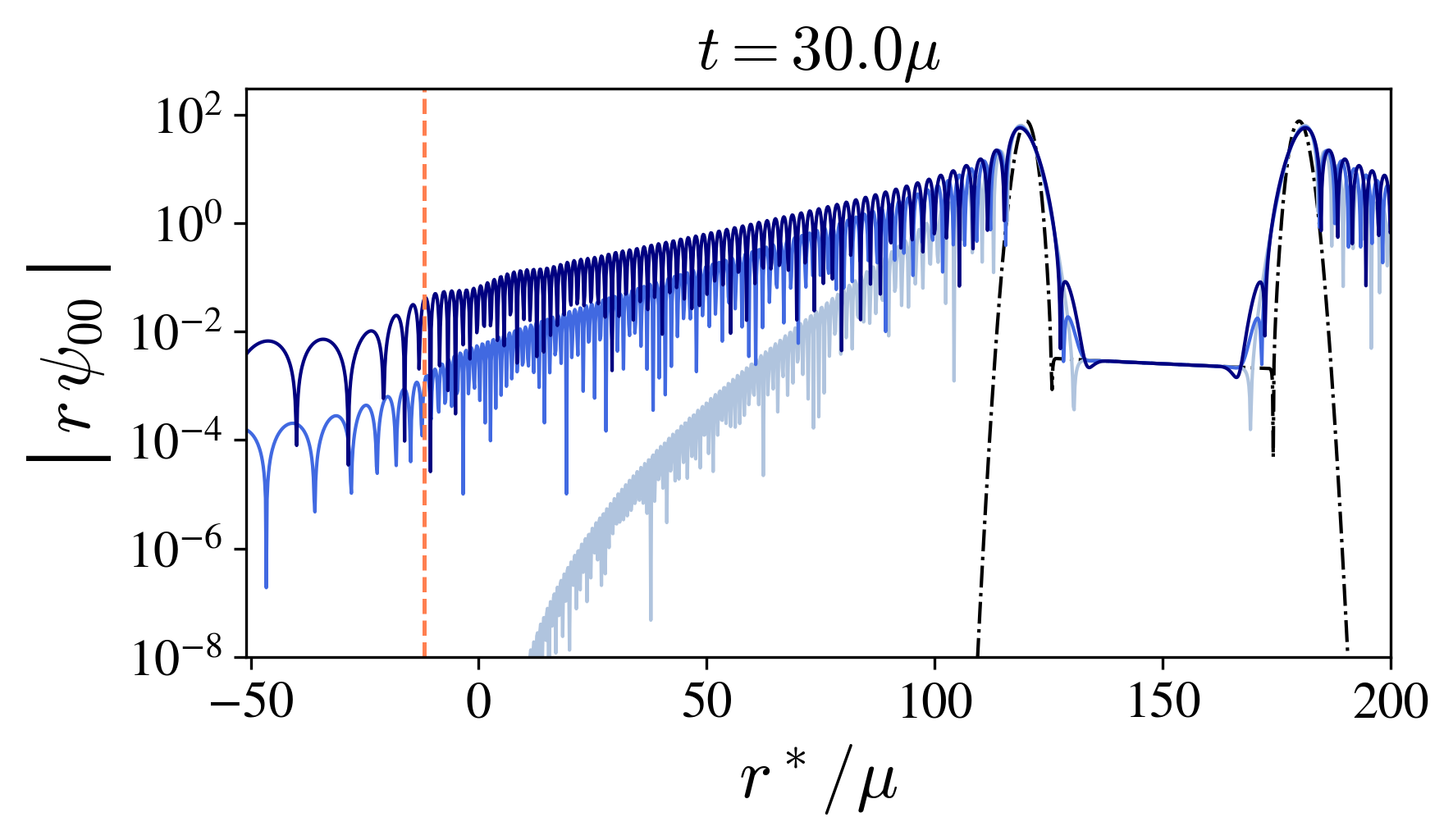}
\includegraphics[width=7.5cm]{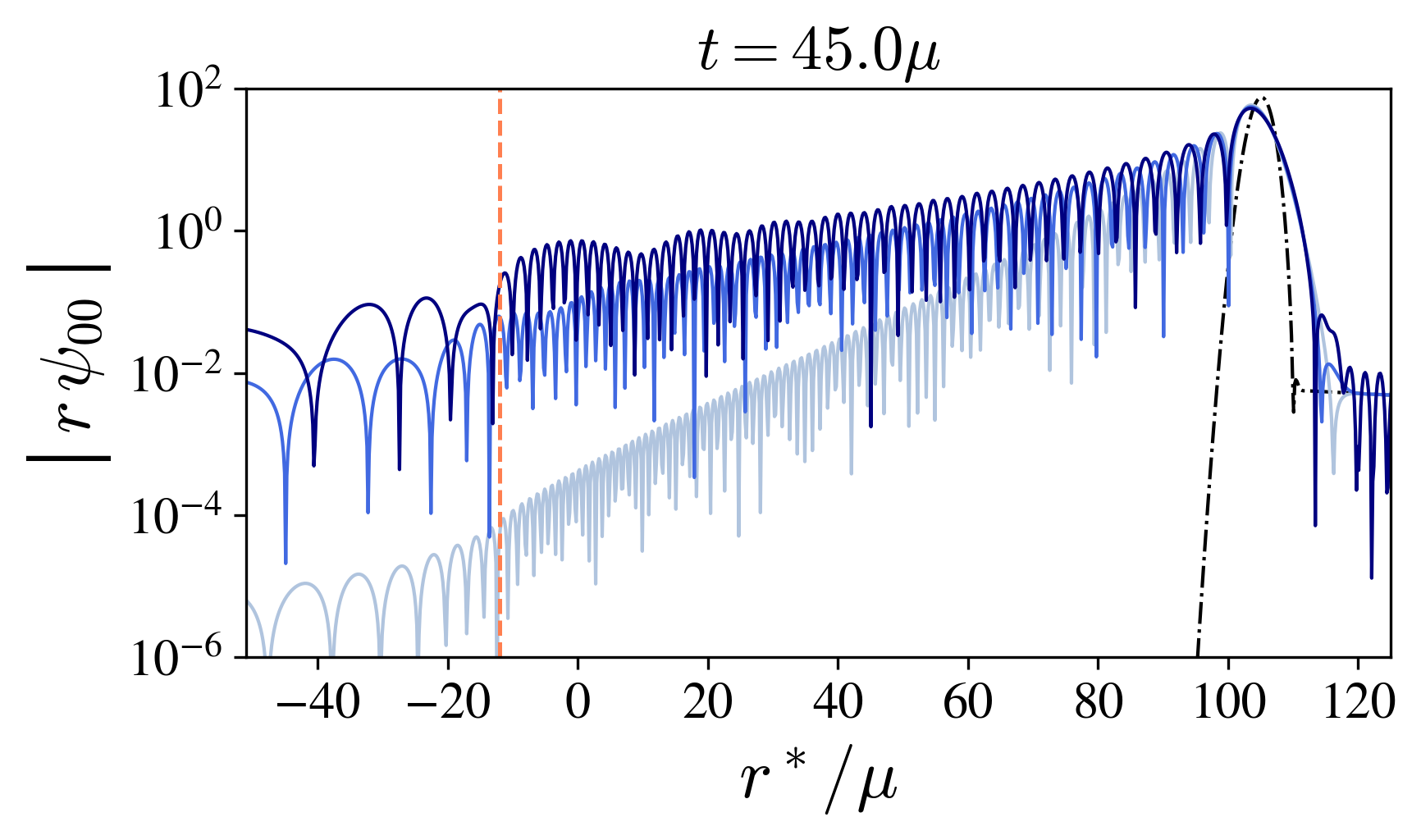}
\includegraphics[width=7.5cm]{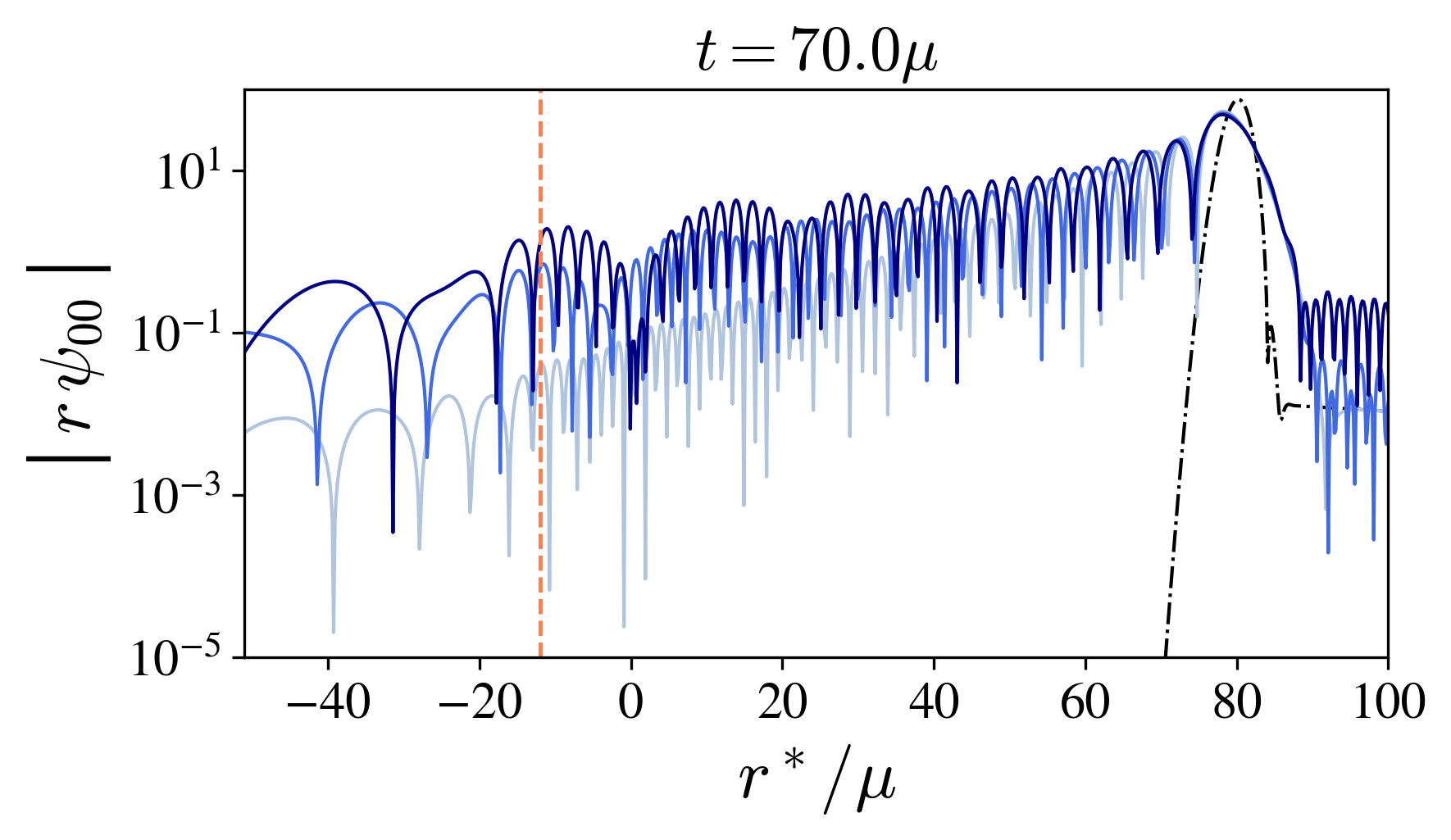}
\includegraphics[width=7.5cm]{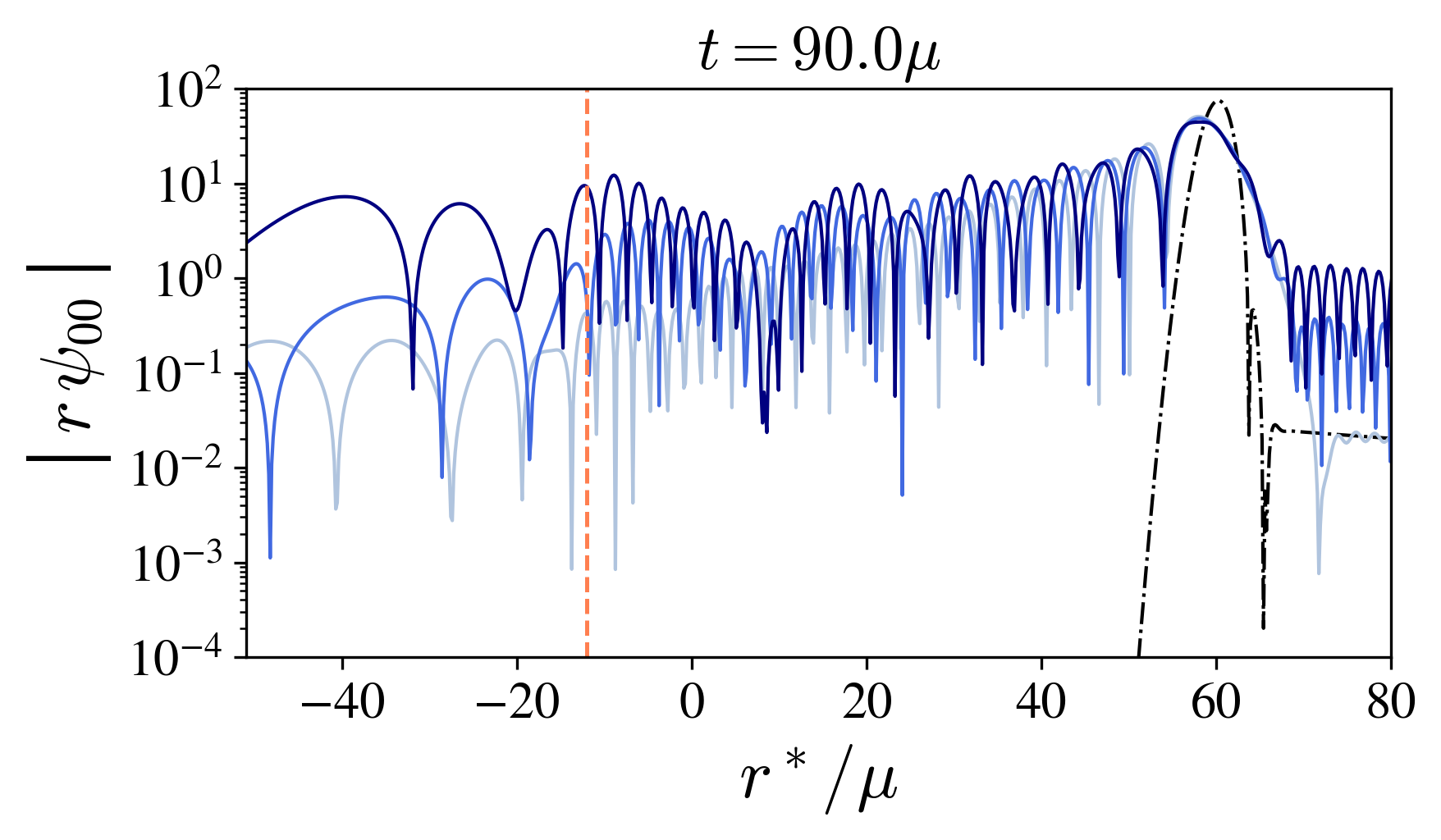}
\includegraphics[width=7.5cm]{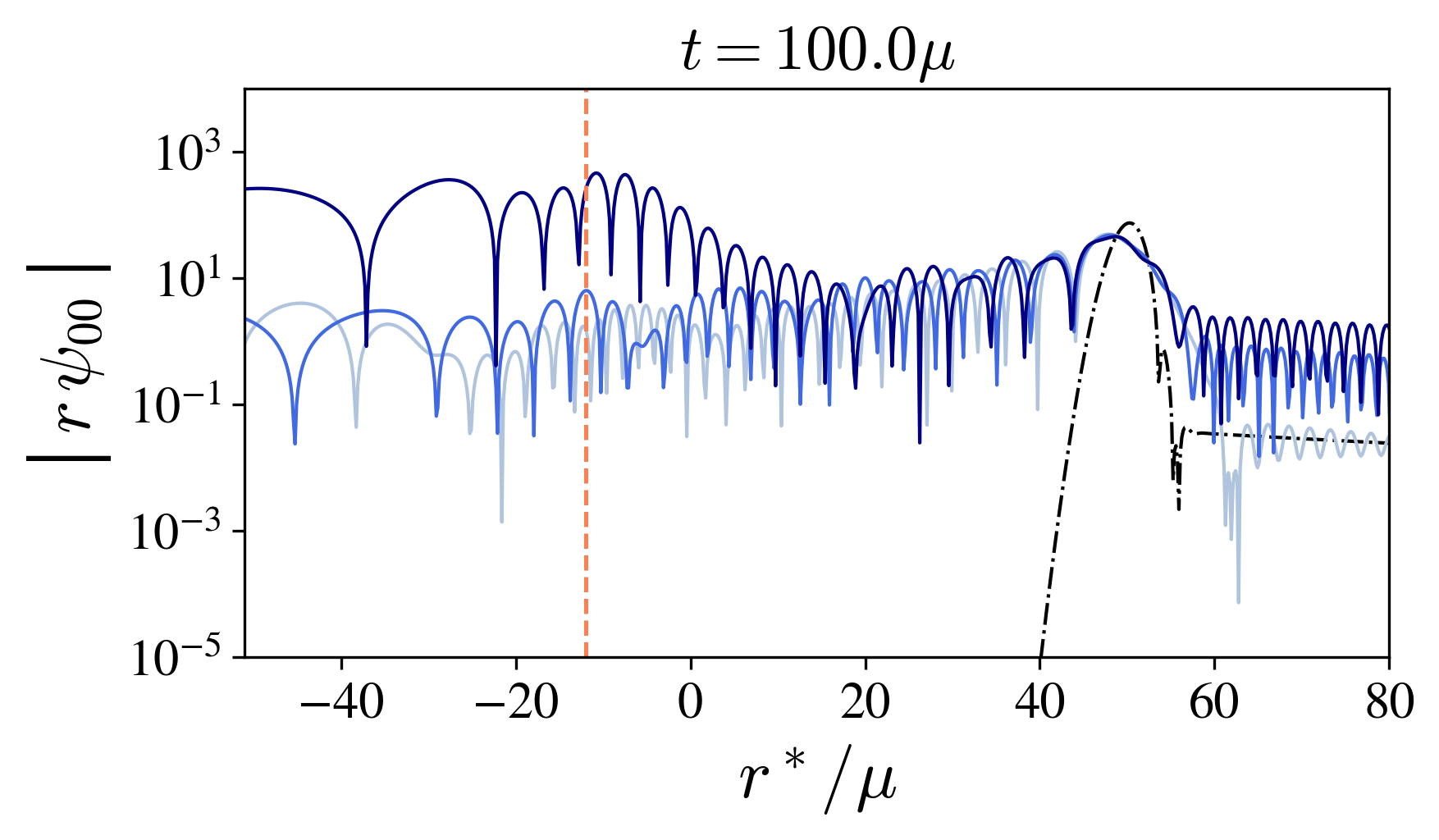}

\caption{\textit{Dynamics of the Lifshitz field (ID type I)}. Snapshots of the evolution of the $\ell=m=0$ mode, for $\kappa_2=0.1$ and different values of $\kappa_3$: $\kappa_3=0.01$ (light blue), $\kappa_3=0.05$ (blue) and $\kappa_3=0.1$ (dark blue). The initial data is the ID type I given by Eq. \eqref{inidat-type1} (a static pulse far from the horizons), with parameters $a_0 = 1$, $r^*_c = 150\mu$ and $\sigma = 2$.
As $\kappa_3$ increases, a faster cascade of high frequency modes grows towards the horizon (orange dotted vertical line). This behavior can be compared with the corresponding evolution of the wave equation (i.e., by setting $\kappa_2=\kappa_3=0$), represented by the dotted black profile.}
\label{plot-evo-lif-ID1-vark3}
\end{figure}

We first evolve the initial data ID type I (static pulse) from $t=0$ to $t=100\mu$, as shown in figure \ref{plot-evo-lif-ID1-vark3}. We set $\kappa_2=0.1$ and vary $\kappa_3$, taking the values $\kappa_3 = \{0.01,0.05,0.1\}$ (from light to dark blue in the plot panels). As it can be seen from the figure, we observe the formation of a rapid cascade of modes nearing the black hole as  time increases. This is produced by wave modes travelling at different speeds depending on their frequency, a behavior that is expected to occur due to the dispersive character of the equation induced by the higher derivative terms.

\begin{figure}
\centering
\includegraphics[width=7.5cm]{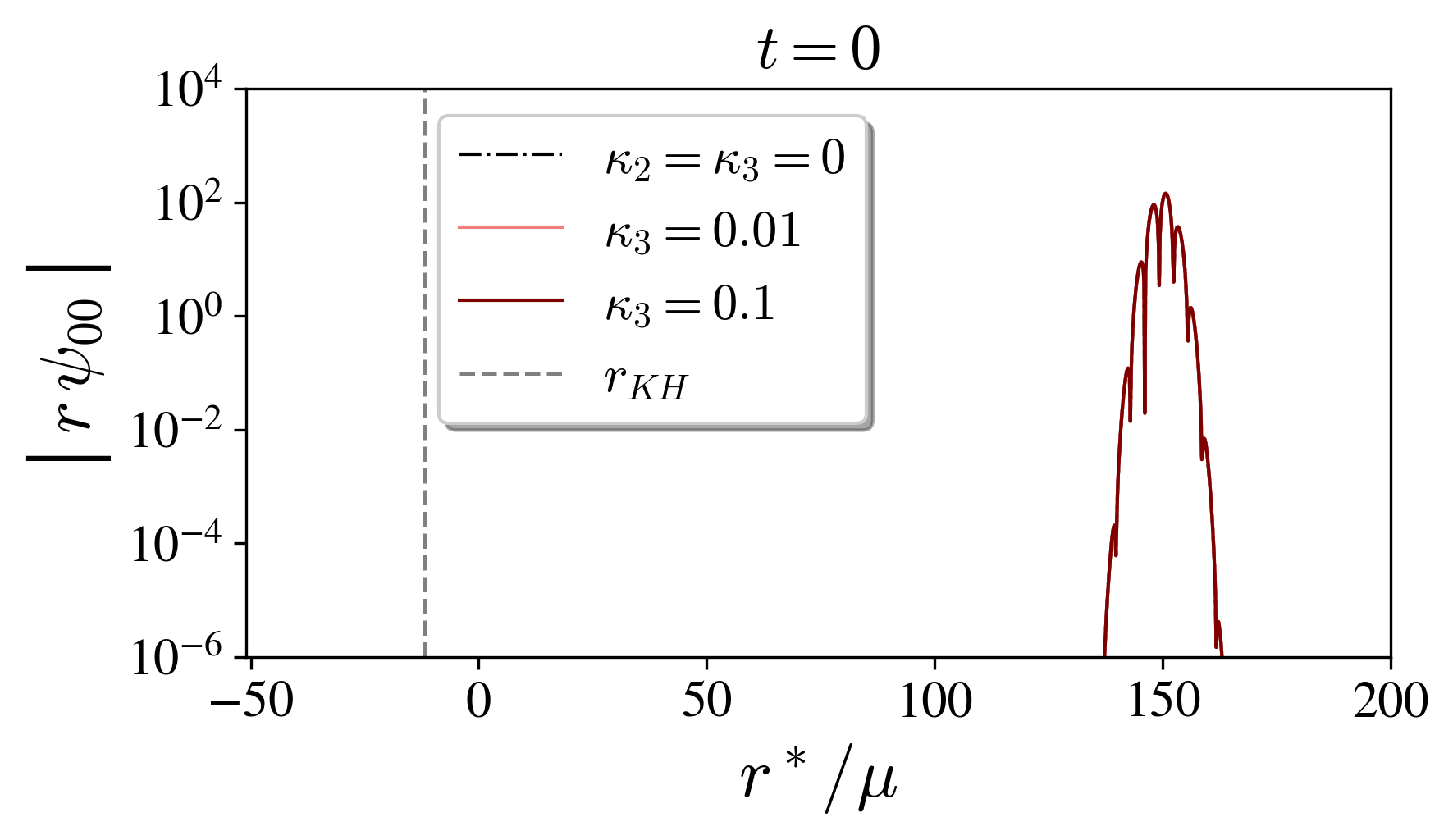}
\includegraphics[width=7.5cm]{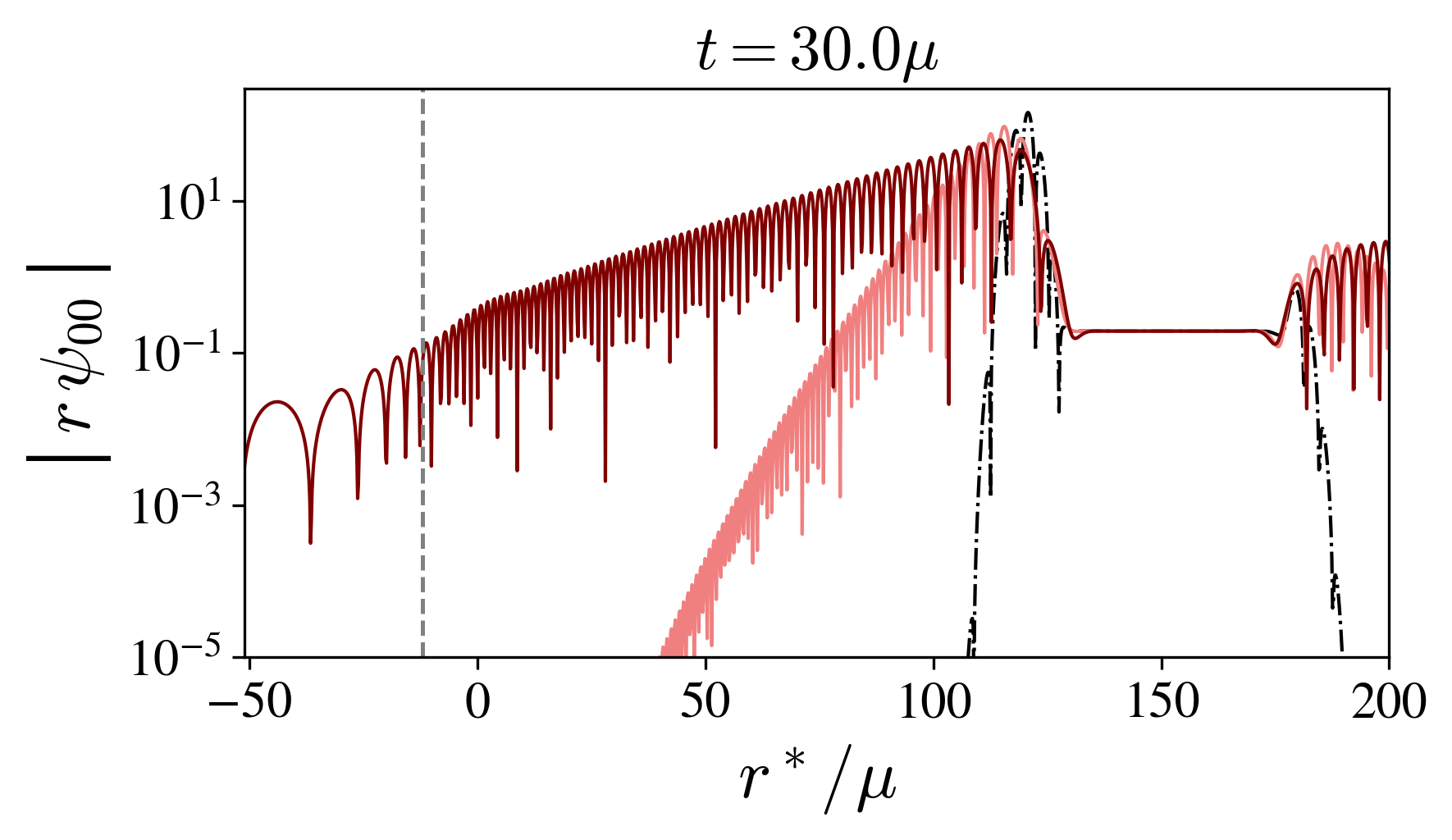}
\includegraphics[width=7.5cm]{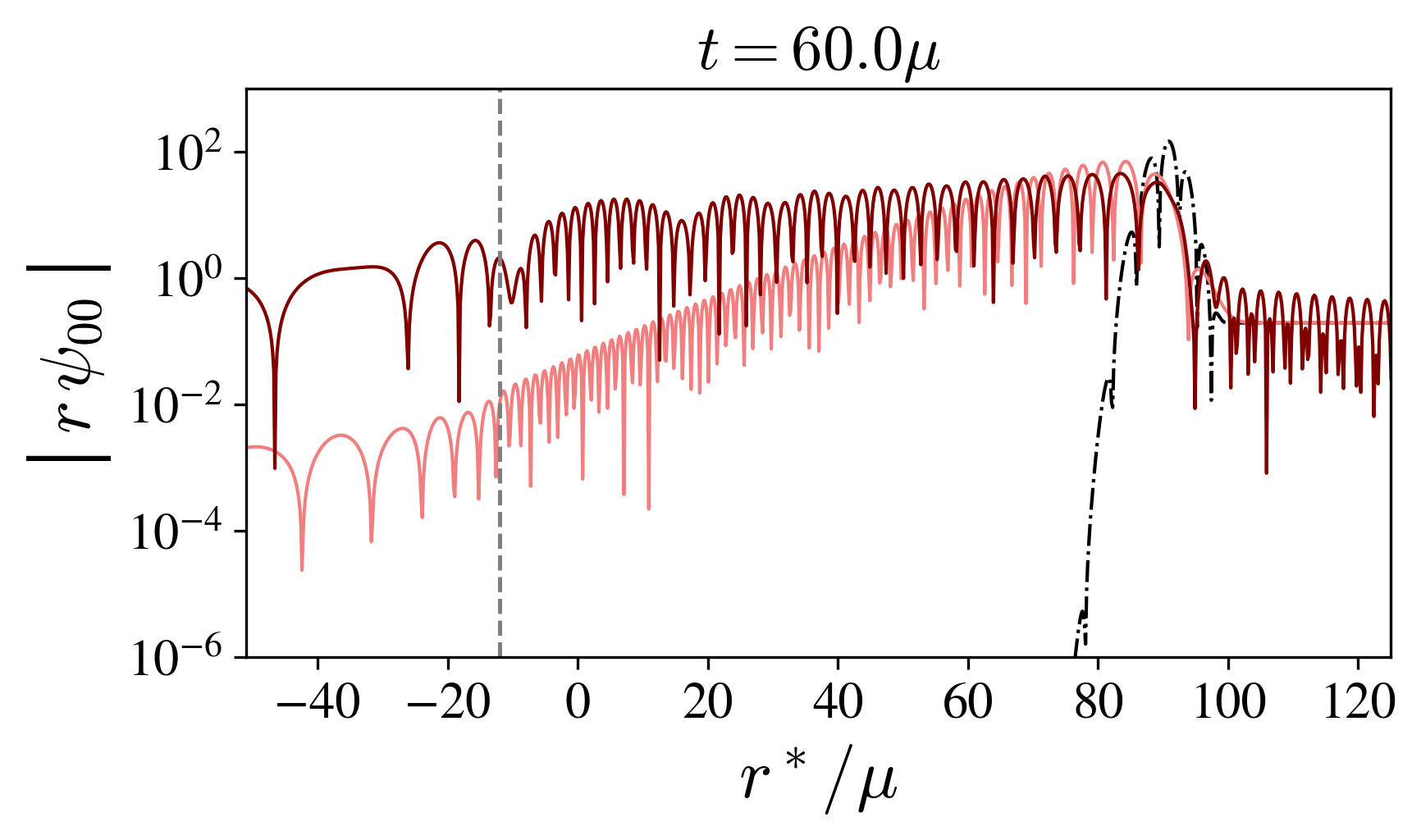}
\includegraphics[width=7.5cm]{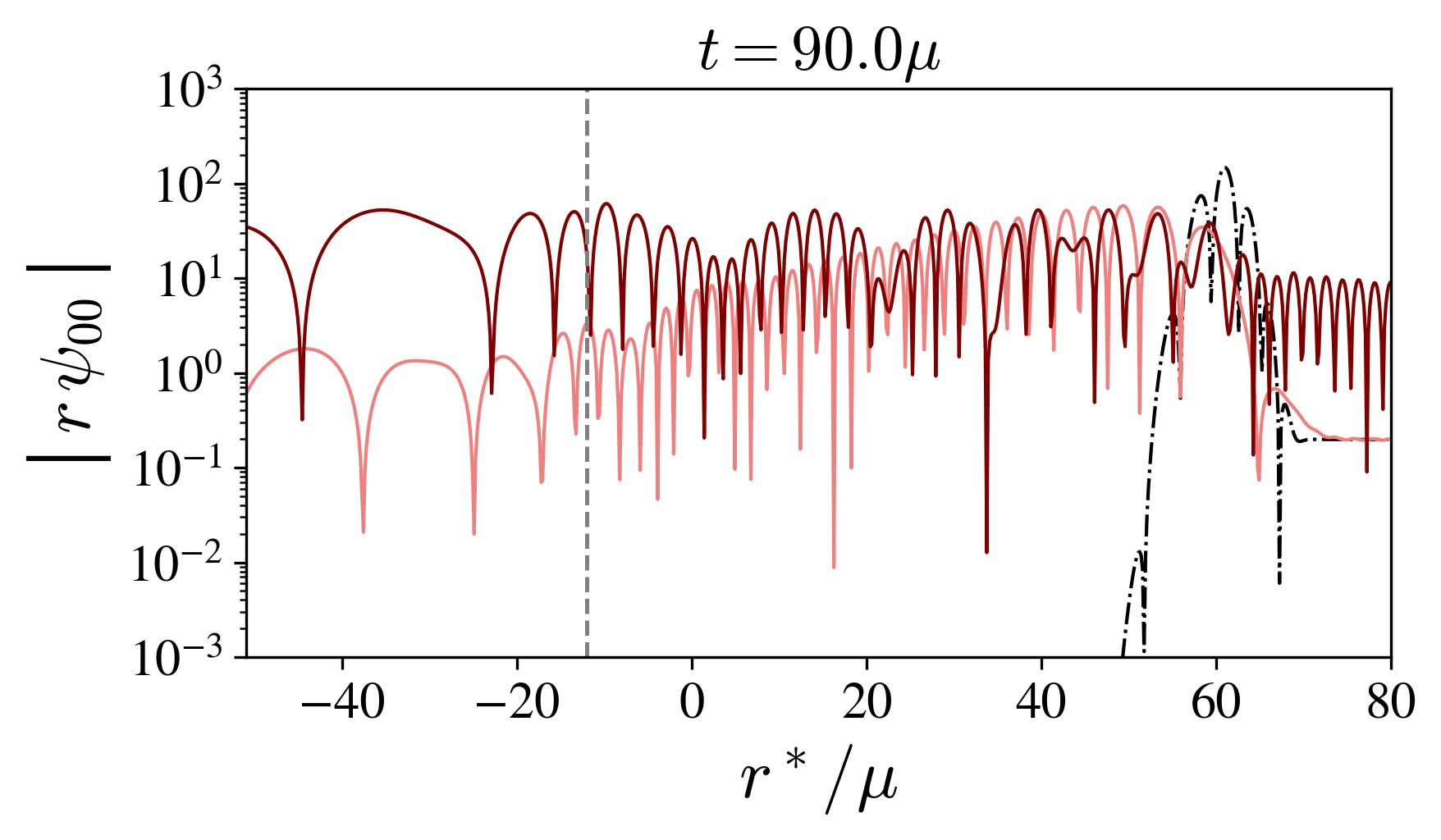}

\caption{\textit{Dynamics of the Lifshitz field (ID type II)}. Evolution of the $\ell=m=0$ mode from the ID type II (ingoing pulse) given in Eq. \eqref{inidat-type2}, with parameters $r^*_c=150\mu$, $\sigma=3$ and $\omega=1$. We fixed $\kappa_2=0.1$, and set $\kappa_3=0.01$ (light red curve) and $\kappa_3=0.1$ (brown curve). A cascade towards the Killing horizon (represented by the gray dotted vertical line) behaves similarly to the previous case shown in Figure \ref{plot-evo-lif-ID1-vark3}. The evolution of the wave equation is also shown in dotted black, for comparison.}
\label{plot-evo-lif-ID2-vark3}
\end{figure}

From early evolution times\footnote{See Appendix C for more details on early-time dynamics.}, modes with faster speeds rapidly escape the numerical domain, unlike slower ones, which stay longer within the physical region. Moreover, we clearly observe that the magnitude of the modes within the cascade grows faster for larger values of $\kappa_3$. A comparison with the standard wave equation evolution (that is, when $\kappa_2=\kappa_3=0$) is shown in the dotted black profile. As expected, the solution does not exhibit a dispersive behavior in this case, as propagation speeds are bounded and independent of the frequency. Notice also that due to the Lorentz violating character of the equation, the solution can smoothly penetrate the Killing horizon, travelling towards the universal horizon, which is located at negative infinity. Finally, we observe a bump in the neighborhood of the Killing horizon forming at later times, around $t\sim 80\mu$, whose magnitude increases faster for larger values of $\kappa_3$. A similar behavior is found in the case of the initial data ID Type II (ingoing pulse) for the same choice of the parameters $\kappa_2$, and $\kappa_3$, as shown in figure \ref{plot-evo-lif-ID2-vark3}.

\begin{figure}
\centering
\includegraphics[width=7.5cm]{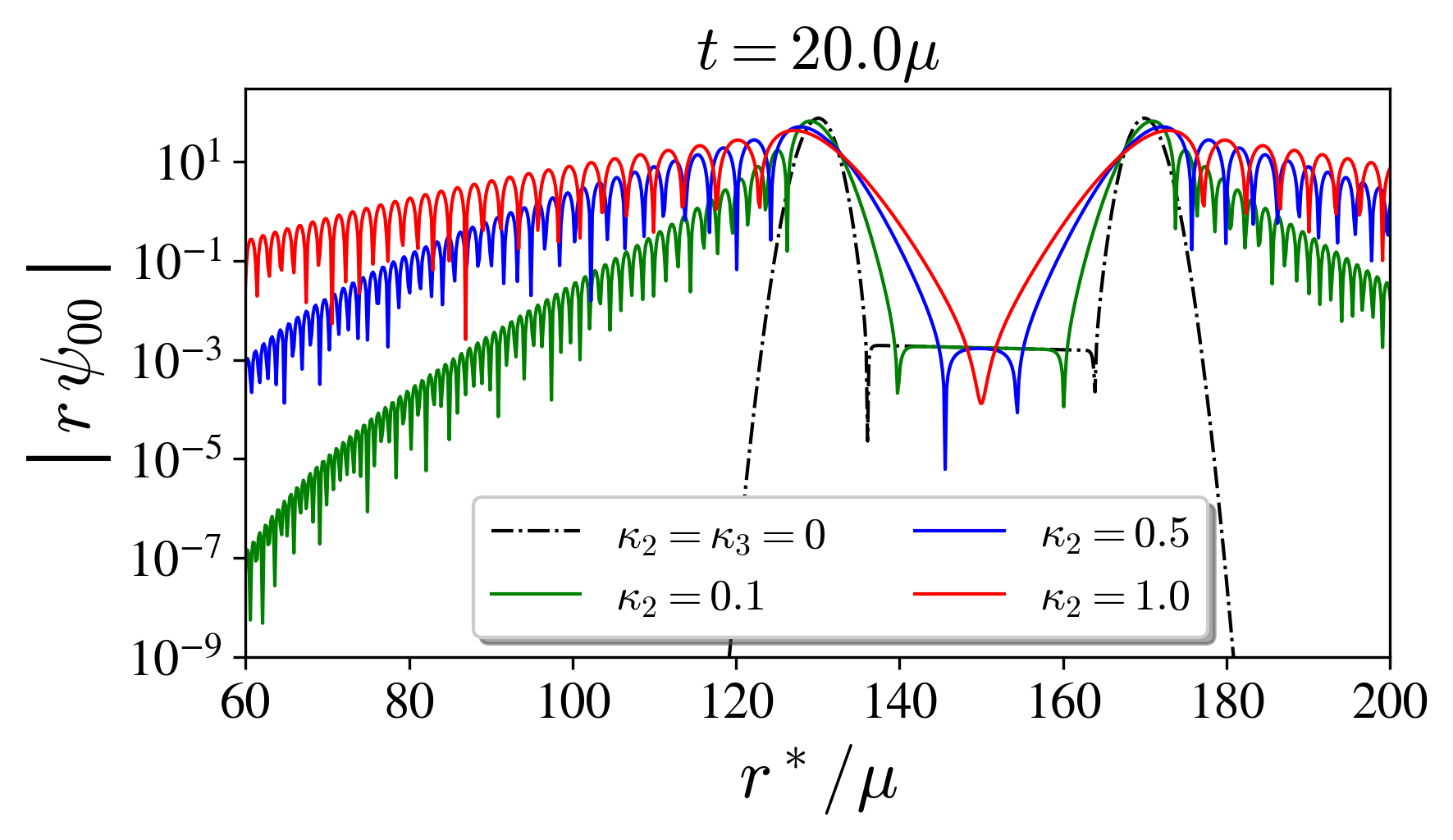}
\includegraphics[width=7.5cm]{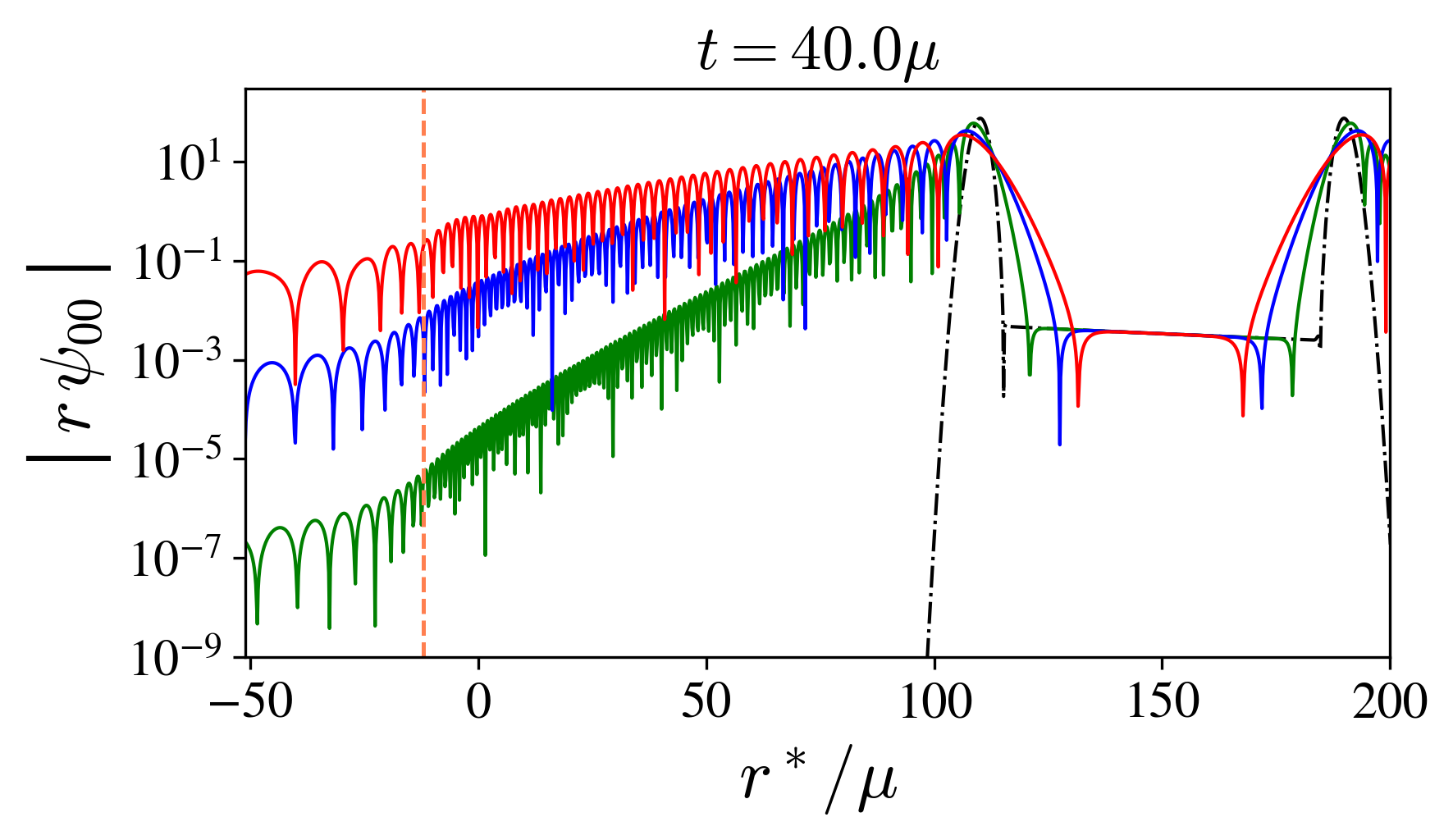}
\includegraphics[width=7.5cm]{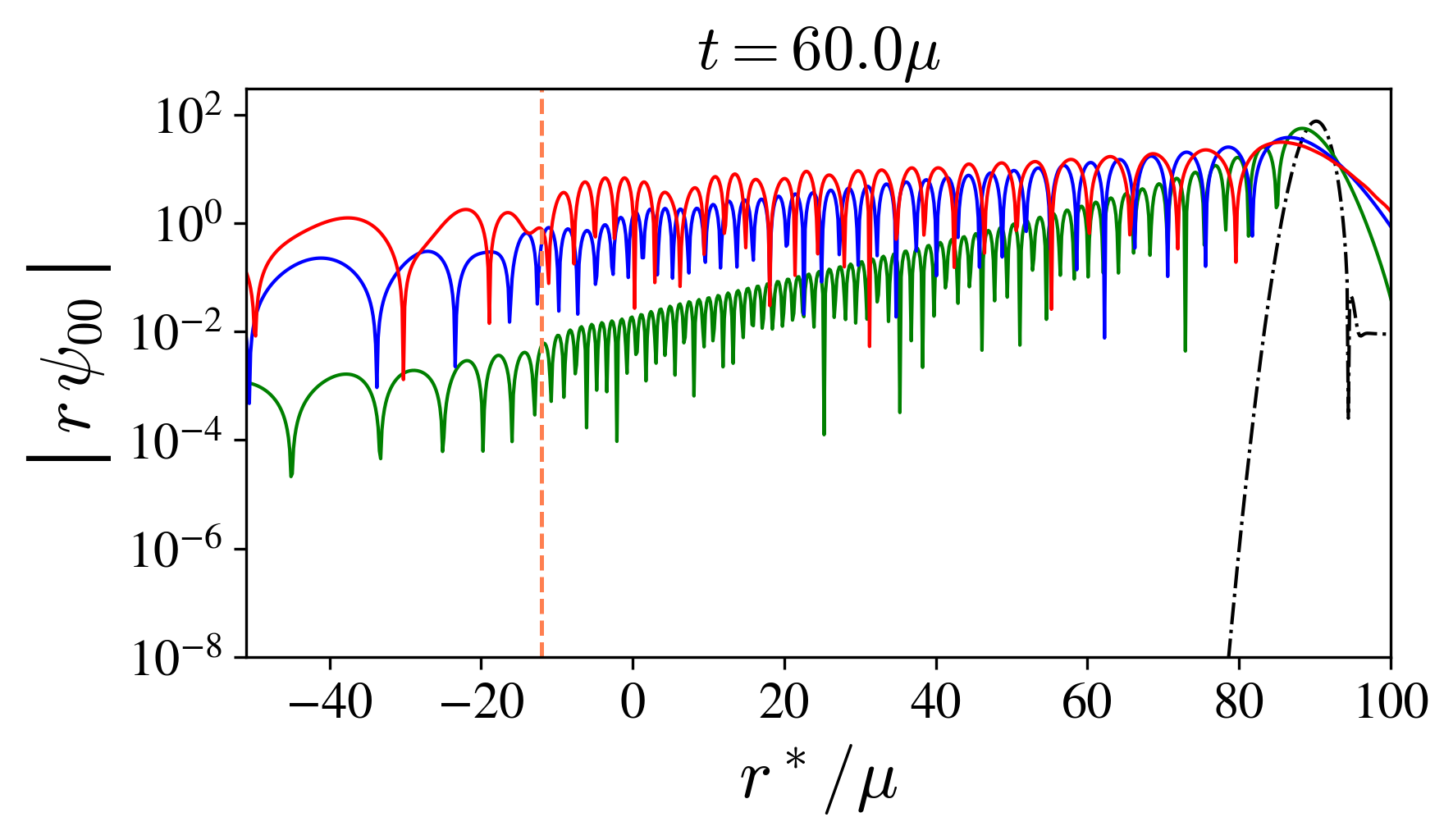}
\includegraphics[width=7.5cm]{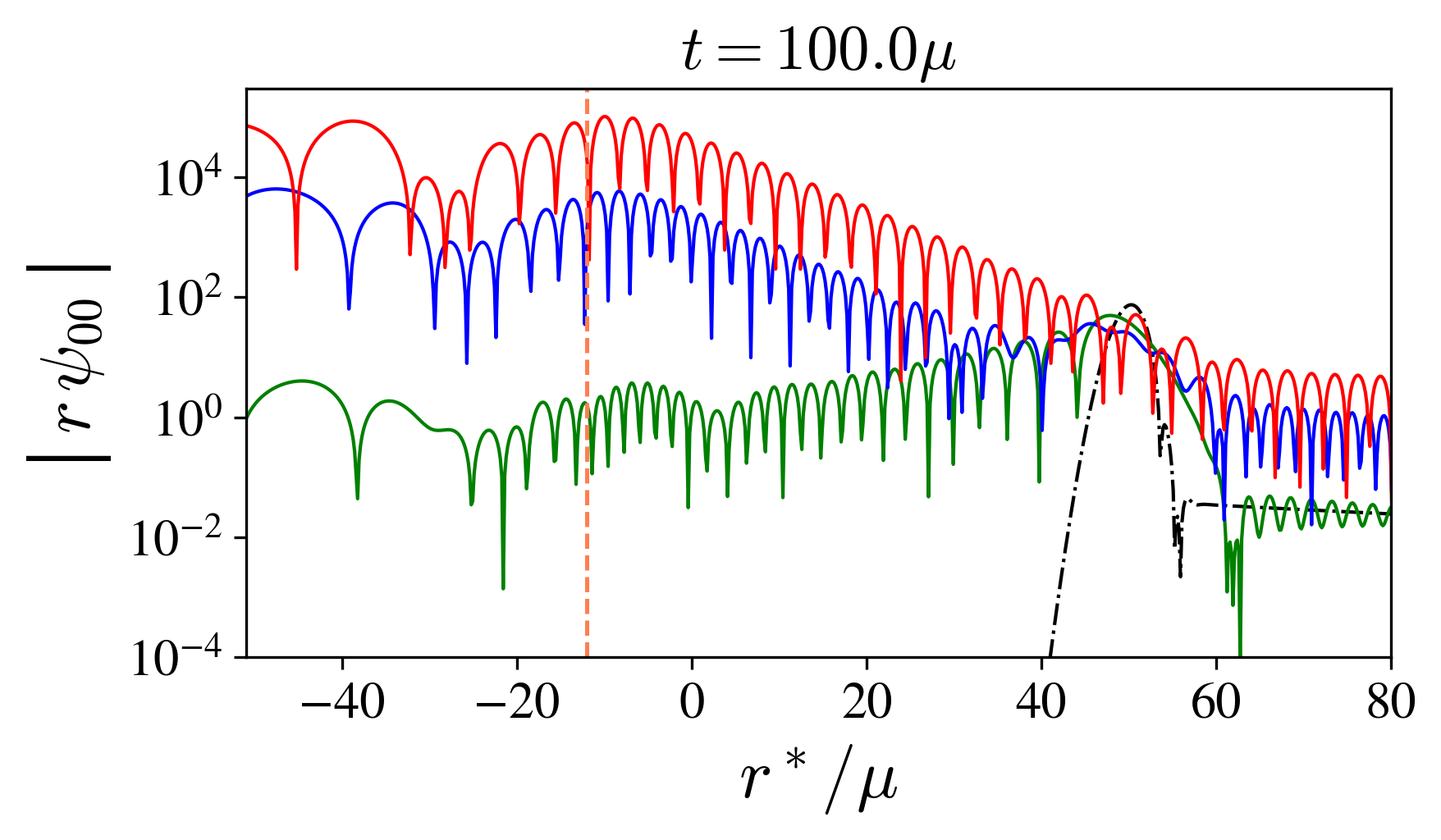}
\caption{Lifshitz evolution of a static pulse, for different values of $\kappa_2$ and fixing $\kappa_3=0.01$. The green curve corresponds to $\kappa_2=0.1$, the blue one to $\kappa_2=0.5$, while the red curve corresponds to $\kappa_2=1$. At early times, the propagation speed of the slowest modes decreases when increasing $\kappa_2$. At late times, instead, the bump  at the Killing horizon (orange dotted vertical line) grows when increasing $\kappa_2$. The evolution of the wave equation from the same static pulse is shown in dotted black. The parameters of the initial static pulse are the same as those considered in Figure \ref{plot-evo-lif-ID1-vark3}.}
\label{plot-evo-lif-ID1-vark2}
\end{figure}

\begin{figure}
\centering
\includegraphics[width=7.5cm]{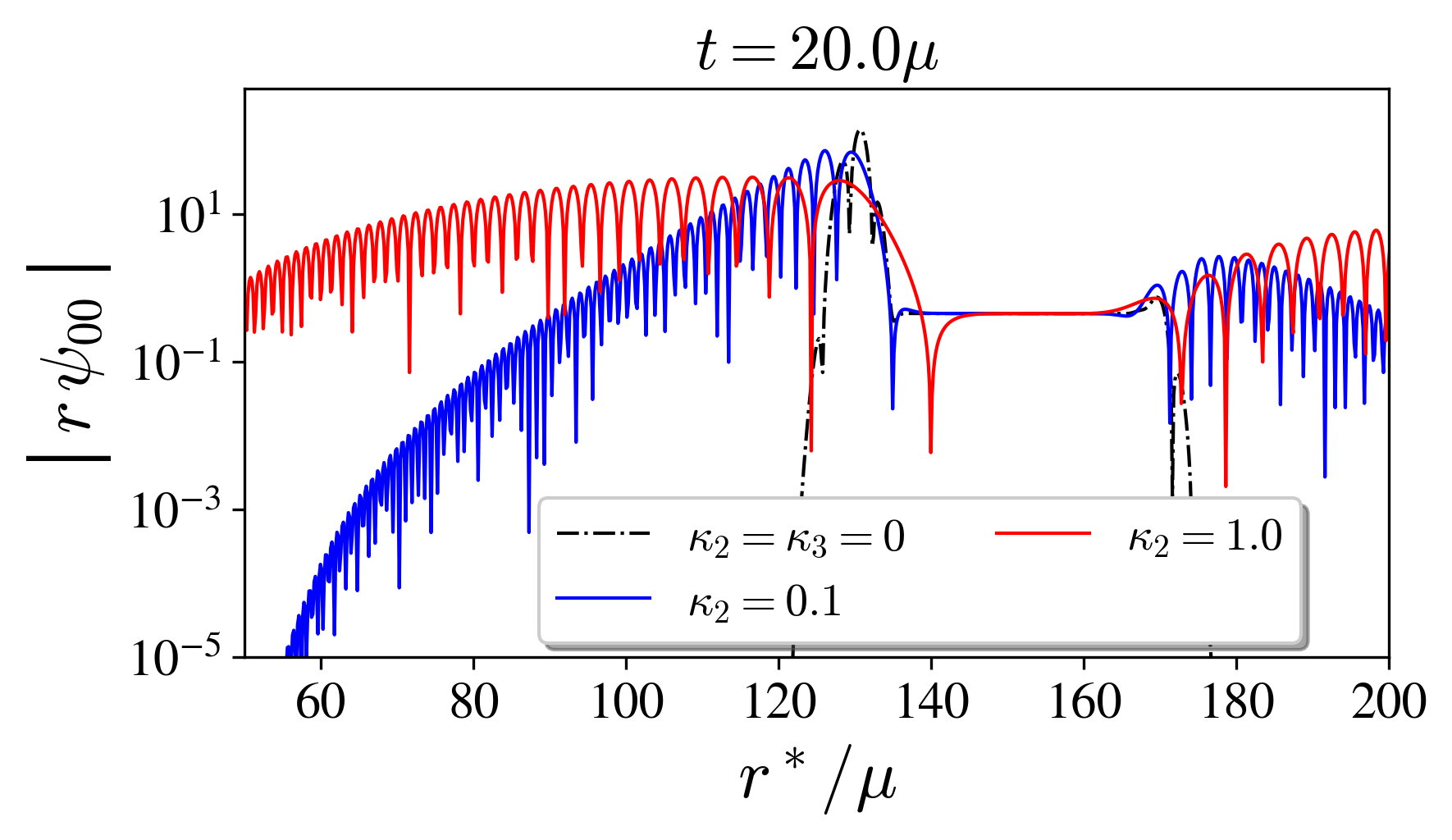}
\includegraphics[width=7.5cm]{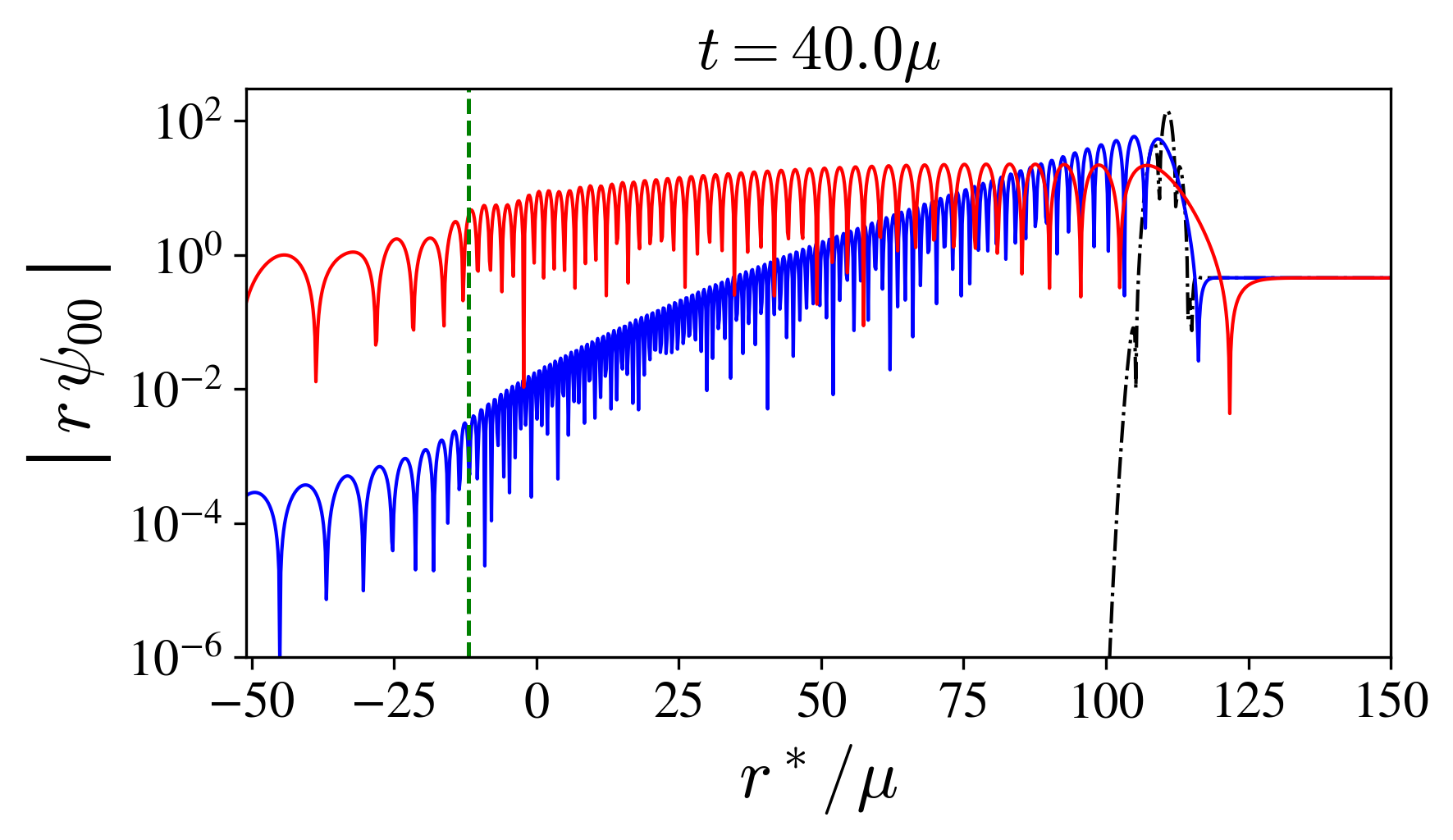}
\includegraphics[width=7.5cm]{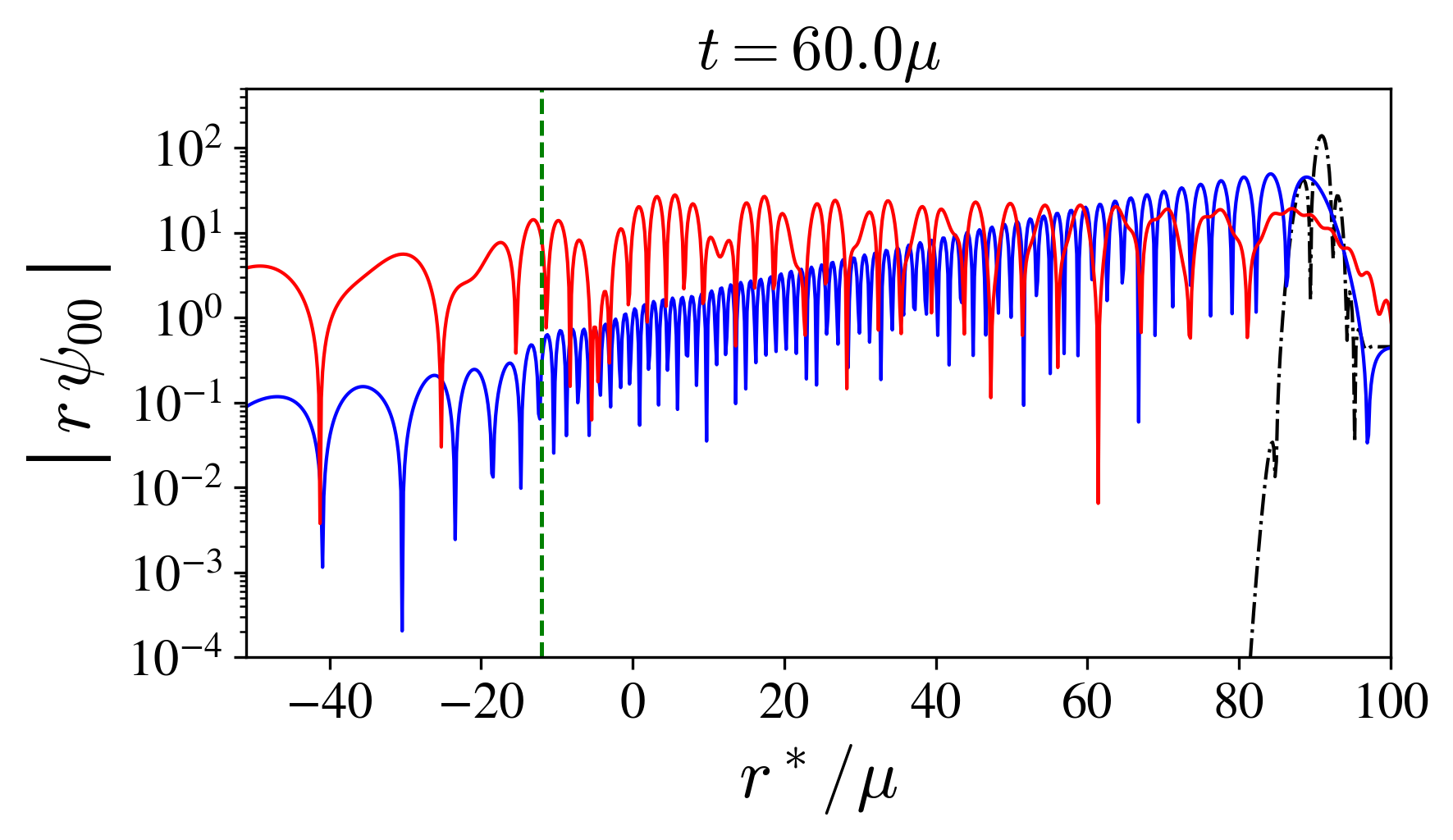}
\includegraphics[width=7.5cm]{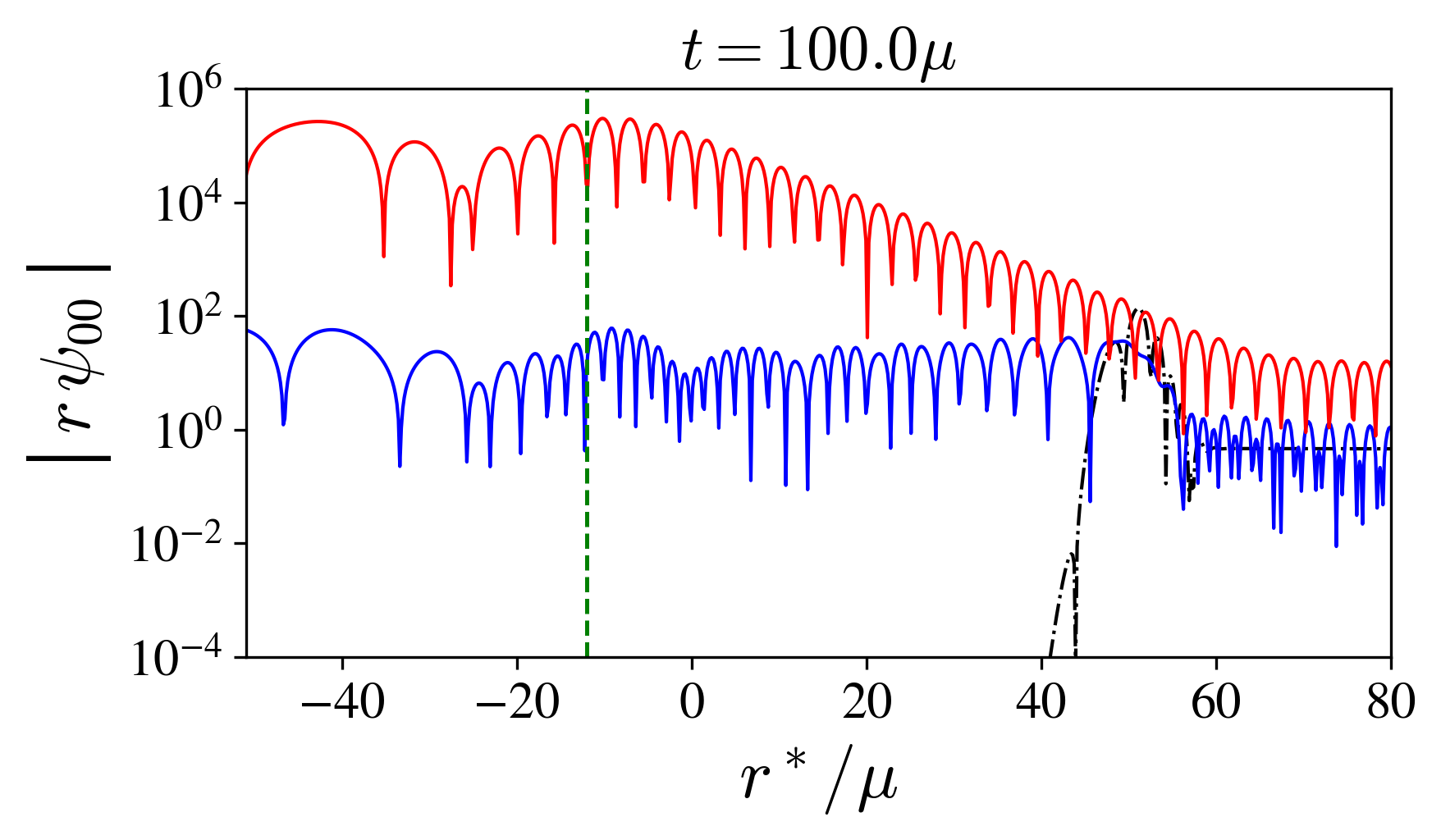}
\caption{Lifshitz evolution of an (approximately) ingoing pulse, for fixed $\kappa_3=0.01$, $\kappa_2=0.1$ (blue curve) and $\kappa_2=1$ (red curve). The dotted black profile corresponds to the solution of the wave equation, for comparison. The initial data parameters are the same as those considered in Figure \ref{plot-evo-lif-ID2-vark3}.}
\label{plot-evo-lif-ID2-vark2}
\end{figure}

Furthermore, in figures \ref{plot-evo-lif-ID1-vark2} and \ref{plot-evo-lif-ID2-vark2} we display the evolution of the solution for fixed $\kappa_3$,  varying instead $\kappa_2$. In particular, we set $\kappa_3=0.01$ and consider $\kappa_2=\{0.1,0.5,1.0\}$. A cascade also develops in this case, but we observe a decrease in the propagation speed of the slowest modes when increasing $\kappa_2$, at least at early times. At later times ($t\sim 60\mu$), and once the solutions have reached the Killing horizon, the order of magnitude of the wave profiles with $\kappa_2=0.5$ and $\kappa_2=1.0$ remain similar, until a bump around the horizon forms and starts growing faster as we increase $\kappa_2$.

The appearance of such a bump in the region  between the universal horizon and the Killing horizon is interesting, since it might signal an instability of the space-time background solution under certain assumptions\footnote{The case of $\kappa_3=1$ is also reported and analysed in Appendix C.}. In particular, let us note that Ho\v rava gravity propagates a scalar degree of freedom together with the usual transverse traceless graviton perturbation \cite{Blas:2010hb}. When expanded around the background solution considered here, the dynamics of the scalar mode must exhibit, for consistency, a Lifshitz scaling of the form \eqref{eq:lifshitz_scaling}, and hence the equation of motion for scalar perturbations must unavoidably take the form \eqref{eq:equation_final}, with the parameters $c_2$ and $c_3$ somehow related to the couplings in the gravitational action. Hence, we can conjecture that, if the bump found in our numerical experiments is a generic feature of \eqref{eq:equation_final}, it will also develop in the gravitational case, therefore signaling a linear instability of the universal horizon \citep{instability-UH}.


\subsection{Effect of the potential barrier}

\begin{figure}
\centering
\includegraphics[width=7.5cm]{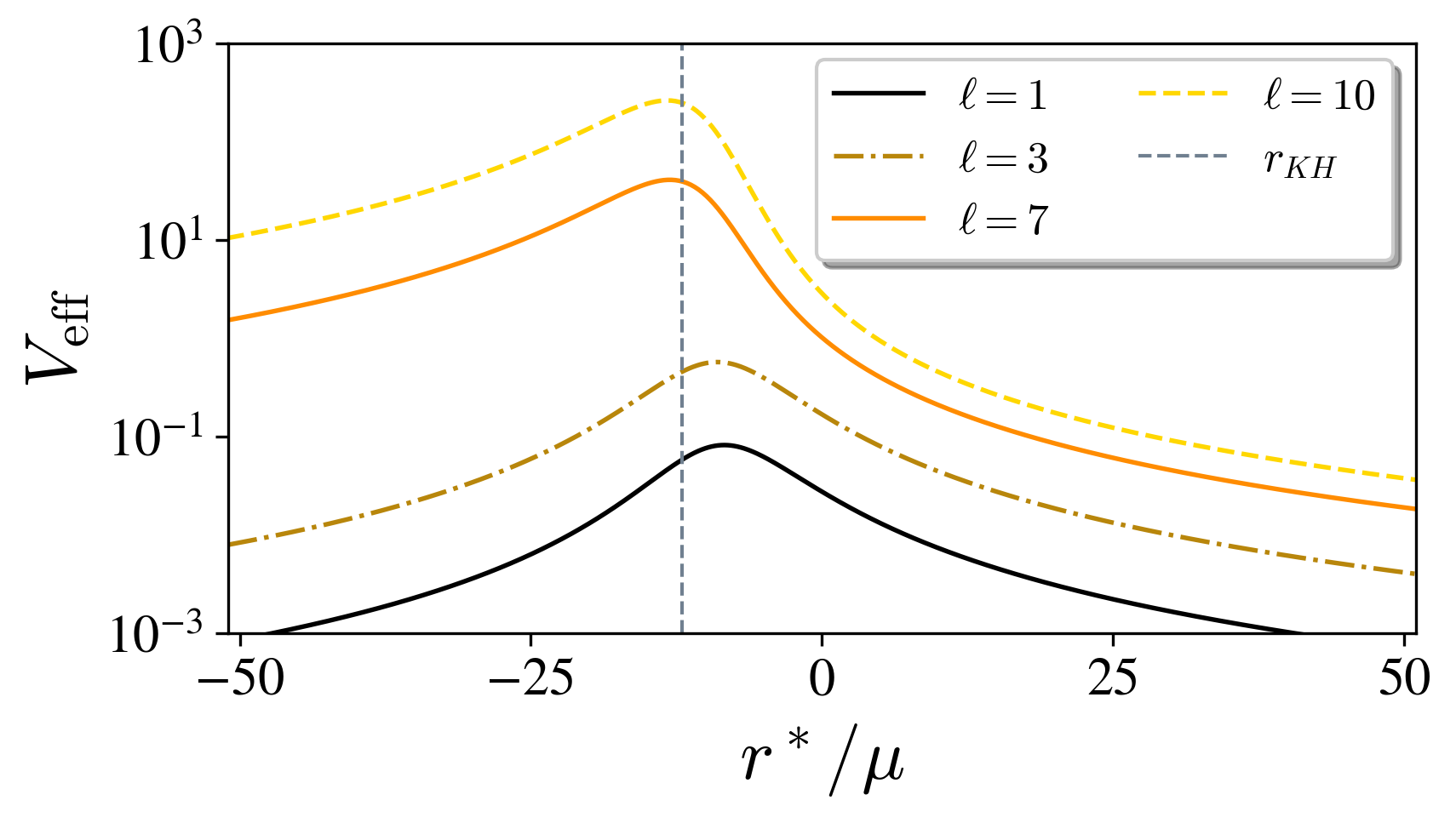}
\includegraphics[width=7.5cm]{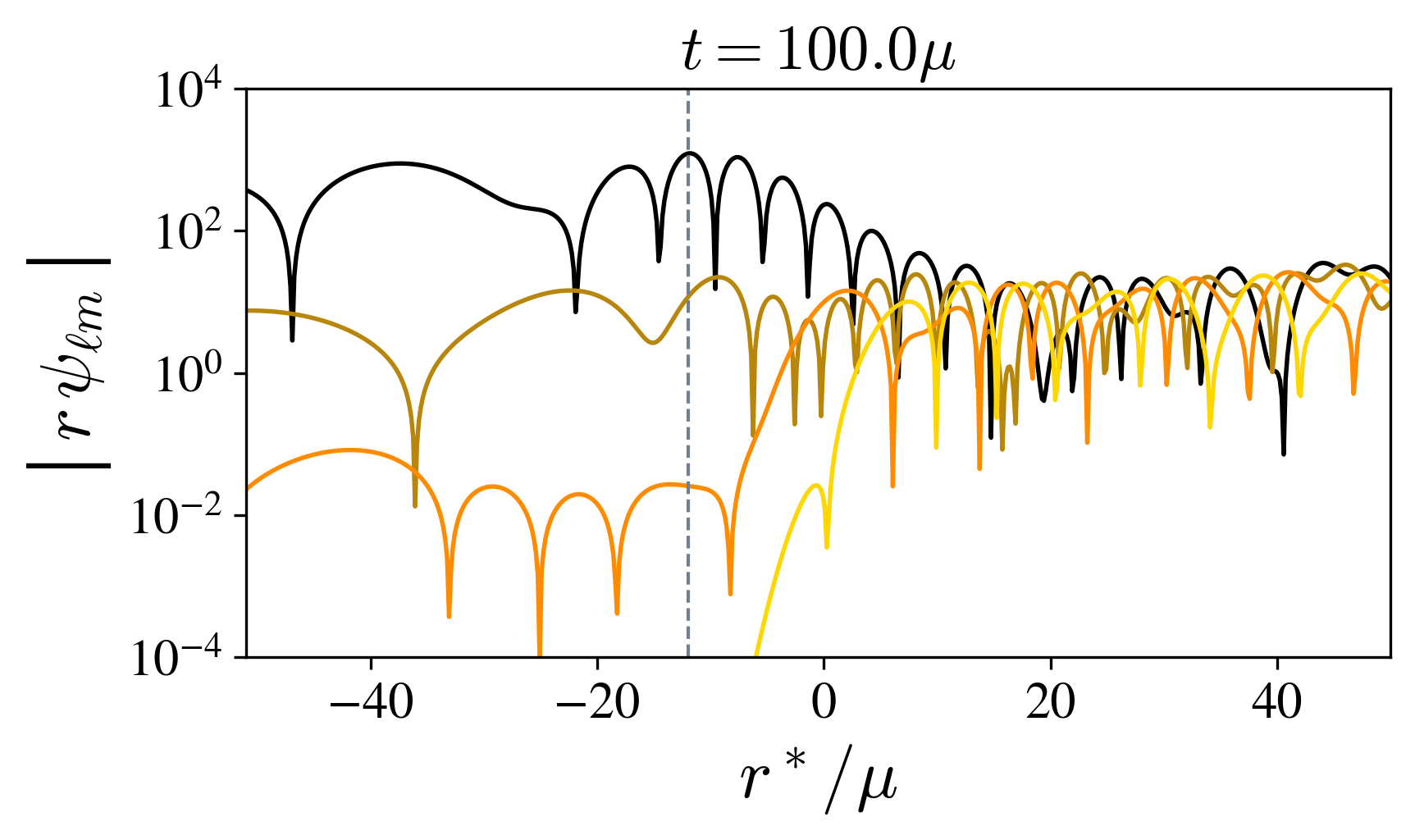}
\caption{\textit{Effect of the centrifugal barrier}. \textbf{Left panel:} Plot of the effective potential $V_{\rm eff}$ (in log-scale) as a function of $r^*$ (in units of the black hole's mass, $\mu$), for angular numbers $\ell = 1$ (black), $\ell = 3$ (dotted golden), $\ell = 7$ (orange) and $\ell = 10$ (dotted yellow). As $\ell$ increases, the maximum of the potential grows by several orders of magnitude. \textbf{Right panel:} Evolution of different wave modes of the Lifshitz field, with the same angular numbers and colors as in the left panel. For higher harmonics, the maximum of the effective potential is larger, acting as a stronger centrifugal barrier, thus preventing the corresponding modes from penetrating the horizon.}
\label{plot-effect-potential-barrier}
\end{figure}

The coefficient multiplying the term linear in $\psi_{\ell m}$ in \eqref{evo-equation-modes} acts as an effective centrifugal barrier for modes with $\ell\neq 0$. Its explicit expression reads
\begin{equation} \label{eq:Veff}
\begin{aligned}
    V_{\text{eff}}(r) {} & = \frac{H^2 \ell (\ell+1)}{32 A^9 r^6} \left(8 A^7 \left(r^4+\kappa_2 \ell (\ell+1) r^2+\kappa_3 \ell^2 (\ell+1)^2\right)+4 A^5 H \left(r H' \left(\kappa_2 r^2+3 \kappa_3 \ell
   (\ell+1)\right) \right.\right. \\
   & \left.\left. -H \left(\kappa_2 r^2+7 \kappa_3 \ell (\ell+1)\right)\right)+A^3 H \kappa_3 \left(12 H^3-r^3 \left(H'\right)^3+H r^2 H' \left(11 H'-4 r H''\right)-H^2 r \left(r^2 H^{'''} \right.\right.\right. \\
   & \left.\left.\left. -6 r H''+22
   H'\right)\right)+15 H^4 \kappa_3 r^3 \left(A'\right)^3-4 A^4 H^2 r A' \left(\kappa_2 r^2+3 \kappa_3 \ell (\ell+1)\right)+A H^3 \kappa_3 r^2 A' \left(A' \left(23 H \right.\right.\right.\\
   & \left.\left.\left. -25 r H'\right)-10 H r A''\right)+A^2 H^2
   \kappa_3 r \left(11 r^2 A' \left(H'\right)^2+H r \left(7 r A' H''+H' \left(7 r A''-34 A'\right)\right)+H^2 \left(r^2 A^{'''} \right.\right.\right. \\
   & \left.\left.\left. -6 r A''+22 A'\right)\right)\right),
\end{aligned}
\end{equation}
where the prime denotes differentiation with respect to $r$, and we have omitted the argument in the functions $A(r)$ and $H(r)$ for the sake of simplicity. Despite this highly non-linear expression, $V_{\rm eff}$ has a single maximum for fixed $\ell$, in the vicinity of the Killing horizon. Its value grows quickly with $\ell$, as can be noticed from the left panel of figure \ref{plot-effect-potential-barrier}. 

The effect of such a potential in the dynamics of the Lifshitz field is stronger for larger values of the angular momentum eigenvalue. This behavior has been verified numerically, setting the coupling constants to $\kappa_2=0.5$ and $\kappa_3=0.01$. The results are reported in the right panel of figure \ref{plot-effect-potential-barrier}. As  can be seen, higher harmonics are pushed away by the barrier, which prevents them from penetrating the Killing horizon at late times. This suggests that the physics of the field within the interior region is captured by the first modes.

\subsection{Convergence tests}

In order to validate our numerical code, we performed several convergence tests, aiming to assess the corresponding accuracy order of the simulations. A simple analytic calculation for Eq. \eqref{master-equation} in flat space --- which matches the characteristic structure of the full equation at high frequencies --, shows that the difference between an exact solution and the numerical approximation using the implicit scheme \eqref{num-scheme-lif1}-\eqref{num-scheme-lif2}, with second-order accurate finite-difference operators, is given by
\begin{equation}
    |\phi_{\text{exa}}-\phi_{\text{num}}|= \alpha_1 (\Delta r^*)^2 + \alpha_2 (\Delta t)^2 + \alpha_3 (\Delta r^*) (\Delta t)+\mathcal{O}(\Delta^3),
\end{equation}
where $\alpha_1,\alpha_2$, and $\alpha_3$ are functions on the grid, and $\mathcal{O}(\Delta^3)$ denote all terms that are cubic in time and spatial steps, like $(\Delta t)(\Delta r^*)^2$. Thus, by rescaling $\Delta r^*\to\lambda\Delta r^*$ and $\Delta t\to\lambda\Delta t$ for some $\lambda\in\mathbb{R}$, one gets 
\begin{equation}\label{conv-scaling}
    |\phi_{\text{exa}}-\phi_{\text{num}}|=\mathcal{O}(\lambda^2),
\end{equation}
which indicates that the scheme is second-order accurate.

\begin{figure}
\centering
\includegraphics[width=7.5cm]{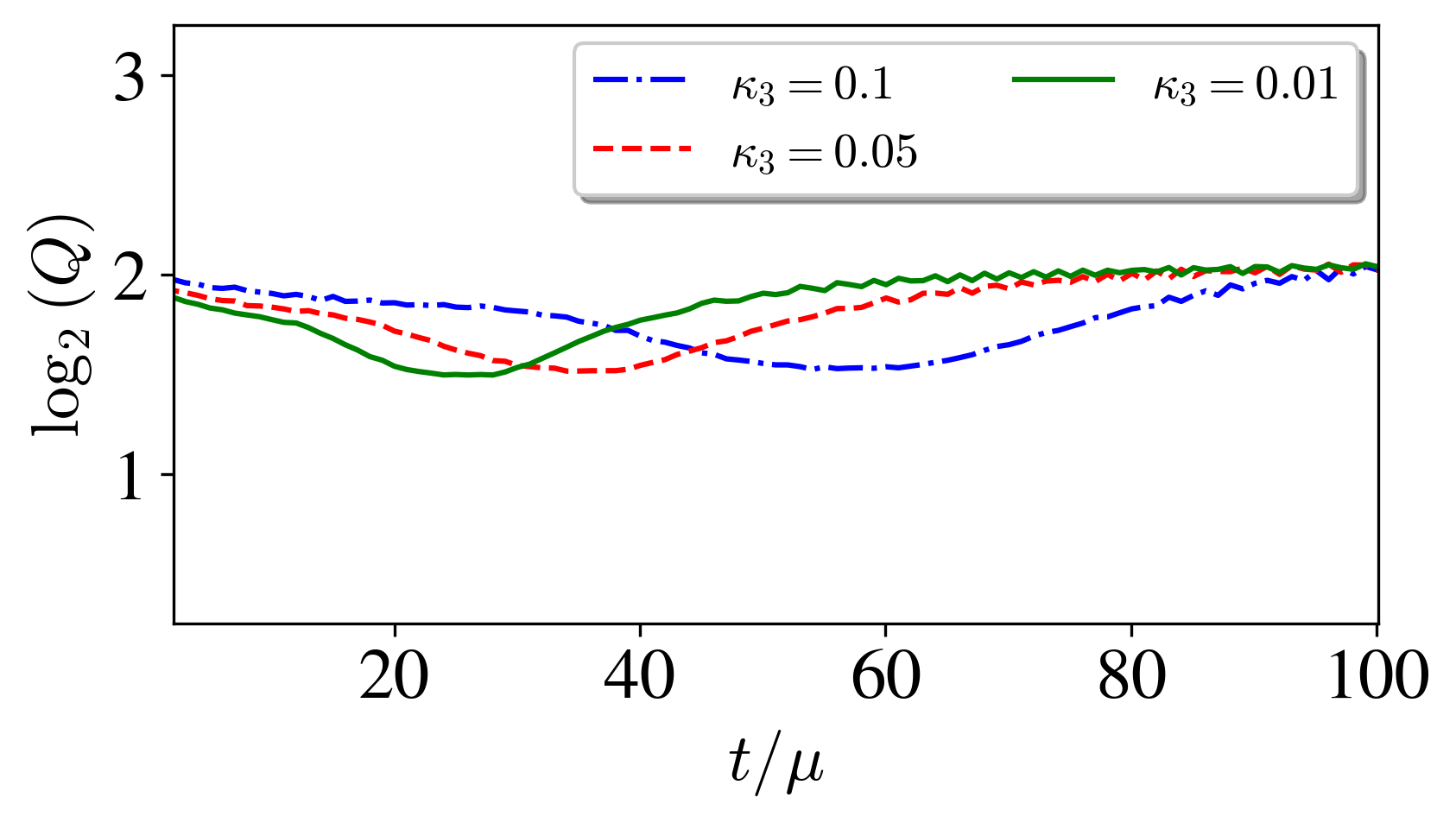}
\caption{\textit{Convergence tests.} Accuracy order of the implicit scheme, with coupling parameters $\kappa_2 = 0.1$ and $\kappa_3 = 0.01$ (green continuous curve); $\kappa_3 = 0.05$ (red dotted curve) and $\kappa_3 = 0.1$ (blue dotted curve), as a function of time (in units of the black hole mass, $\mu$). The simulations are performed from the initial data ID type I, with initial steps $\Delta t = 0.1$ and $\Delta r^* = 0.3$, taking $\lambda = 1$ (low resolution), $\lambda = 0.5$ (medium resolution) and $\lambda = 0.25$ (high resolution).}
\label{plot-convergence}
\end{figure}

In order to confirm that Eq.~\eqref{conv-scaling} approximately holds during the numerical evolution, we perform three different runs with $\lambda=1$ (low resolution), $\lambda=0.5$ (medium resolution) and $\lambda=0.25$ (high resolution). Then, we define the ratio \cite{NumRecipes}
\begin{equation}
    Q(t)=\frac{\lVert\phi_{\text{low}}-\phi_{\text{med}} \rVert}{\lVert\phi_{\text{med}}-\phi_{\text{high}} \rVert},
\end{equation}
which behaves as $Q(t)\sim 2^p$, with $p$ the accuracy order of the desired numerical scheme (i.e. $p=2$ in our case). The result of these tests is shown in Figure \ref{plot-convergence}, where the value of $p$ is computed as a function of time for different values of $\kappa_3$ (fixing $\kappa_2=0.1$). We can see that the method is approximately second-order accurate, as expected, with convergence improving for smaller values of $\kappa_3$. Finally, and as a last consistency check of the numerical code, an independent residual evaluator test was also done, whose results are reported in Appendix C.

\begin{figure}
\centering
\includegraphics[width=7.5cm]{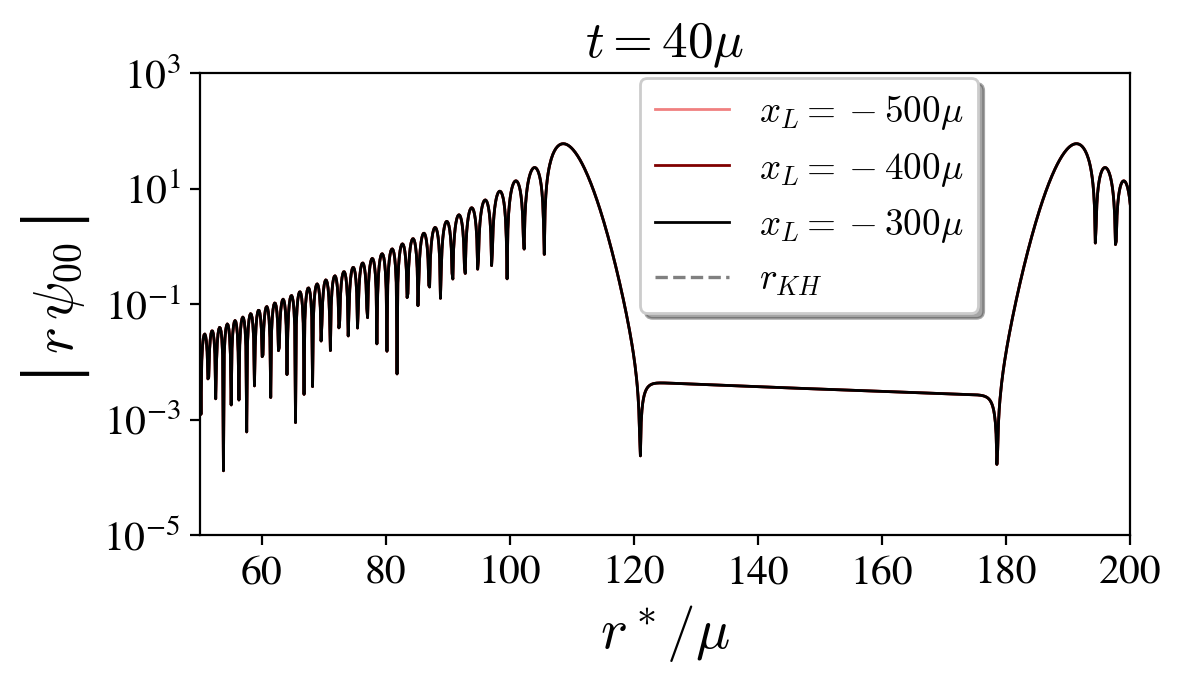}
\includegraphics[width=7.5cm]{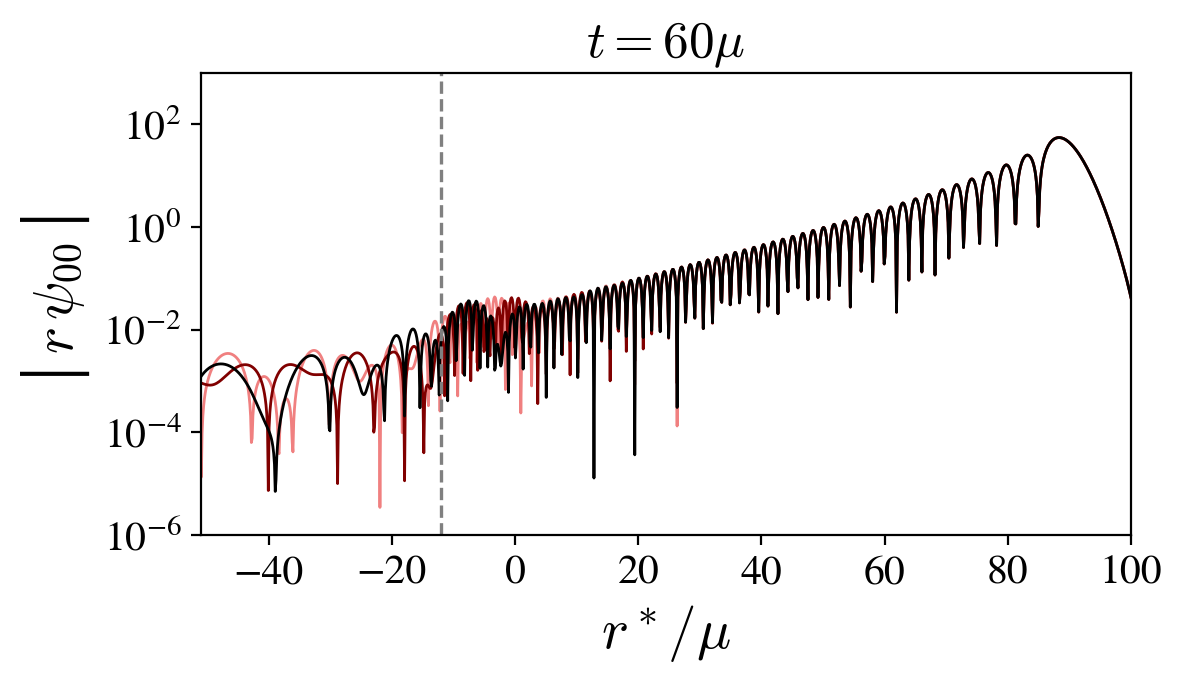}
\includegraphics[width=7.5cm]{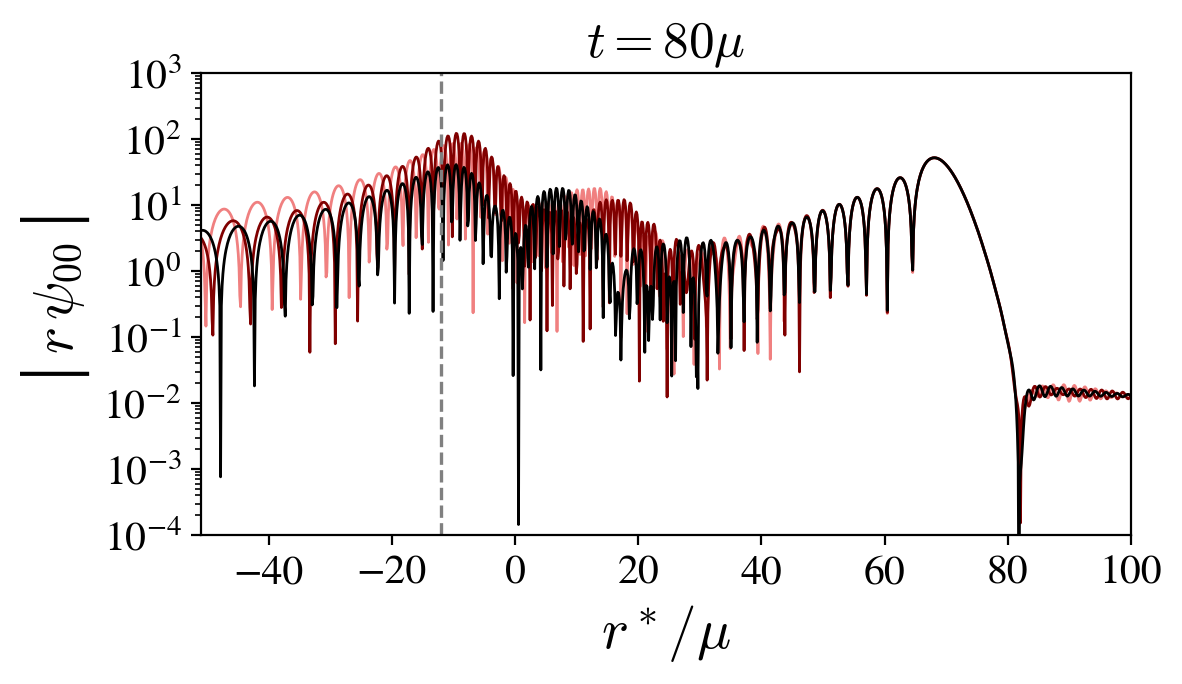}
\includegraphics[width=7.5cm]{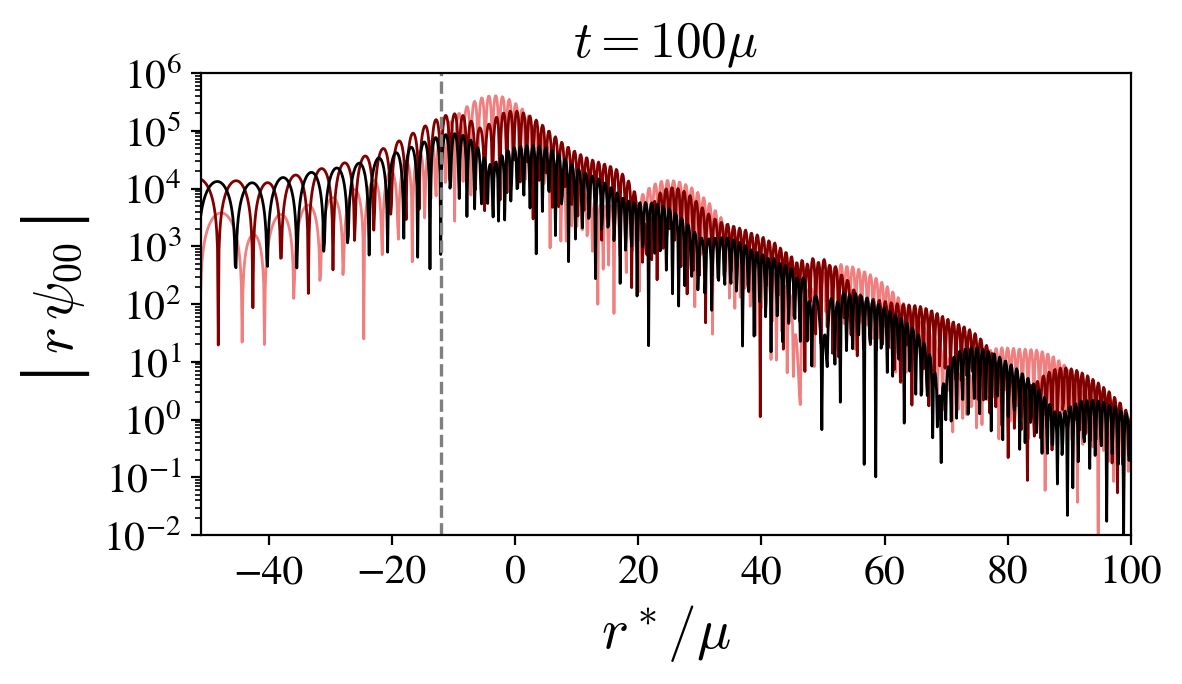}
\caption{\textit{Effect of the dissipative layer.} Evolution of the Lifshitz field for different positions of the left side of the ADL: $x_L=-300\mu$ (black); $x_L=-400\mu$ (maroon) and $x_L=-500\mu$ (coral). At early times, the evolution remains almost unchanged, until $t\sim 60\mu$, when small differences start  showing up. Nevertheless, the global behavior of the field is the same in all cases, even at late times. Notice in particular the appearance of the bump close to the Killing horizon. The simulations were run with $\kappa_2 = 0.1$, $\kappa_3=0.01$, $\Delta t = 0.1$, and $\Delta r^*=0.125$, using the initial data \textit{ID type I}. The position of the right side of the layer is fixed at $x_R = 300\mu$.}
\label{plot-indep-layer}
\end{figure}

We also explore the dependence of our results on the position of the ADL. For doing so, we perform three runs with the same initial parameters, using the initial data ID Type I, and setting $\kappa_2 =0.1$, and $\kappa_3 =0.01$. We test three different positions of the left wall of the layer, namely $x_{\rm L}=\{-300\mu,-400\mu,-500\mu\}$, while keeping the right wall fixed far from the horizon, at $x_{\rm R} = 300\mu$. Results are shown in figure \ref{plot-indep-layer}. We notice that there is a minimum distance between the left wall of the ADL and the Killing horizon, above which the dynamical features of the field remain almost unchanged. If the layer is close enough to the boundary of the physical region of interest, the dynamics at long times is altered by its presence, contrary to what happens if the layer is placed far away. Such a minimum position depends on the final evolution time, with the layer having to be placed further and further away if longer evolution times are required. Nevertheless, the computational cost of placing the layer far from the physical region -- i.e., the number of grid points needed in order to keep the same spatial resolution -- is kept reasonable thanks to the implicit character of the method, as larger time steps are allowed, unlike in standard explicit methods. 

Focusing on the particular simulations considered here (where we evolve until $t \sim 100 \mu$), this analysis suggests that choosing $x_{\rm L} \sim -300 \mu$ is optimal. Although there may be small effects from the layer -- no matter how far it is placed --, the general features of the solution remain robust. This allows us to draw solid conclusions on the features of the Lifshitz field \textit{independently} of the layer, as discussed above.

\section{Conclusions}
\label{sec:conclusions}

In this paper we have introduced an implicit numerical scheme that allows us to solve evolution equations with an anisotropic scaling between time and space of the form \eqref{eq:lifshitz_scaling}. Our approach is based on a generalization of the Crank-Nicolson method for diffusion equations, replacing the usual discretization of spatial derivatives by finite-difference operators with averages of them evaluated in two consecutive time steps. This allows for evading the stringent stability bound implied by the CFL condition for standard explicit methods, removing obstructions to high-resolution and long-time evolutions. Our implicit scheme is free 
from the CFL constraint, and can therefore be evolved for substantially long times with small grid sizes. Let us highlight that this method is equally valid for linear and non-linear equations, as long as they feature some form of anisotropic scaling.

As an application and proof of concept of our numerical scheme, we have studied the case of a Lifshitz scalar field in four dimensions, propagating in a spherically symmetric and static black-hole space-time, solution to the equations of motion of Ho\v rava gravity at low energies. In contrast to the case of fields with a relativistic dispersion relation, the Lifshitz scalar field can probe the region enclosed by the Killing horizon, freely escaping from it. Instead, it is the universal horizon, sitting at a smaller radius, that represents the inner semi-permeable boundary for the propagation of the field within the background geometry. 

We have performed simulations with varying values of the higher derivative couplings, $\kappa_2$ and $\kappa_3$, with two classes of initial data: a static Gaussian pulse and an approximately ingoing wave-packet. Being unable to disentangle ingoing and outgoing modes exactly due to the dispersive character of the equations, we have implemented boundary conditions by introducing an artificial dissipative layer, essentially absorbing the modes far from the physical region of interest. 

Our results show a consistent picture, where UV modes of the field solution develop a cascade near the Killing horizon, growing stronger with larger values of the couplings accompanying higher derivatives. At late times, this cascade accumulates in the region between the Killing and universal horizons, producing a bump in the amplitude, which grows exponentially. This may have important implications for the fate of the universal horizon, which has so far been studied only in the low energy limit of Ho\v rava gravity. The effect of higher derivatives on its structure and stability  is therefore unknown so far. Provided that one could model the scalar mode contained in the dynamical degrees of freedom of Ho\v rava gravity as a Lifshitz field, our results seem to strongly indicate a linear instability of the universal horizon. We have also shown that a large centrifugal barrier with peak outside the Killing horizon develops for higher harmonics, so that only the first few modes can effectively penetrate the Killing horizon.
 
Finally, we have shown that our methods retain an approximately second-order convergence along time evolution, and that the effect of the dissipative layer is negligible within the physical region, which allows us to trust on the robustness of our results.

The research presented here constitutes a first step within a larger program. Although here we have focused on the dynamics of the Lifshitz scalar field, which is linear, our implicit method is equally suitable for non-linear equations. In particular, this seems the way to approach the non-linear and non-perturbative evolution of the fully general equations of motion of Ho\v rava gravity, which also present a dispersive character with anisotropic scaling. We hope that our method can provide stable simulations even in this more challenging situation.


\section*{Acknowledgements}
We are grateful to Stefano Liberati, Shinji Mukohyama, Marc Schneider, and Toby Wiseman for discussions throughout the realization of this work. M. H-V. wants to thank the APP group at SISSA for their hospitality during the completion of this work. M. R., A. K., M. B. and E. B. acknowledge support from the European Union’s H2020 ERC Consolidator Grant “GRavity from Astrophysical to Microscopic Scales” (Grant No. GRAMS-815673) and the EU Horizon 2020 Research and Innovation Programme under the Marie Sklodowska-Curie (Grant Agreement No. 101007855). The work of M. H-V has been supported by the Spanish State Research Agency MCIN/AEI/10.13039/501100011033 and the EU NextGenerationEU/PRTR funds, under grant IJC2020-045126-I; and by the Departament de Recerca i Universitats de la Generalitat de Catalunya, Grant No 2021 SGR 00649. IFAE is partially funded by the CERCA program of the Generalitat de Catalunya.


\appendix

\section{Black holes in Einstein-\AE ther and Ho\v rava Gravity}\label{app:EA_gravity}

As discussed in the main text, the background space-time \eqref{eq:bg_sol} is solution to the equations of motion of Einstein-\AE ther gravity \cite{Jacobson:2000xp}, with action
\begin{align}\label{eq:action_EA}
    S_{\rm EA}=\frac{1}{16\pi G}\int \mbox{d}^4x \sqrt{|g|} \left(-R+{\cal L}_{\rm U} + \eta(U_\m U^\m-1)\right),
\end{align}
where $R$ is the Ricci scalar, $\eta$ is a Lagrange multiplier implementing the unit norm condition of the \ae ther,  and ${\cal L}_{\rm U}=K^{\alpha\beta}_{\m\n}\nabla_\alpha U^\m \nabla_\beta U^\n$, with $K^{\alpha\beta}_{\m\n}=c_1 g^{\alpha\beta}g_{\m\n}+c_2 \delta^\alpha_\m \delta^\beta_\n +c_3 \delta^\alpha_\n \delta^\beta_\m + c_4 U^\alpha U^\beta g_{\m\n}$ and couplings $c_i\in\mathbb{R}$. 

At the two derivative level, this is the most general covariant action that leads to Lorentz violations and modified dispersion relations for all degrees of freedom in the gravitational action -- the usual graviton, plus  vector and scalar modes contained in $U^\m$ \cite{Jacobson:2004ts} --, as well as for generic matter fields coupling to them -- perhaps through renormalization group flow in the latter case. As such, it includes specific models as particular cases of the coefficients $c_i$. If one imposes the \ae ther to be hypersurface orthogonal at the level of the action, and identifies $c_1=c_3$, $c_2=(1-\beta')(\lambda-1)-\beta'$, $\beta'\equiv c_1+c_3= 1-\eta^{-1}$, and $c_1+c_4=\alpha$, \eqref{eq:action_EA} reproduces the low energy limit of Ho\v rava Gravity, also known as khronometric gravity \cite{Blas:2010hb} -- fully described by three couplings $(\alpha,\lambda,\eta)$.

Spherically symmetric and static solutions to \eqref{eq:action_EA} can be searched for with the ansatz \cite{Berglund:2012bu,Barausse:2011pu}
\begin{align}
    \ud s^2=f(r)\ud \tau^2-\frac{B(r)^2 }{f(r)}\ud r^2-r^2\ud \Omega^2, \quad   U_\m \ud x^\m=\frac{H(r)}{2A(r)}\ud t+\frac{B(r)(1-f(r)A(r)^2)}{2A(r)f(r)}\ud r,
\end{align}
where $H(r)=1+f(r)A(r)^2$, and the form of $U^\m$ is chosen for convenience, automatically satisfying $U_\m U^\m=1$. Accidentally, spherical symmetry and staticity automatically impose hypersurface orthogonality, and hence all solutions to Einstein-\AE ther gravity with these isometries are solutions to Ho\v rava gravity as well (and vice versa). Hence, both theories are generally studied together when discussing their features and phenomenology in static spherically symmetric configurations. This is not true when any of the conditions above are relaxed. In particular, axisymmetric solutions in Einstein-\AE ther gravity are not hypersurface orthogonal \cite{Barausse:2015frm,Adam:2021vsk}. We also do not expect them to be equivalent when higher derivatives are included in the gravitational action, even in the spherically symmetric and static case.

The explicit form of the functions $f(r), B(r)$, and $A(r)$ must be found solving the equations of motion explicitly. Although numerical solutions can always be achieved at any generic point of the parameter space~\cite{Barausse:2011pu}, there exist two corners where analytic solutions can be attained \cite{Berglund:2012bu}. These correspond to $c_{14}=0$ and $c_{123}=0$, where we are using the notation $c_{ij\dots k}=c_i+c_j+\dots+c_k$. The latter case, although leading to simpler functions, is incompatible with observational bounds constraining the parameter space of the theory \cite{Gupta:2021vdj}. The former case, instead, is perfectly compatible with current bounds and leads to
\begin{align}
    &f(r)=1-\frac{2\mu}{r}-c_{13} \frac{r_{\text{\ae}}^4}{r^4}, \\
    &B(r)=1,\\
    &A(r)=\frac{1}{f(r)}\left(-\frac{r_{\text{\ae}}^2}{r^2}+\sqrt{f(r)+\frac{r_{\text{\ae}}^4}{r^4}}\right).
\end{align}

The parameter $r_{\text{\ae}}$ is in principle arbitrary. However, the solution displays a singularity at the universal horizon, unless 
\begin{align}
     r_{\text{\ae}}=\frac{\mu}{2}\left(\frac{27}{1-c_{13}}\right)^{1/4}.
\end{align}
This value is therefore chosen to ensure regularity of the solution everywhere except for the central singularity, sitting at $r=0$.

At high energies, Ho\v rava gravity departs from action \eqref{eq:action_EA} by higher derivative spatial terms \cite{Blas:2010hb}. The contribution of these operators -- which are critical for achieving power-counting renormalizability --, and in particular their backreaction onto space-times of the form \eqref{eq:bg_sol}, are unknown. Although here we assume that such effects are small and that gravitational perturbations are negligible, we leave this as an open question for the future.

\section{Coefficients of the evolution equation}
\label{app-coeffs-explicit}
In this appendix, we explicitly give the form of the coefficients in Eq.~\eqref{evoeq-expl}.

\begin{equation} \nonumber
\begin{aligned}
    \zeta_{01} = {} & \frac{1}{64 A^{11} H r^5}\left\{16 \left(6 \kappa_3 \left(\ell^2+\ell\right)^2 H^3+r \left(((6-H) H-8) r^4+2 H^2 \kappa_2 \ell (\ell+1) r^2+3 H^2 \kappa_3 \left(\ell^2+\ell\right)^2\right) H'\right) A^9 \right.\\
    & \left. -16 H r
   \left(4 H r^4-4 r^4+H^2 \left(-r^4+2 \kappa_2 \ell (\ell+1) r^2+3 \kappa_3 \left(\ell^2+\ell\right)^2\right)\right) A' A^8-4 H^2 \left(12 \kappa_3 \ell (\ell+1) H^3 \right.\right.\\
   & \left. \left. +r \left(\left(2
   \kappa_2 r^2-8 \kappa_3 \ell (\ell+1)\right) H'+r \left(2 \left(\kappa_2 r^2+\kappa_3 \ell (\ell+1)\right) H''+r \left(\kappa_2 r^2+3 \kappa_3 \ell (\ell+1)\right)
   H^{(3)}(r)\right)\right) H^2 \right.\right.\\
   & \left. \left. -2 r^3 H' \left(\kappa_2 r H'+\left(\kappa_2 r^2+3 \kappa_3 \ell (\ell+1)\right) H''\right) H+r^3 \left(\kappa_2 r^2+3 \kappa_3 \ell (\ell+1)\right)
   \left(H'\right)^3\right) A^7+4 H^3 r \left(\left(\left(2 \kappa_2 r^2 \right.\right.\right.\right.\\
   & \left.\left.\left.\left. -8 \kappa_3 \ell (\ell+1)\right) A'+r \left(2 \left(\kappa_2 r^2+\kappa_3 \ell (\ell+1)\right) A''+r \left(\kappa_2 r^2+3
   \kappa_3 \ell (\ell+1)\right) A^{(3)}(r)\right)\right) H^2+r \left(r \left(\kappa_2 r^2 \right.\right.\right.\right.\\
   & \left.\left.\left.\left. +3 \kappa_3 \ell (\ell+1)\right) H' A''+A' \left(4 \kappa_3 \ell (\ell+1) H'+r \left(\kappa_2
   r^2+3 \kappa_3 \ell (\ell+1)\right) H''\right)\right) H-r^2 \left(\kappa_2 r^2 \right.\right.\right.\\
   & \left.\left.\left. +3 \kappa_3 \ell (\ell+1)\right) A' \left(H'\right)^2\right) A^6+H^2 r \left(\kappa_3 r^4 \left(H'\right)^5-2 H
   \kappa_3 r^3 \left(H'+2 r H''\right) \left(H'\right)^3+2 H^2 r^2 \left(\kappa_3 \left(2 \left(H'\right)^2 \right.\right.\right.\right.\\
   & \left.\left.\left.\left. +r \left(2 H''+r H^{(3)}(r)\right) H'+2 r^2 \left(H''\right)^2\right)-2 \left(\kappa_2 r^2+3 \kappa_3
   \ell (\ell+1)\right) \left(A'\right)^2\right) H'+H^3 r \left(-8 \left(\kappa_2 r^2 \right.\right.\right.\right.\\
   & \left.\left.\left.\left. +2 \kappa_3 \ell (\ell+1)\right) \left(A'\right)^2-16 r \left(\kappa_2 r^2+3 \kappa_3 \ell
   (\ell+1)\right) A'' A'+\kappa_3 \left(-H'' \left(6 H''+5 r H^{(3)}(r)\right) r^2-8 \left(H'\right)^2 \right.\right.\right.\right.\\
   & \left.\left.\left.\left. +H' \left(H^{(4)}(r) r^3+6 H'' r\right)\right)\right)+H^4 \kappa_3 \left(4 H'+r \left(r \left(H^{(5)}(r) r^2+4
   H^{(4)}(r) r+2 H^{(3)}(r)\right)-4 H''\right)\right)\right) A^5 \right.\\
   & \left. -H^3 r \left(\kappa_3 \left(4 A'+r \left(A^{(5)}(r) r^3+4 A^{(4)}(r) r^2+2 A^{(3)}(r) r-4 A''\right)\right) H^4+\kappa_3 r \left(r \left(r \left(H''
   \left(12 A''+5 r A^{(3)}(r)\right) \right.\right.\right.\right.\right.\\
   & \left.\left.\left.\left.\left. +5 r A'' H^{(3)}(r)\right)+2 H' \left(3 A^{(4)}(r) r^2+8 A^{(3)}(r) r+6 A''\right)\right)+2 A' \left(r \left(3 H^{(4)}(r) r^2+8 H^{(3)}(r) r+6 H''\right) \right.\right.\right.\right.\\
   & \left.\left.\left.\left. -12 H'\right)\right) H^3-r^2
   \left(12 \left(\kappa_2 r^2+3 \kappa_3 \ell (\ell+1)\right) \left(A'\right)^3+\kappa_3 \left(-24 \left(H'\right)^2+2 r \left(8 H''+r H^{(3)}(r)\right) H' \right.\right.\right.\right.\\
   & \left.\left.\left.\left. +11 r^2 \left(H''\right)^2\right) A'+\kappa_3 r H'
   \left(r A'' H''-2 H' \left(2 A''+3 r A^{(3)}(r)\right)\right)\right) H^2+2 \kappa_3 r^4 \left(H'\right)^2 \left(H' A''+5 A' H''\right) H \right.\right.\\
   & \left.\left. -3 \kappa_3 r^4 A' \left(H'\right)^4\right) A^4+H^4 \kappa_3 r^2 \left(\left(3
   A'' \left(6 A''+5 r A^{(3)}(r)\right) r^2+A' \left(11 A^{(4)}(r) r^2+32 A^{(3)}(r) r+18 A''\right) r \right.\right.\right.\\
   & \left.\left.\left. -16 \left(A'\right)^2\right) H^3+r \left(4 \left(4 H^{(3)}(r) r^2+7 H'' r+12 H'\right) \left(A'\right)^2+r \left(A''
   \left(80 H'+29 r H''\right)+46 r H' A^{(3)}(r)\right) A' \right.\right.\right.\\
   & \left.\left.\left. +25 r^2 H' \left(A''\right)^2\right) H^2-2 r^2 A' H' \left(A' \left(6 H'+8 r H''\right)-11 r H' A''\right) H+6 r^3 \left(A'\right)^2 \left(H'\right)^3\right)
   A^3 \right.\\
   & \left. -2 H^5 \kappa_3 r^3 A' \left(\left(14 \left(A'\right)^2+2 r \left(31 A''+15 r A^{(3)}(r)\right) A'+35 r^2 \left(A''\right)^2\right) H^2+r A' \left(85 r H' A''+A' \left(32 H' \right.\right.\right.\right.\\
   & \left.\left.\left.\left. +15 r H''\right)\right) H+5 r^2
   \left(A'\right)^2 \left(H'\right)^2\right) A^2+3 H^6 \kappa_3 r^4 \left(A'\right)^3 \left(A' \left(26 H+35 r H'\right)+70 H r A''\right) A-105 H^7 \kappa_3 r^5 \left(A'\right)^5\right\}
\end{aligned}
\end{equation}

\begin{equation} \nonumber
\zeta_{10} = \frac{(H-2) H r A'+A (2-H) r H'}{4 A^3 r}, \qquad \zeta_{11} = 2-\frac{4}{H}
\end{equation}

\begin{equation}\nonumber
\begin{aligned}
    \zeta_{02} = {} & \frac{1}{16 A^8 H^2 r^4}\left\{-16 A^8 \left(H^2 \left(2 \kappa_2 \ell (\ell+1) r^2+3 \kappa_3 \left(\ell^2+\ell\right)^2-r^4\right)+4 H r^4-4 r^4\right)+4 A^6 H^2 \left(6 H^2 \kappa_3 \ell (\ell+1) \right.\right.\\
    & \left.\left. +7 r^2
   \left(H'\right)^2 \left(\kappa_2 r^2+3 \kappa_3 \ell (\ell+1)\right)-2 H r \left(2 r H'' \left(\kappa_2 r^2+3 \kappa_3 \ell (\ell+1)\right)+H' \left(4 \kappa_2 r^2+\kappa_3 \ell
   (\ell+1)\right)\right)\right) \right.\\
   & \left. +35 H^6 \kappa_3 r^4 \left(A'\right)^4-A^4 H^2 r \left(H^2 r \left(4 \left(A'\right)^2 \left(\kappa_2 r^2+3 \kappa_3 \ell (\ell+1)\right)+\kappa_3 \left(22 r^2
   \left(H''\right)^2+6 \left(H'\right)^2 \right.\right.\right.\right.\\
   & \left.\left.\left.\left. +r H' \left(21 r H^{(3)}(r)+62 H''\right)\right)\right)+31 \kappa_3 r^3 \left(H'\right)^4-4 H \kappa_3 r^2 \left(H'\right)^2 \left(17 r H''+10 H'\right)+2 H^3 \kappa_3 \left(2
   H' \right.\right.\right.\\
   & \left.\left.\left. -r \left(3 r^2 H^{(4)}(r)+11 r H^{(3)}(r)+2 H''\right)\right)\right)-2 A H^6 \kappa_3 r^3 \left(A'\right)^2 \left(35 r A''+24 A'\right)+8 A^5 H^3 r \left(H \left(2 r A'' \left(\kappa_2 r^2 \right.\right.\right.\right.\\
   & \left.\left.\left.\left. +3 \kappa_3 \ell
   (\ell+1)\right)+A' \left(4 \kappa_2 r^2+\kappa_3 \ell (\ell+1)\right)\right)-3 r A' H' \left(\kappa_2 r^2+3 \kappa_3 \ell (\ell+1)\right)\right)+A^2 H^4 \kappa_3 r^2 \left(8 r^2
   \left(A'\right)^2 \left(H'\right)^2 \right.\right.\\
   & \left.\left. +H r A' \left(3 r A'' H'+A' \left(4 H'-11 r H''\right)\right)+H^2 \left(14 r^2 \left(A''\right)^2+2 \left(A'\right)^2+r A' \left(27 r A^{(3)}(r)+70 A''\right)\right)\right) \right.\\
   & \left. -A^3 H^3
   \kappa_3 r \left(12 r^3 A' \left(H'\right)^3+H r^2 H' \left(5 r A'' H'-A' \left(15 r H''+4 H'\right)\right)+H^2 r \left(A' \left(3 r^2 H^{(3)}(r)+4 r H'' \right.\right.\right.\right.\\
   & \left.\left.\left.\left. -4 H'\right)+r \left(3 r A^{(3)}(r) H'+4 A'' \left(H'-2 r
   H''\right)\right)\right)+H^3 \left(6 r^3 A^{(4)}(r)+22 r^2 A^{(3)}(r)+4 r A''-4 A'\right)\right)\right\}
\end{aligned}
\end{equation}

\begin{equation}\nonumber
\begin{aligned}
    \zeta_{03} = \frac{1}{4 A^3 H r^2} {} & \left\{8 A^3 \left(2 H \textit{k}_2 r-3 H' \left(\textit{k}_2 r^2+3 \textit{k}_3 \ell (\ell+1)\right)\right)+24 A^2 H A' \left(\textit{k}_2 r^2+3 \textit{k}_3 \ell
   (\ell+1)\right) \right. \\
   & \left. +H \textit{k}_3 \left(-90 r^2 A' \left(H'\right)^2+H r \left(45 r A' H''+H' \left(45 r A''+92 A'\right)\right)-\left(H^2 \left(15 r^2
   A^{(3)}+46 r A''+6 A'\right)\right)\right) \right.\\
   & \left. +A \textit{k}_3 \left(90 r^2 \left(H'\right)^3-2 H r H' \left(45 r H''+46 H'\right)+H^2 \left(15 r^2 H^{(3)}+46 r
   H''+6 H'\right)\right)\right\},
\end{aligned}
\end{equation}

\begin{equation} \nonumber
\begin{aligned}
\zeta_{04} = \frac{1}{A^2 H^2 r^2} {} & \left\{4 A^4 \left(\textit{k}_2 r^2+3 \textit{k}_3 \ell (\ell+1)\right)+A^2 \textit{k}_3 r \left(4 H \left(5 r H''+12 H'\right)-65 r \left(H'\right)^2\right)-25 H^2
   \textit{k}_3 r^2 \left(A'\right)^2 \right.\\
   & \left.-2 A H \textit{k}_3 r \left(10 H r A''+3 A' \left(8 H-15 r H'\right)\right)\right\},
\end{aligned}
\end{equation}

\begin{equation} \nonumber
\zeta_{05} = -\frac{12 A \textit{k}_3 \left(5 H r A'+A \left(2 H-5 r H'\right)\right)}{H^3 r}, \qquad \zeta_{06} = -\frac{16 \textit{k}_3 A^4}{H^4},
\end{equation}
where the prime denotes derivative with respect to $r$, and the coefficients are functions of $r^*$ through the transformation $r(r^*)$ given in \eqref{tortoise_coord}. 

\section{Further details of the Lifshitz dynamics}

In this appendix we show in detail some features of the dynamics of the Lifshitz field. These include the formation of the first oscillations at early times, as well as the case $\kappa_3 = 1$, compared with the choices $\kappa_3=0.01$ and $\kappa_3=0.1$ reported in Section \ref{sec:results}. We also report the results of an independent residual evaluator test, which serves as a consistency check of the numerical method, and complements the convergence studies reported in sections \ref{sec:results} and \ref{sec:conclusions}.

\subsection{Early-time dynamics}

\begin{figure}
\centering
\includegraphics[width=5cm]{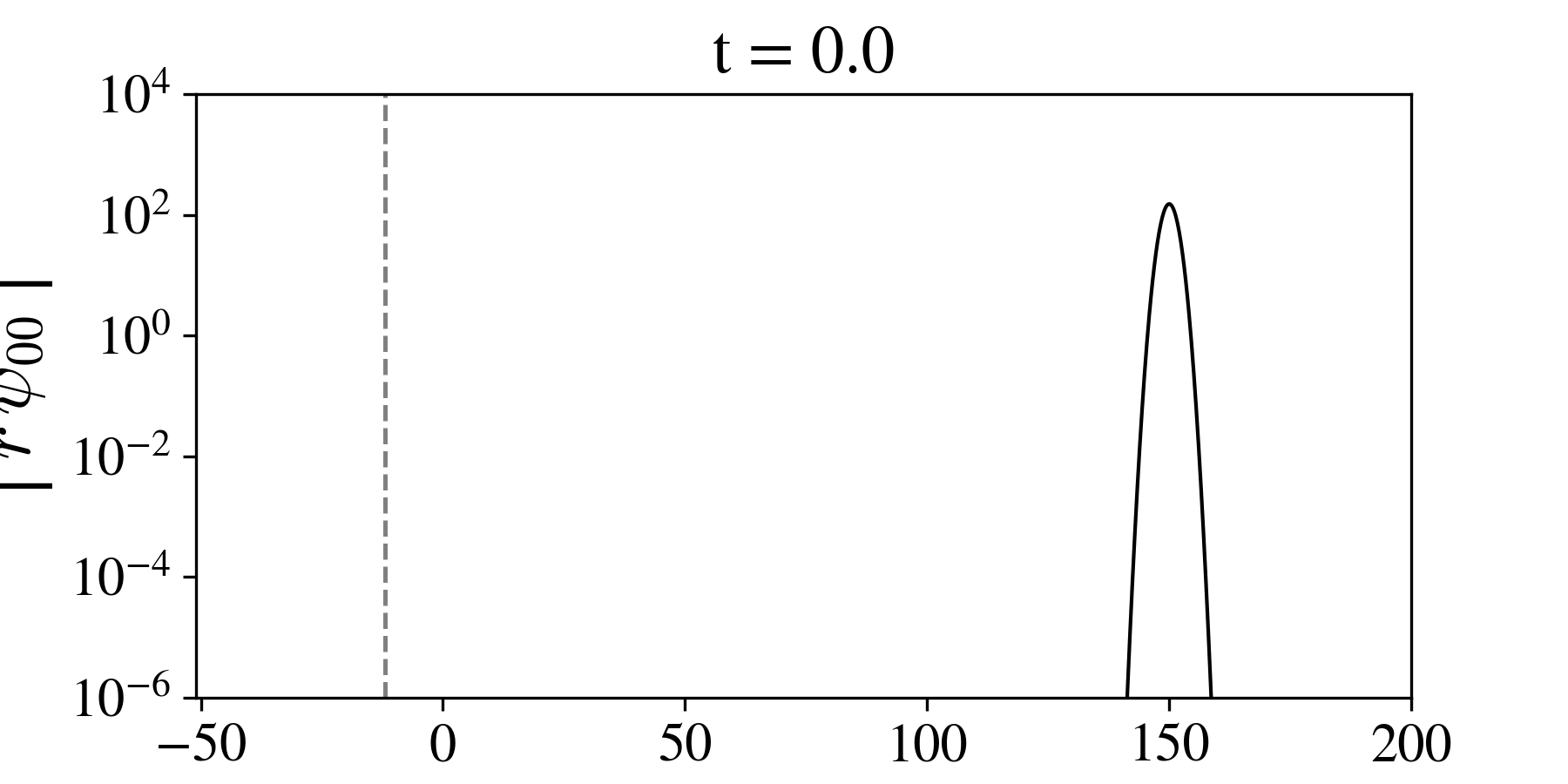}
\includegraphics[width=5cm]{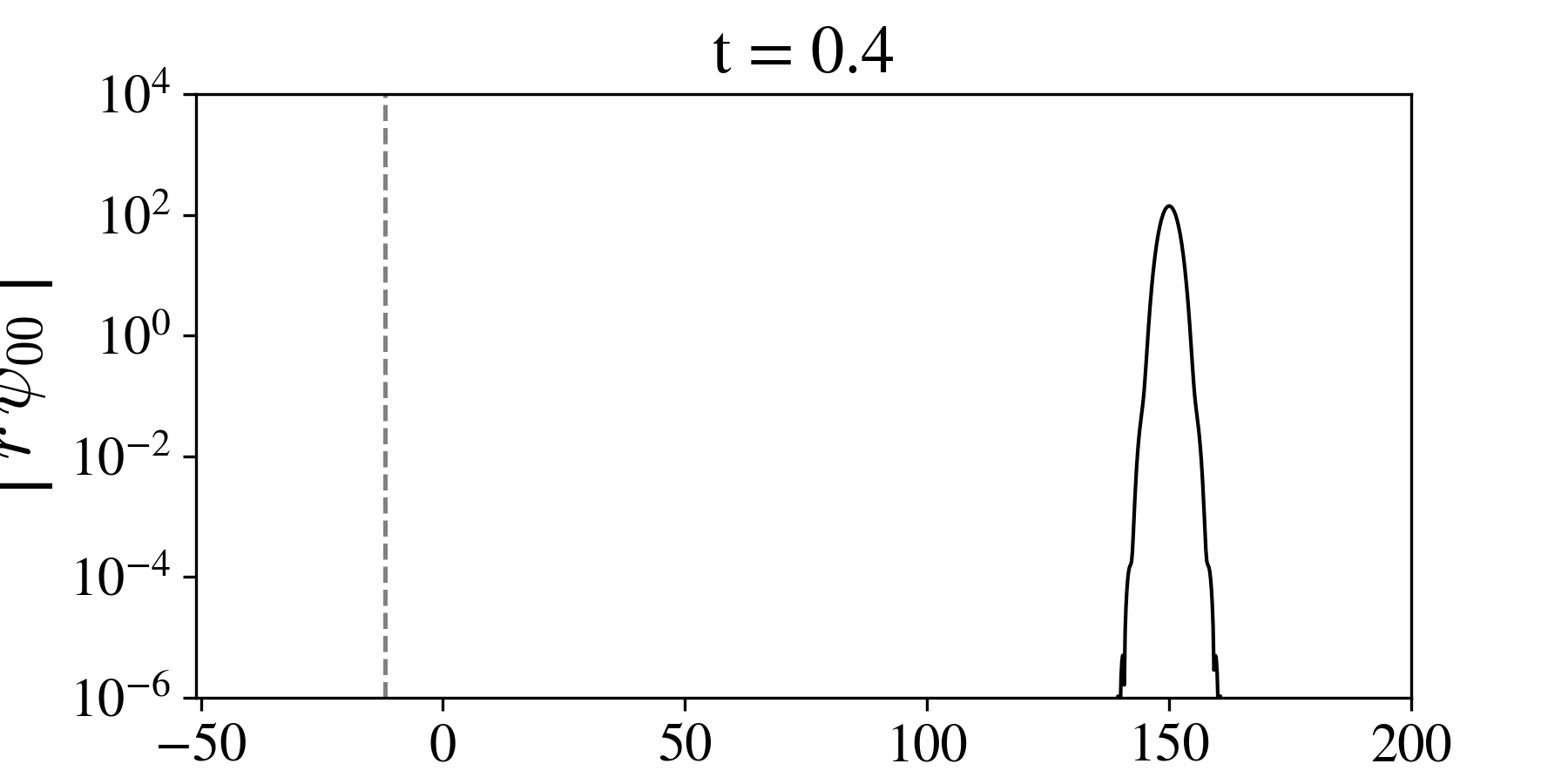}
\includegraphics[width=5cm]{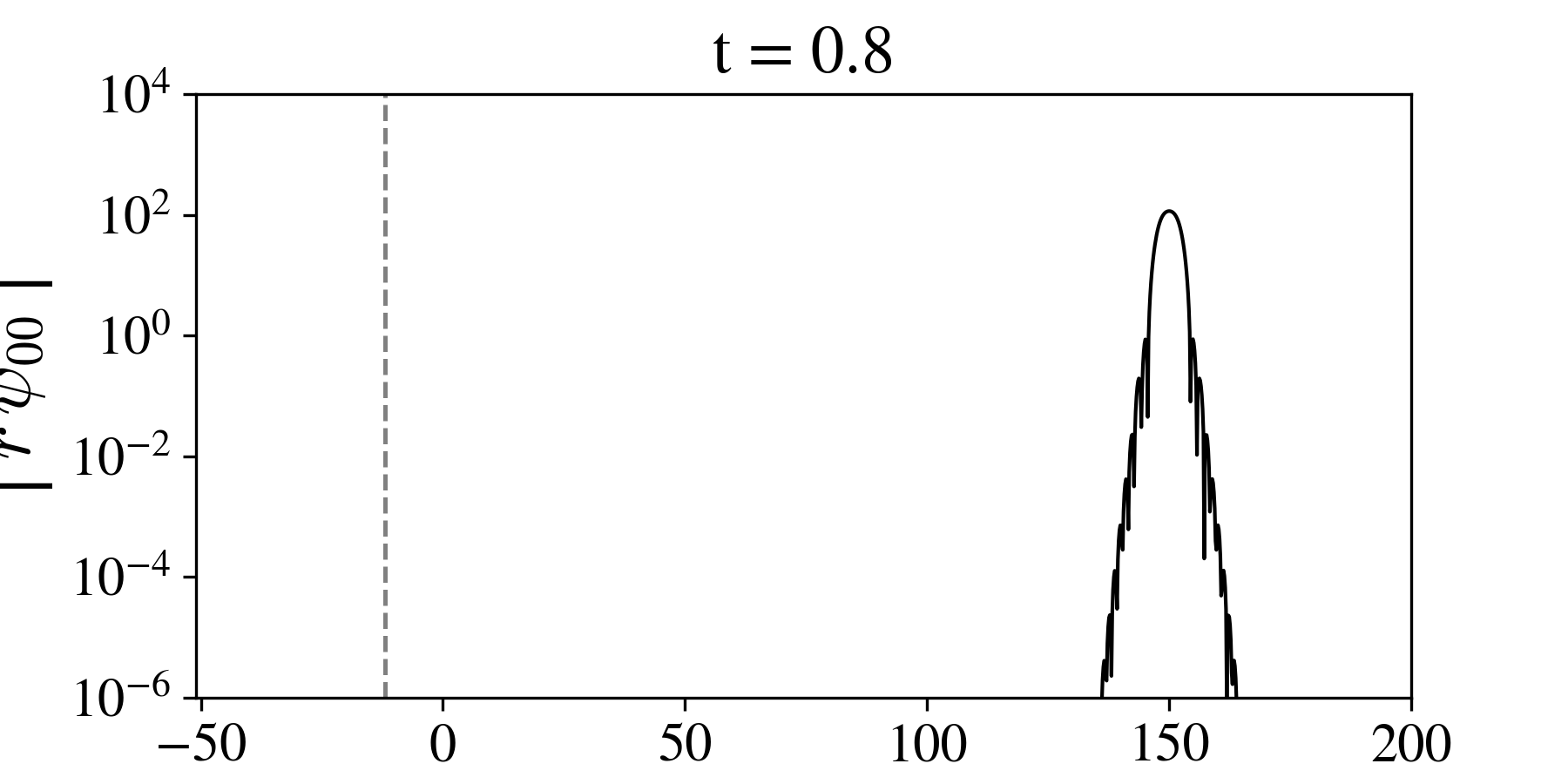}
\includegraphics[width=5cm]{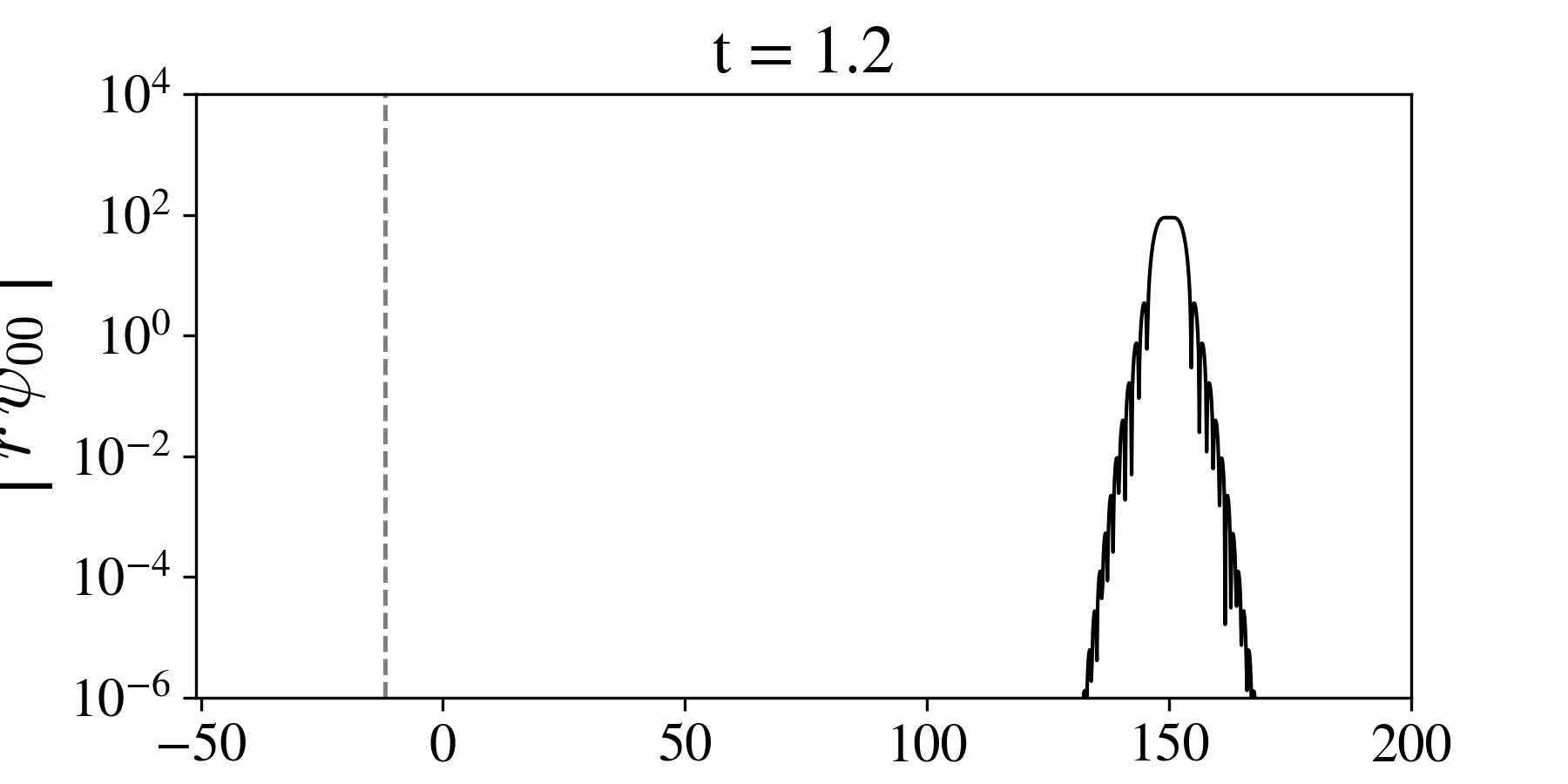}
\includegraphics[width=5cm]{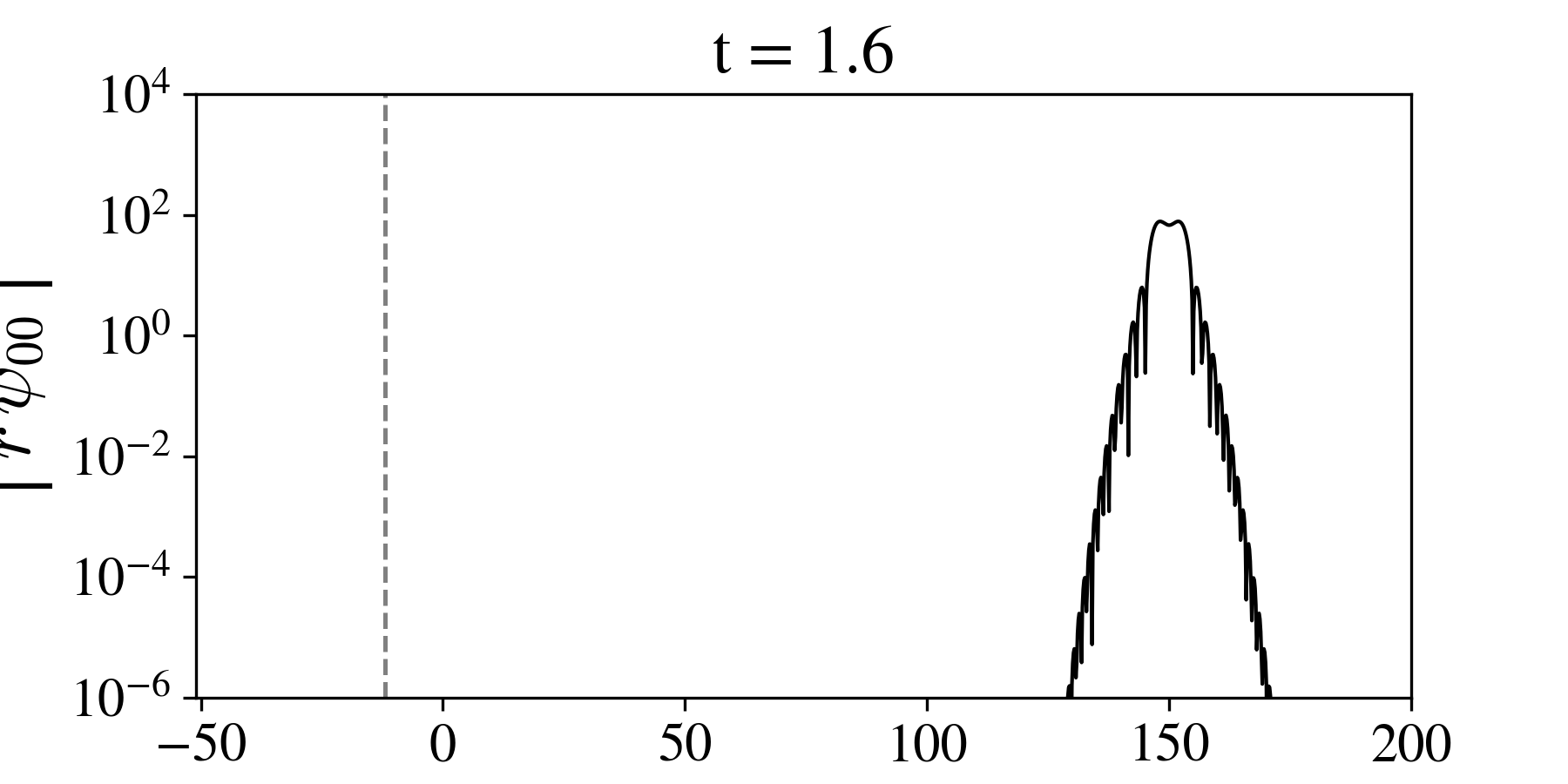}
\includegraphics[width=5cm]{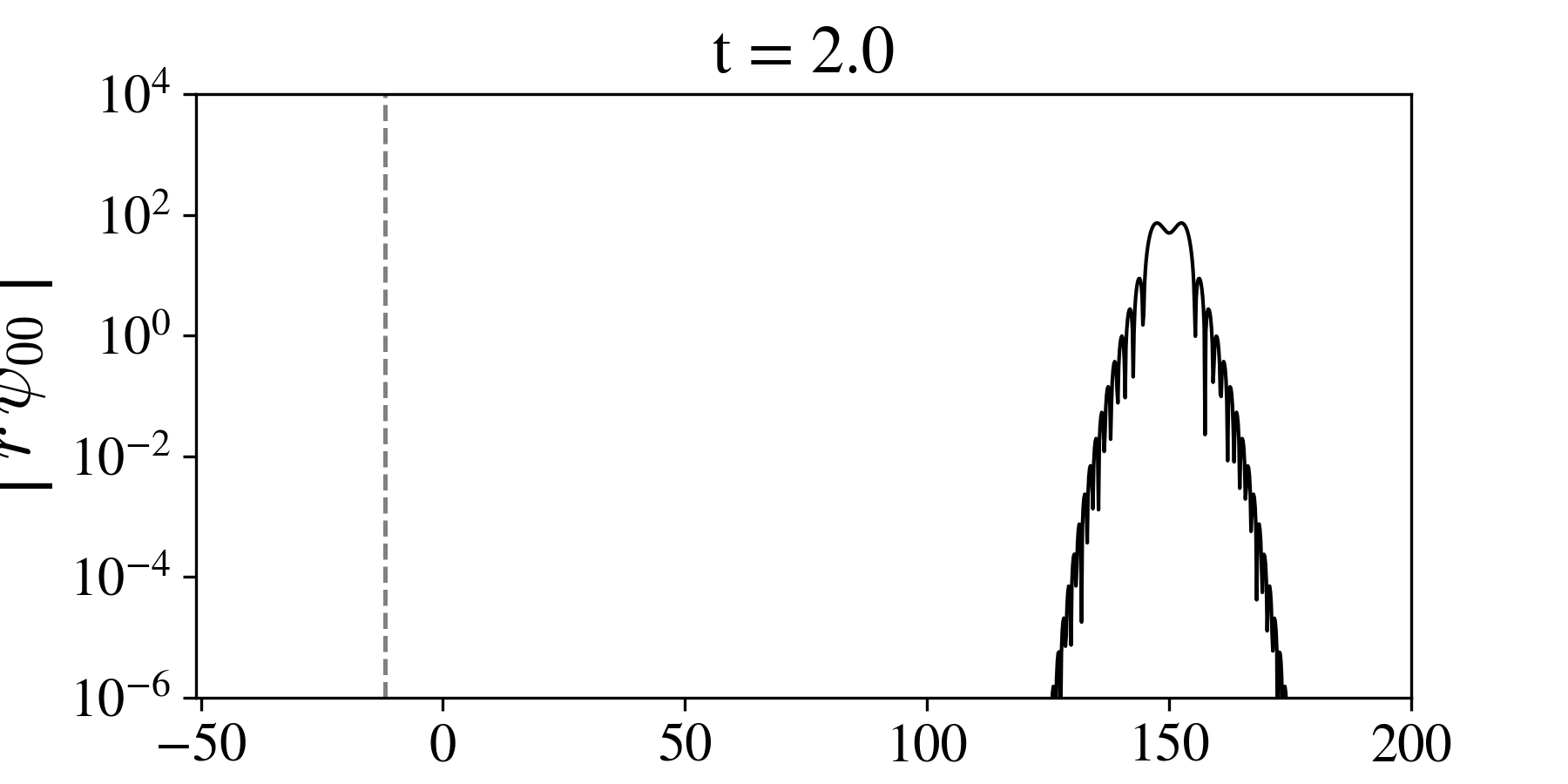}
\includegraphics[width=5cm]{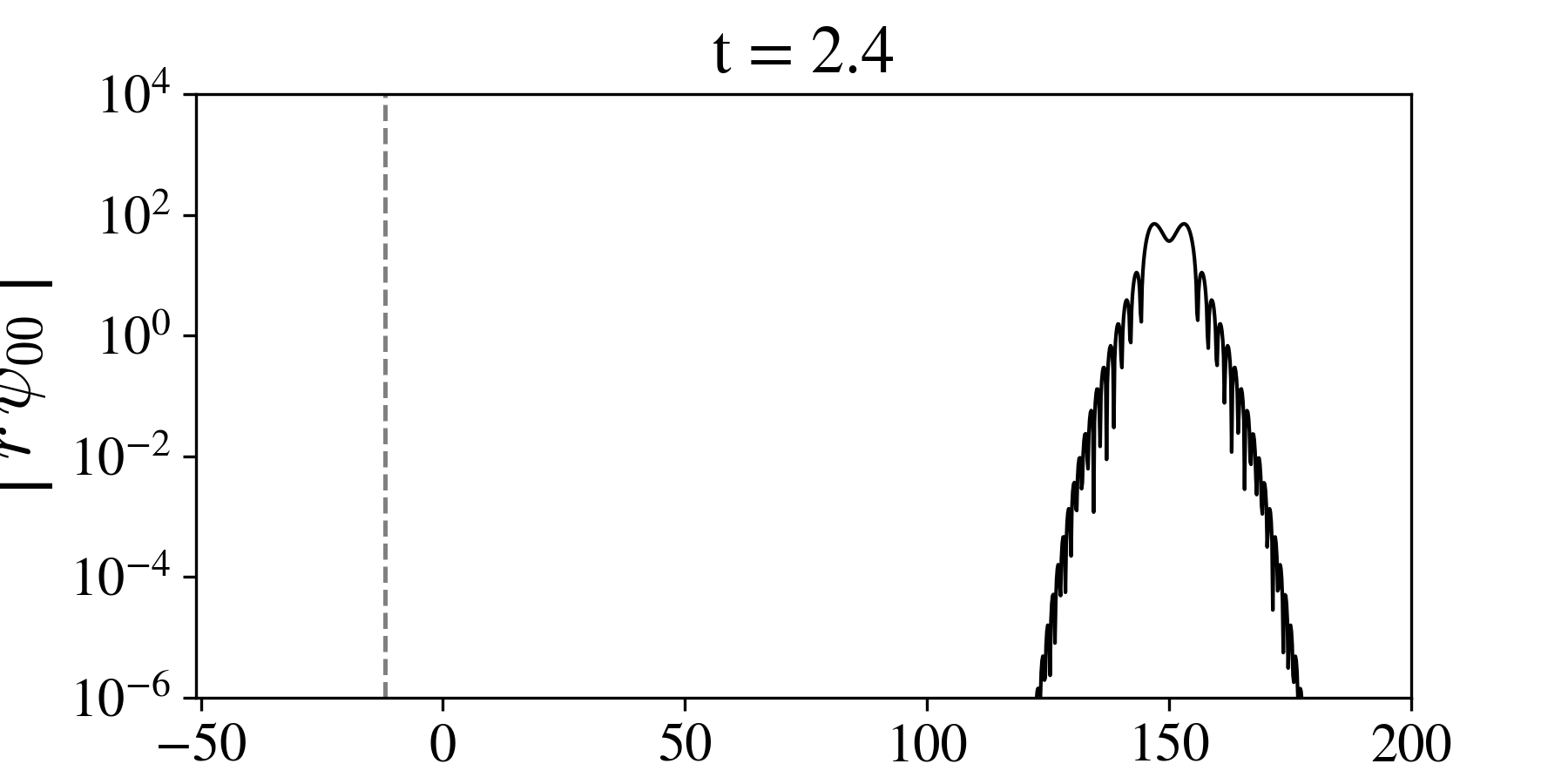}
\includegraphics[width=5cm]{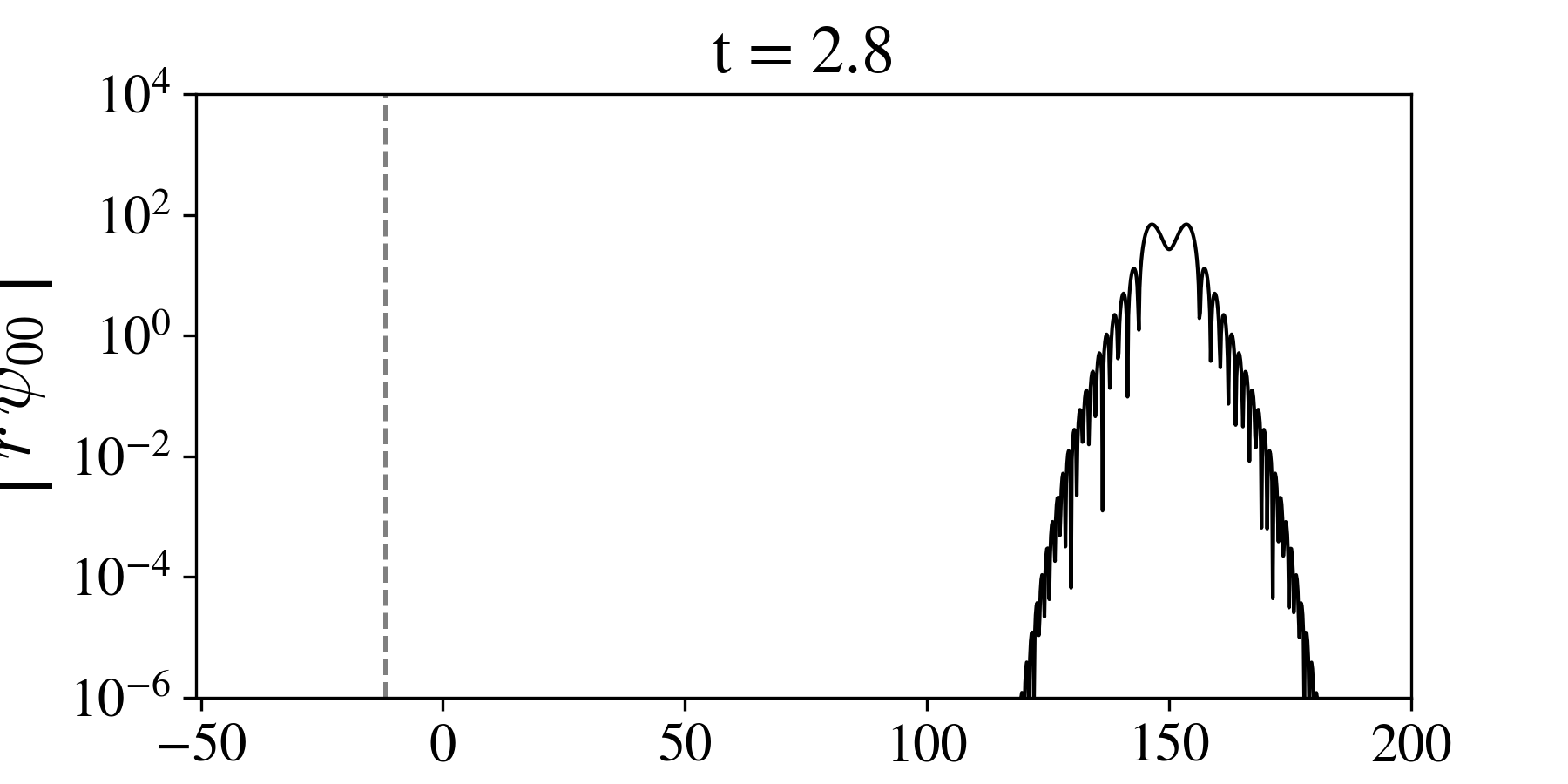}
\includegraphics[width=5cm]{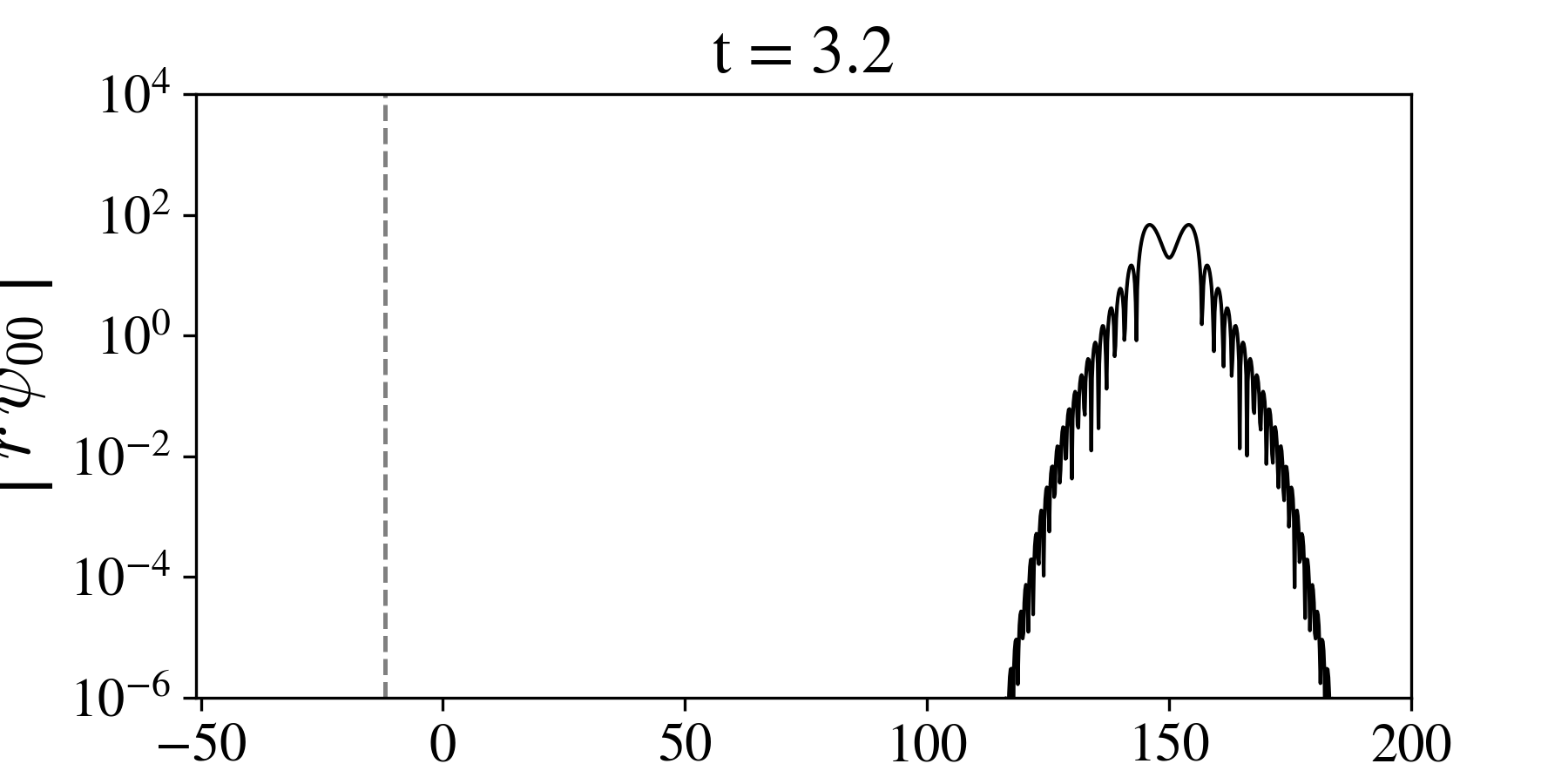}
\includegraphics[width=5cm]{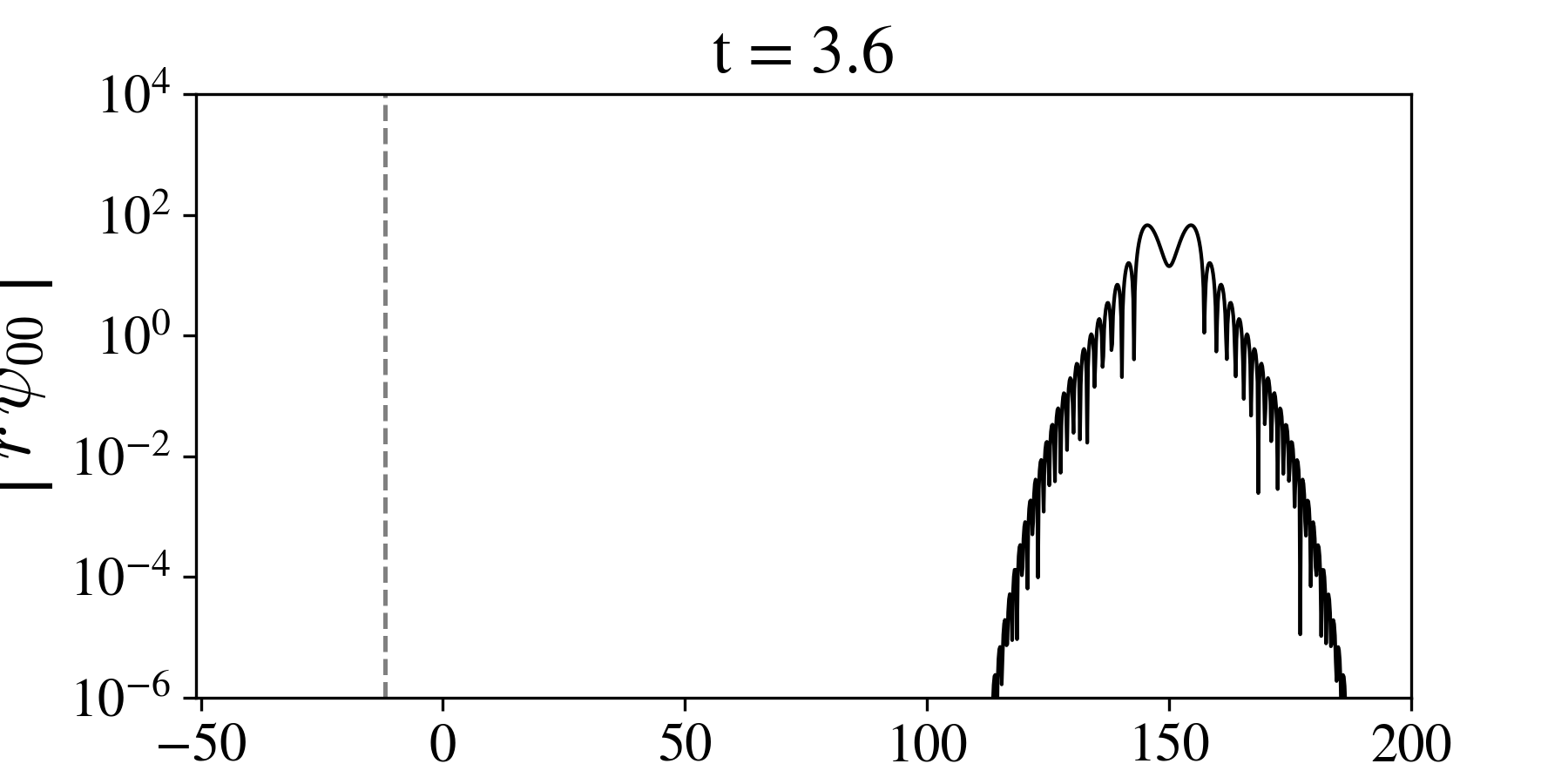}
\includegraphics[width=5cm]{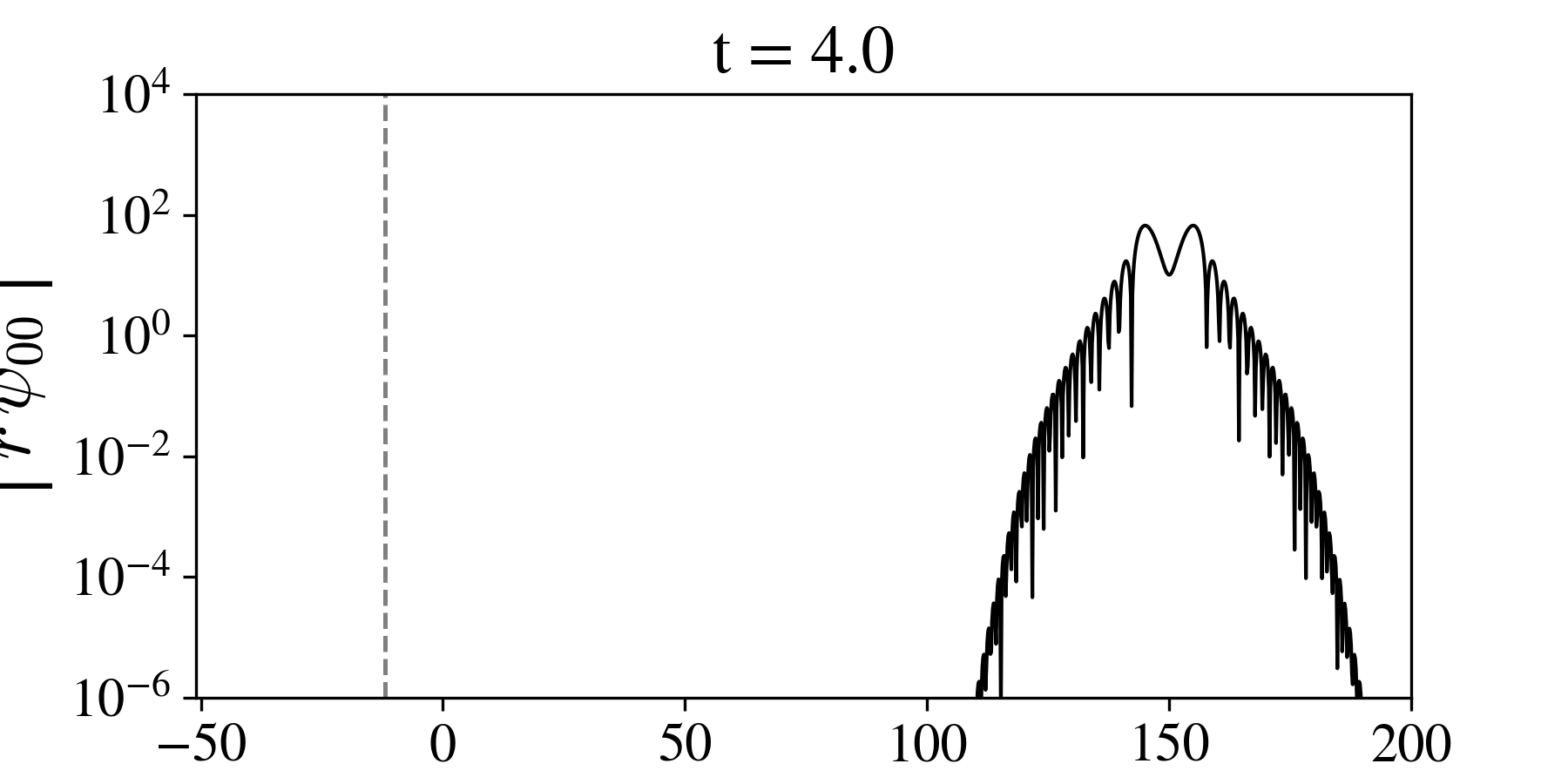}
\includegraphics[width=5cm]{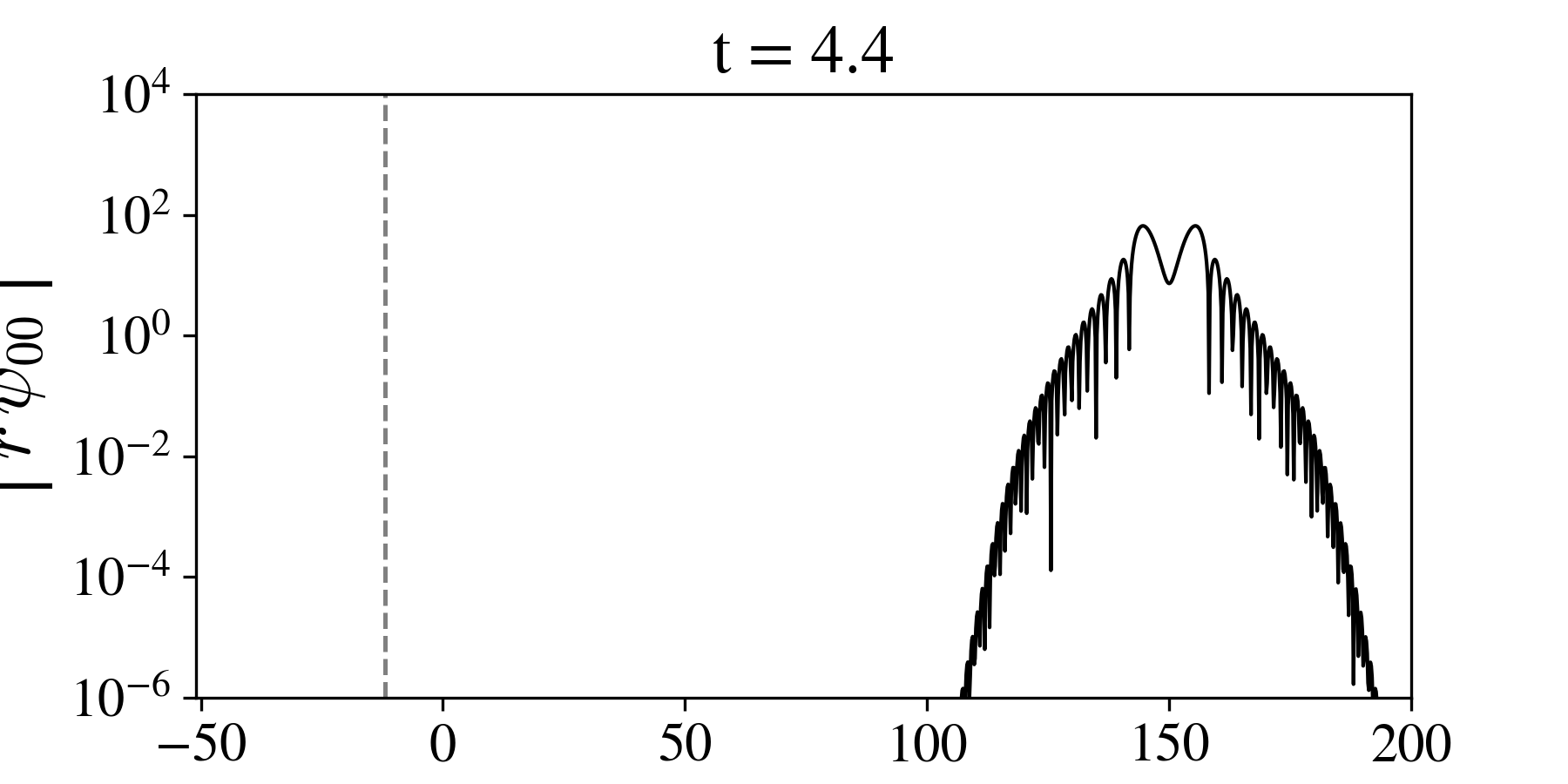}
\includegraphics[width=5cm]{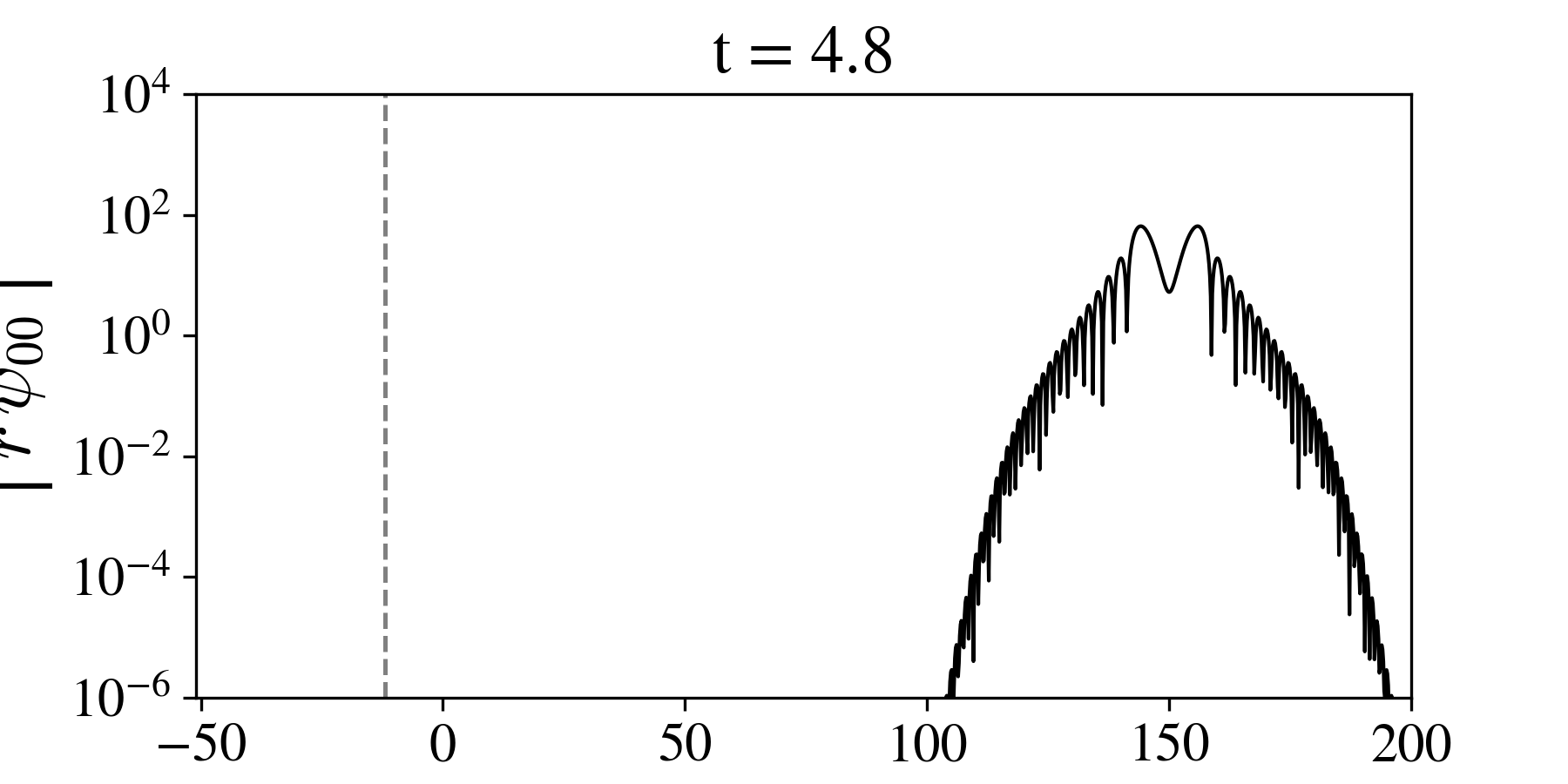}
\includegraphics[width=5cm]{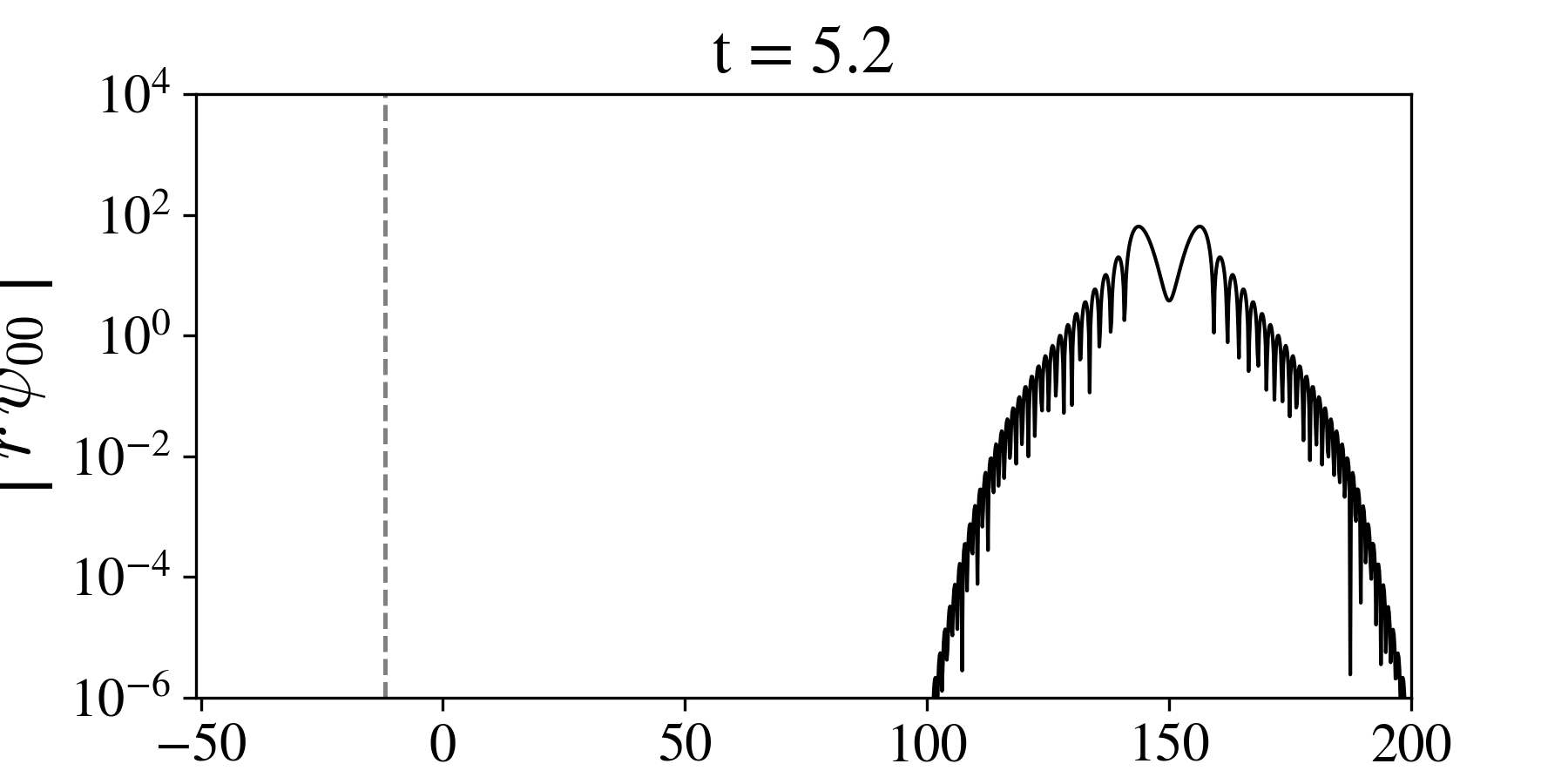}
\includegraphics[width=5cm]{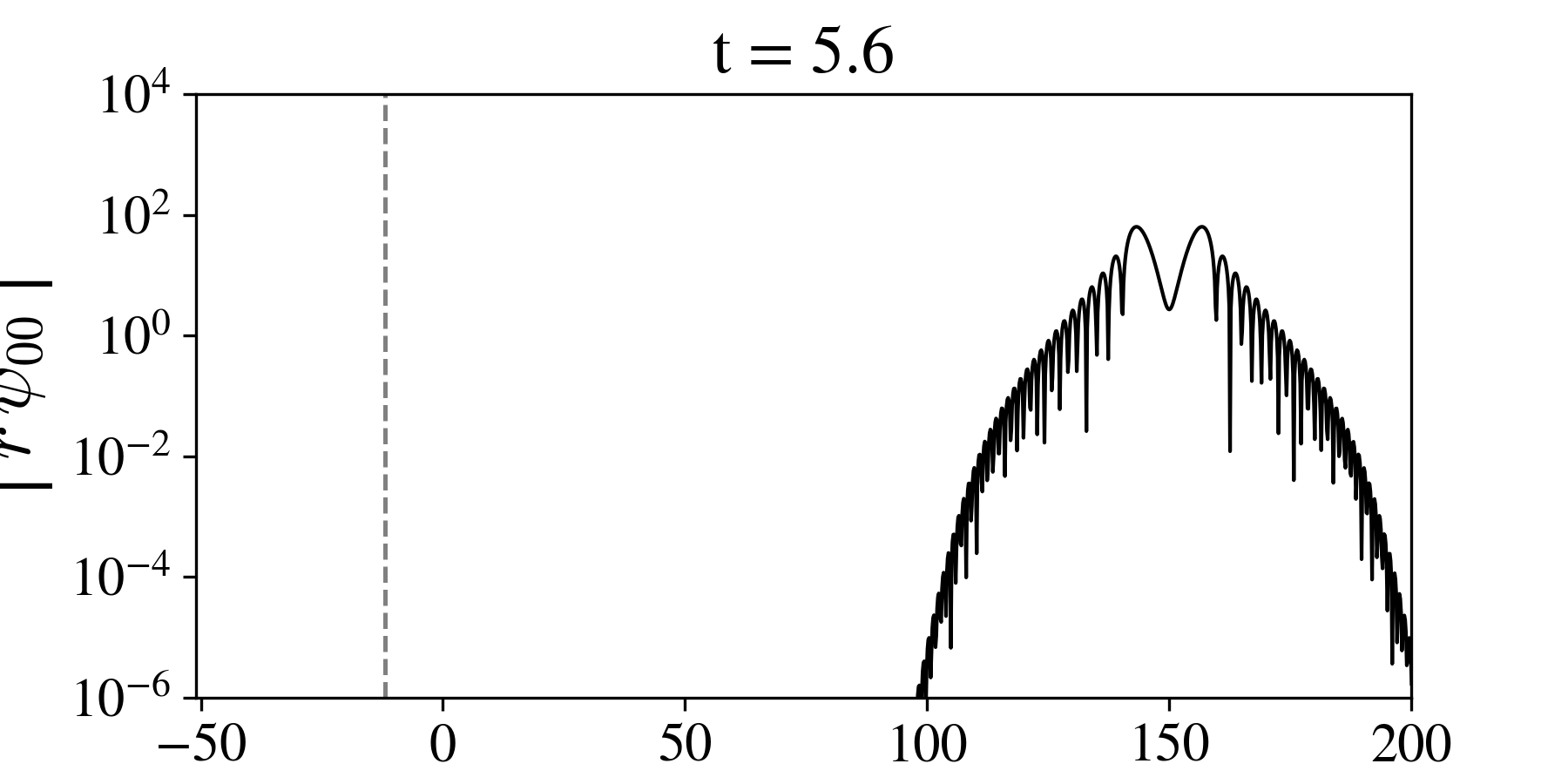}
\includegraphics[width=5cm]{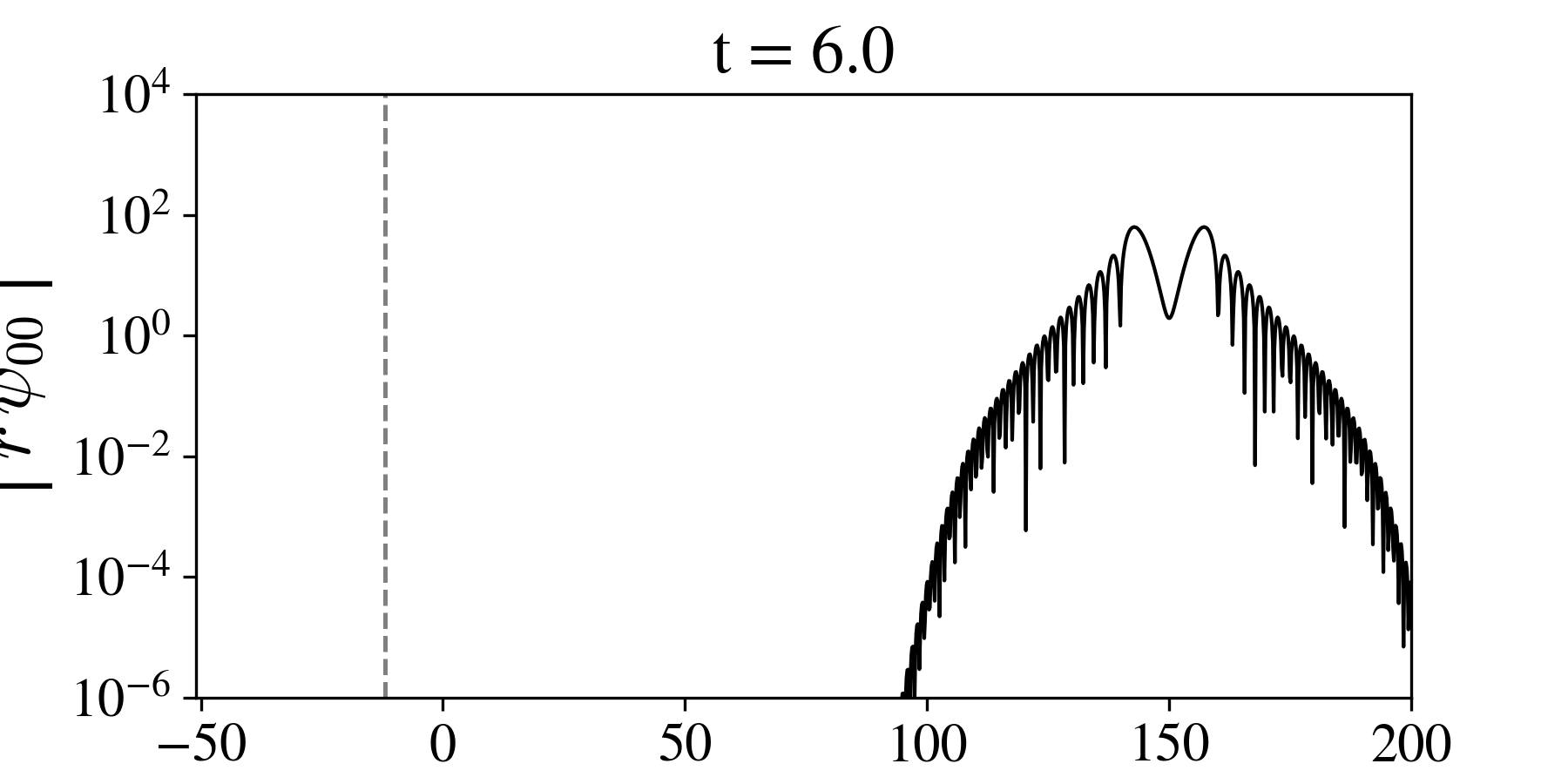}
\includegraphics[width=5cm]{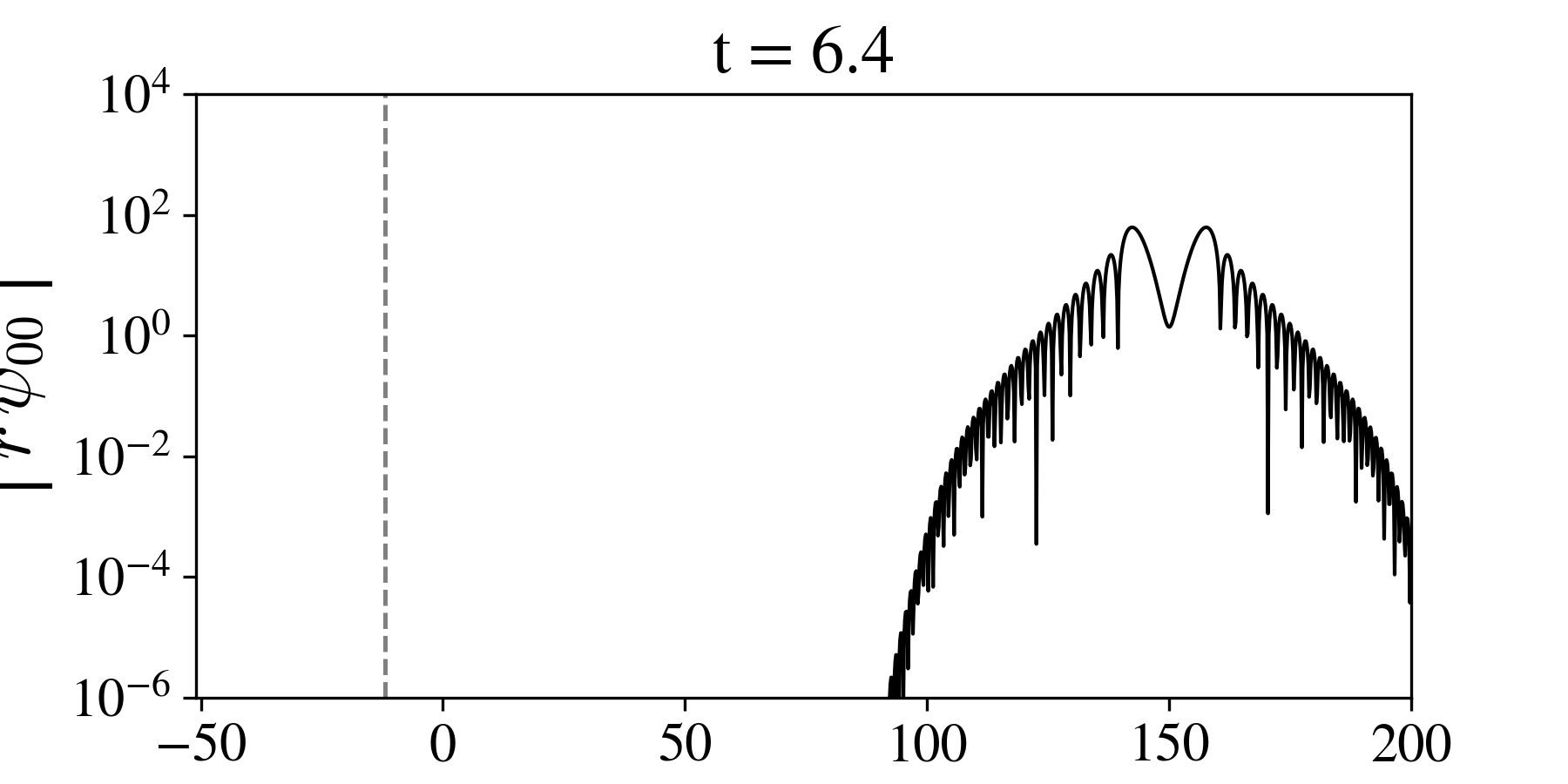}
\includegraphics[width=5cm]{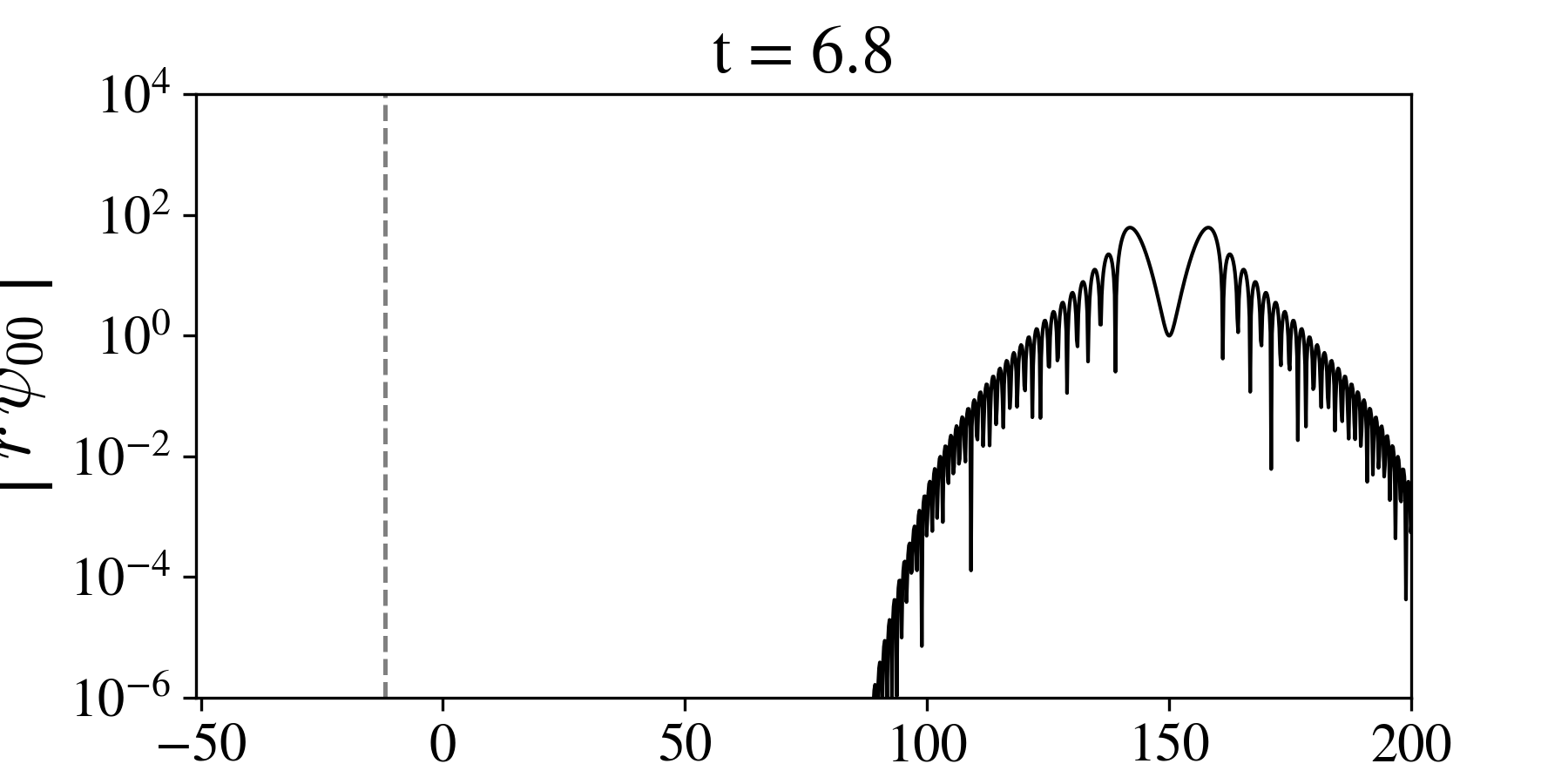}
\includegraphics[width=5cm]{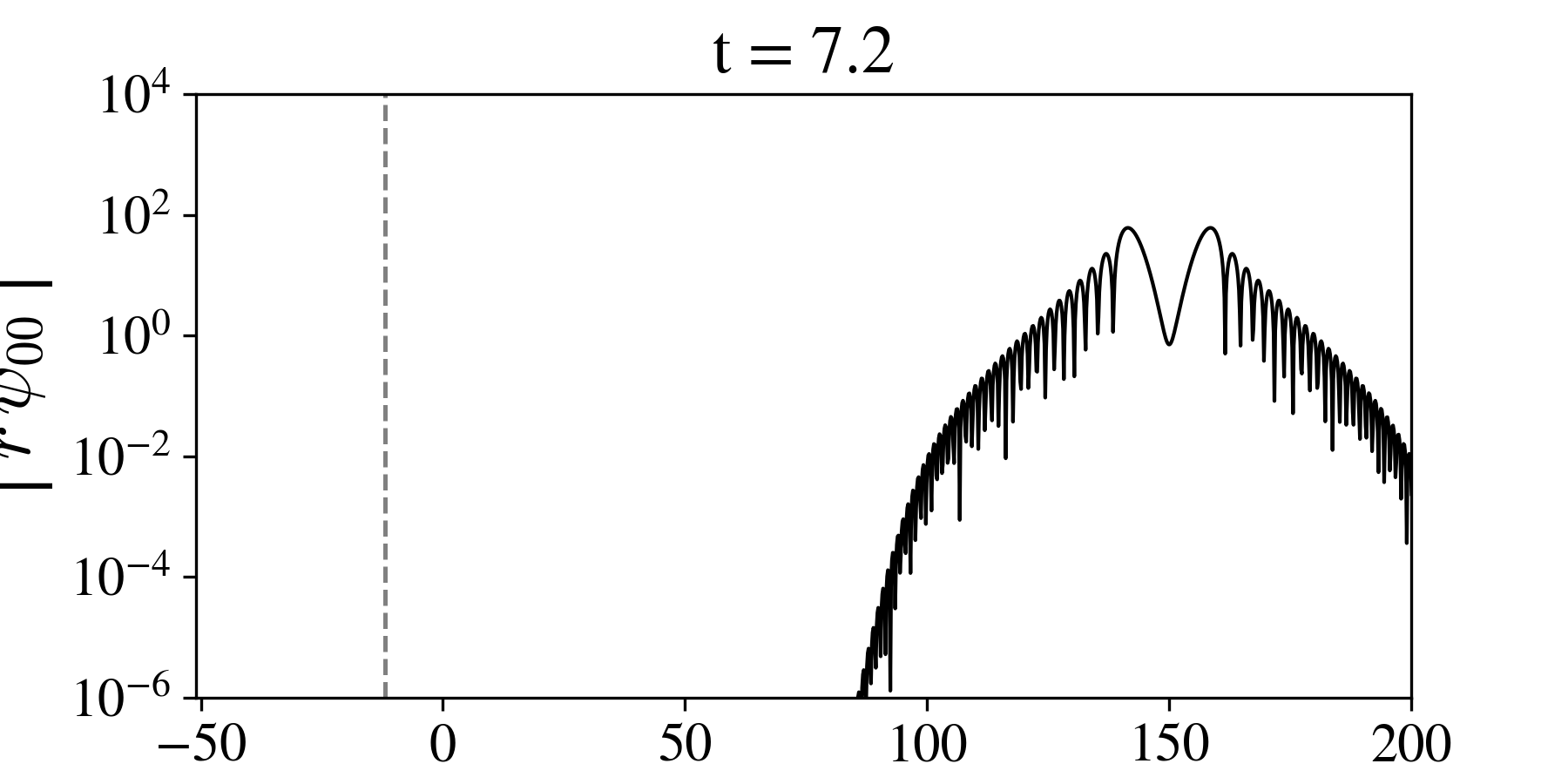}
\includegraphics[width=5cm]{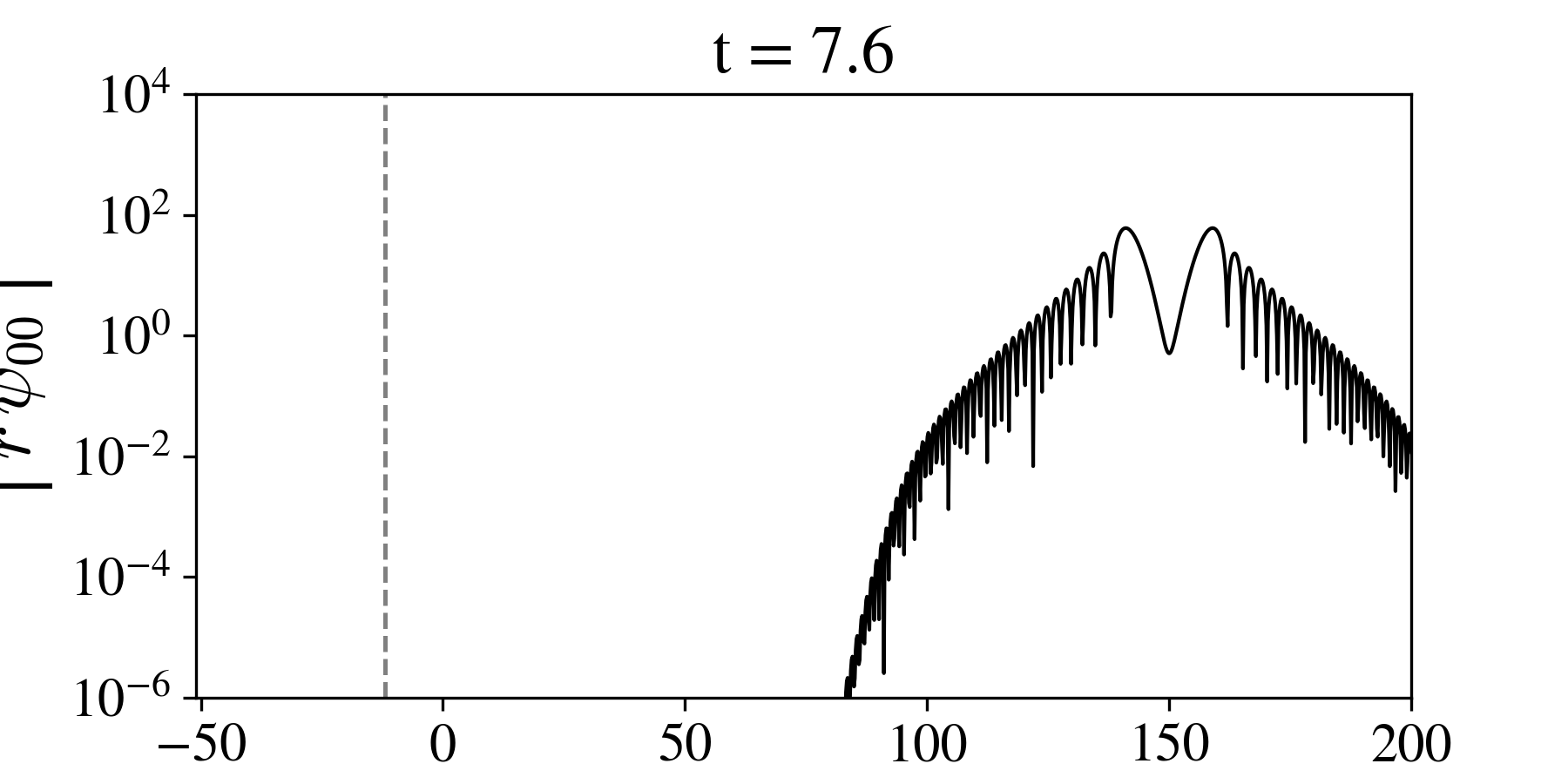}
\includegraphics[width=5cm]{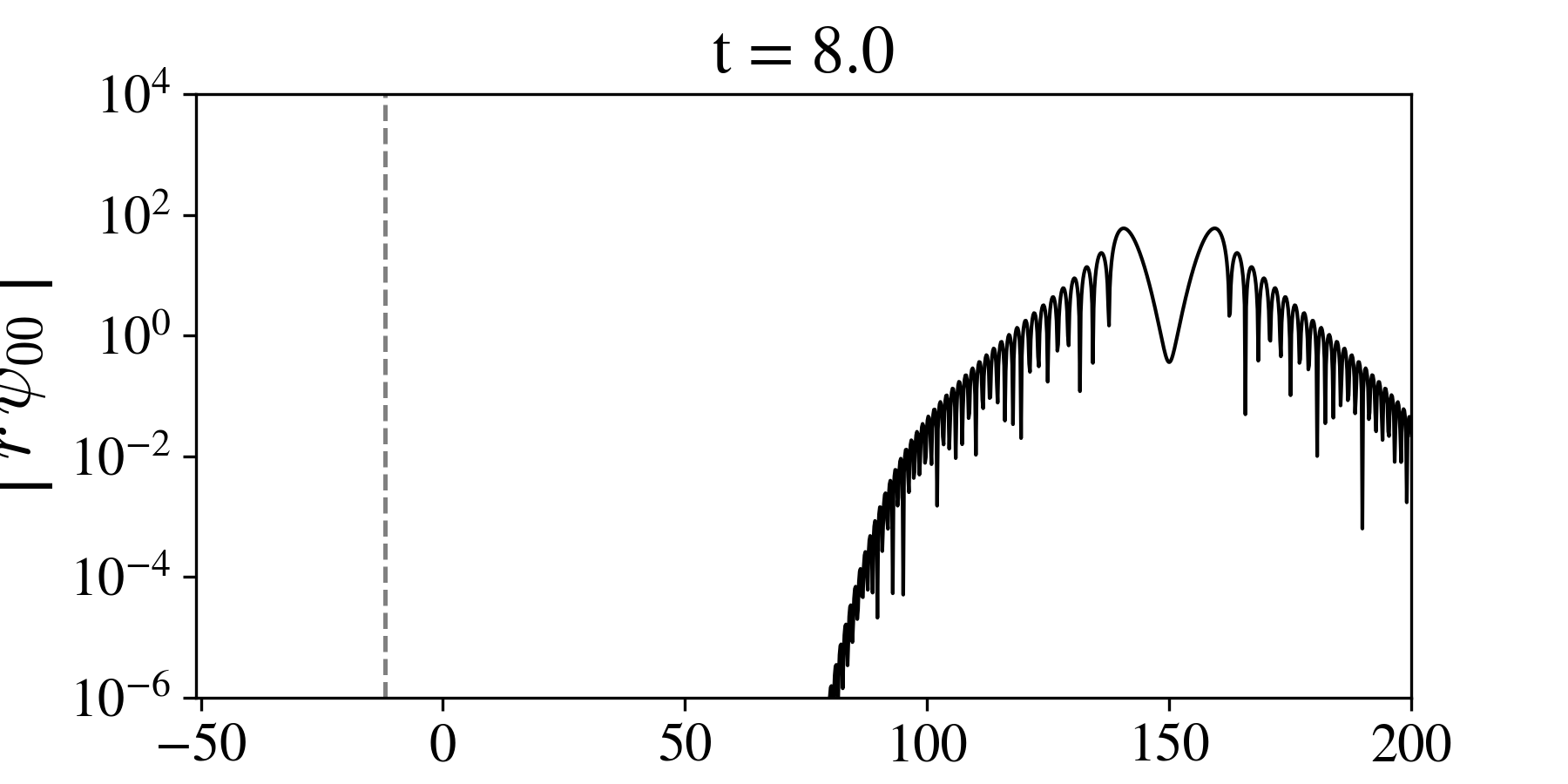}
\caption{\textit{Early-time dynamics.} Lifshitz evolution of the $\ell=0$ mode from $t=0$ to $t=8.0\mu$, for the parameters $\kappa_2=0.1$ and $\kappa_3=0.01$. From $t=0.8\mu$, a ``cascade-like'' effect can be appreciated, as a consequence of the dispersive nature of the evolution equation.}
\label{plot-early}
\end{figure}

In figure \ref{plot-early} we show the early dynamics of the $\ell=0$ mode of the scalar field. There, we can observe the production of modes with different amplitudes and propagation speeds, as a consequence of the dispersive nature of the evolution equation \eqref{eq:equation_final}. Due to the presence of higher spatial derivatives, generic propagating modes exhibit velocities with depend on their wave-number. Indeed, we observe the coexistence of modes with different speeds, which combine to produce a ``cascade'' effect. We also note that the slowest mode present coincides approximately with a solution to the wave equation (i.e. the case $\kappa_2=\kappa_3=0$), which propagates with constant speed.

\subsection{The case $\kappa_3=1$}

\begin{figure}
\centering
\includegraphics[width=8cm]{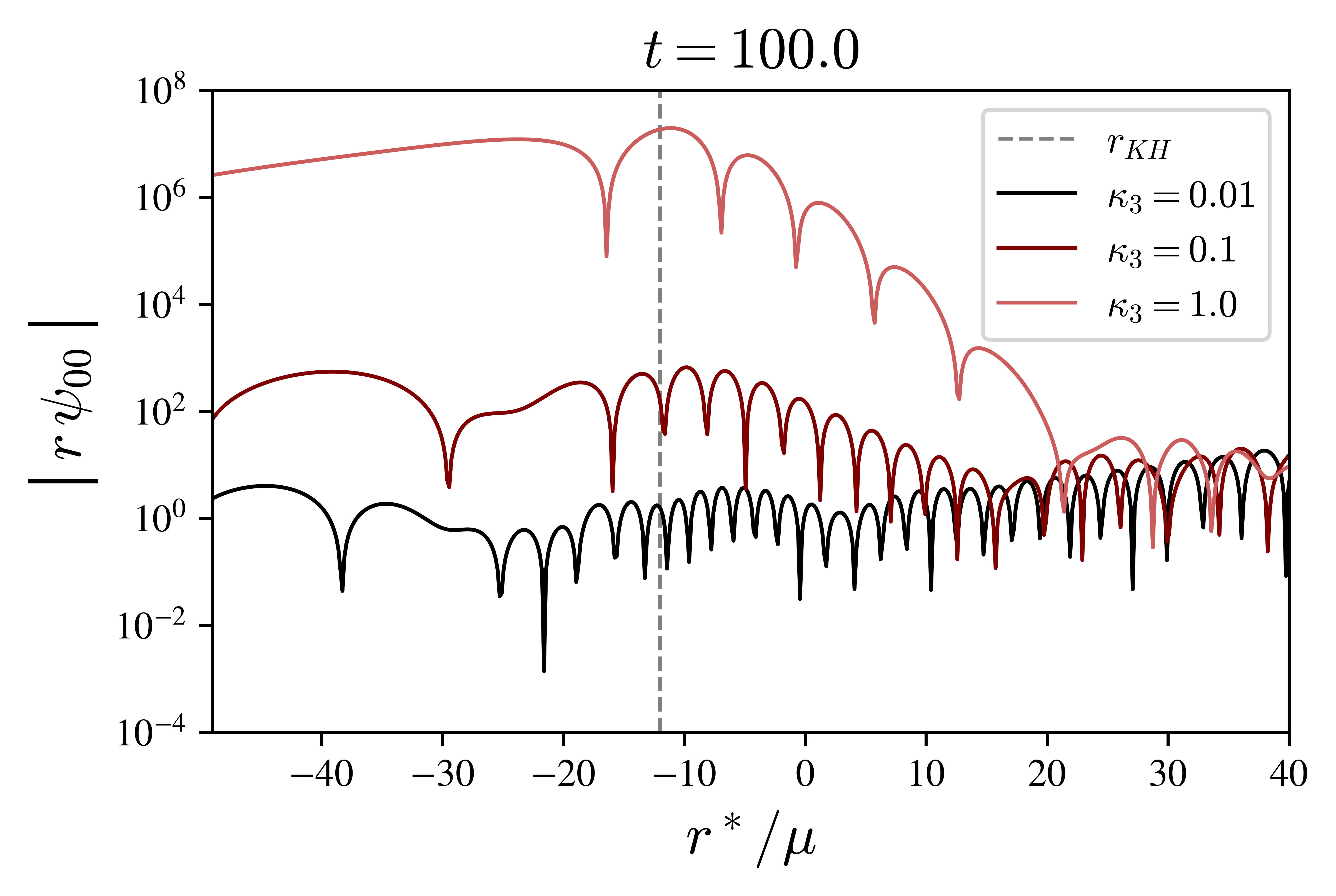}
\caption{\textit{Varying $\kappa_3$.} Snapshot of the evolution for the Lifshitz field at $t=100.0\mu$, for different values of the parameter $\kappa_3$. The growth of the bump around the Killing horizon increases with $\kappa_3$, and the ringing oscillations are still present at large distances.}
\label{plot-vark3}
\end{figure}

In figure \ref{plot-vark3} we display the profile of the scalar field at $t=100\mu$, for $\kappa_2 = 0.1$ and $\kappa_3 =\{0.01,\,0.1,\,1\}$. As  can be seen, the ``bump'' 
between the universal and Killing horizons increases together with $\kappa_3$. One can also observe that the ringing far from the Killing horizon is robust against the choice of this parameter. This rules out the possibility that these features are an artifact of the numerical method presented in Section \ref{sec:numerical}, and supports their presence as genuine physical features of the scalar field evolution.

\subsection{Independent residual evaluator}

As a final consistency check of our implicit scheme, we performed an independent residual evaluator test, by forcing the evolution equation to admit an exact solution $u_{\rm{exact}}(t,r^*)$ at the cost of adding an extra source. Numerically evolving the new ``sourced equation'' with initial data $u_{\rm{exact}}(0,r^*)$ should reproduce the original solution. 

For a given spatial resolution $\Delta r^*$, the numerical approximation $u_{\rm{num}}$ should scale as
\begin{equation}
    u_{\rm{num}} \sim u_{\rm{exact}} + \alpha(\Delta r^*)^p,
\end{equation}
with $p$ the accuracy order of the scheme, and $\alpha$ a scheme-depending coefficient. For two different approximations $u_1$ and $u_2$, with respective resolutions $\Delta r^*_1$ and $\Delta r^*_2$, we thus get
\begin{equation}
    \frac{|u_{\rm{exa}}-u_1|}{|u_{\rm{exa}}-u_2|}\sim \left(\frac{\Delta r^*_1}{\Delta r^*_2}\right)^p.
\end{equation}

\begin{figure}
\centering
\includegraphics[width=8cm]{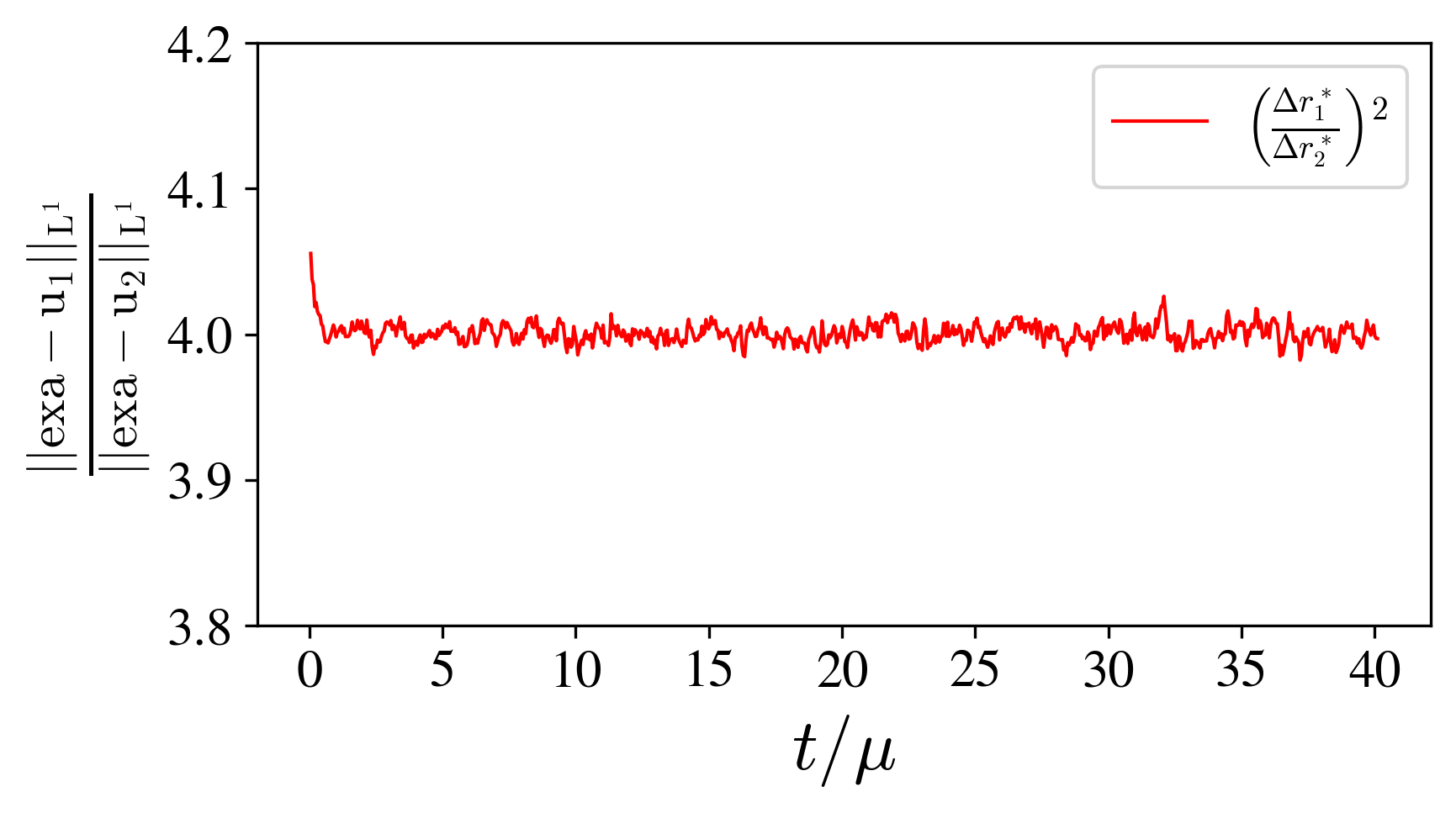}
\caption{\textit{Independent residual evaluator.} Quotient between the difference of the exact solution (exa) with numerical approximations $u_1,\, u_2$ corresponding to $\Delta r^*_1 = 0.01$ and $\Delta r^*_2 = 0.05$. The simulations were done taking $\kappa_2 = 0.1$ and $\kappa_3 = 0.01$.}
\label{plot-ire}
\end{figure}

We verified that the above condition actually holds for our scheme, which is of order $p=2$, by choosing the exact solution
\begin{equation}
    u_{\rm{exact}}(t,r^*) = a(t) \exp{\left[-\frac{(r^*-r^*_c)^2}{\sigma^2}\right]},
\end{equation}
with $a(t)=10e^{-t/100}$, $r^*_c=25$, and $\sigma=5$. We performed numerical integration with resolutions $\Delta r^*_1 = 0.01$ and $\Delta r^*_2 = 0.005$, for which we should get
\begin{equation}
    \left(\frac{\Delta r^*_1}{\Delta r^*_2}\right)^p\sim 4.
\end{equation}
The obtained result is shown in Figure \ref{plot-ire}. As can be easily seen, the numerical approximation converges to the exact solution with the expected convergence rate, ensuring a correct behavior of the numerical scheme, even for small values of $\kappa_3$.

\bibliographystyle{JHEP}
\bibliography{main}{}

\end{document}